\documentclass[aps,preprint,prd,superscriptaddress,showpacs,showkeys,nofootinbib,tightenlines]{revtex4-1}

\usepackage{graphicx}  
\usepackage{amsmath}
\usepackage{bm}  
\usepackage{ulem}
\usepackage{amssymb}
\usepackage{comment}
\usepackage{lineno}
\usepackage{multirow}
\usepackage{makecell}

\newcommand{\bea}{\begin{eqnarray}}
\newcommand{\eea}{\end{eqnarray}}
\newcommand{\beq}{\begin{equation}}
\newcommand{\eeq}{\end{equation}}
\newcommand{\bqa}{\begin{eqnarray}}
\newcommand{\eqa}{\end{eqnarray}}

\begin{document}

\title{
Triangle Singularity\\
 in the Production of $\bm{T_{cc}^+(3875)}$ and a Soft Pion}

\author{Eric Braaten}
\email{braaten.1@osu.edu}
\affiliation{Department of Physics,  The Ohio State University, Columbus, OH\ 43210, USA}

\author{Li-Ping He}
\email{heliping@hiskp.uni-bonn.de}
\affiliation{Helmholtz-Institut f\"ur Strahlen- und Kernphysik and Bethe Center for Theoretical Physics, Universit\"at Bonn, D-53115 Bonn, Germany}

\author{Kevin Ingles}
\email{ingles.27@buckeyemail.osu.edu}
\affiliation{Department of Physics, The Ohio State University, Columbus, OH\ 43210, USA}

\author{Jun Jiang}
\email{jiangjun87@sdu.edu.cn}
\affiliation{School of Physics, Shandong University, Jinan, Shandong 250100, China}

\date{\today}

\begin{abstract}
The double-charm tetraquark meson  $T_{cc}^+(3875)$ can be produced in high-energy proton-proton collisions
by the creation of  the charm mesons $D^{*+} D^0$ at short distances followed by their binding  into  $T_{cc}^+$.
The $T_{cc}^+$ can also be produced by the creation of $D^{*+} D^{*+}$ 
at short distances followed by  their rescattering  into $T_{cc}^+  \pi^+$.
A charm-meson triangle singularity produces a narrow peak in the $T_{cc}^+ \pi^+$ invariant mass distribution
6.1~MeV above the threshold with a width of about 1~MeV.
Well beyond the peak, the differential cross section decreases with the invariant kinetic energy $E$ of $T_{cc}^+ \pi^+$
as $E^{-1/2}$.  The fraction of $T_{cc}^+$  that are accompanied by $\pi^+$ with $E< m_\pi$
is estimated to be roughly 3\%.
The fraction of $T_{cc}^+$ events with  $T_{cc}^+ \pi^+$ in the narrow peak from the triangle singularity 
could be comparable.
\end{abstract}

\smallskip
\maketitle

\section{Introduction}
\label{sec:Introduction}


The discoveries since the beginning of the 21$^\mathrm{st}$ century of dozens of exotic heavy hadrons
not predicted by the quark model have resulted in a second revolution in hadron spectroscopy \cite{Guo:2017jvc,Ali:2017jda,Olsen:2017bmm,Karliner:2017qhf,Yuan:2018inv,Liu:2019zoy,Brambilla:2019esw}.
This second revolution began with the  discovery of the $X(3872)$ 
(also known as $\chi_{c1}(3872)$ or, more concisely, $X$)
by the Belle collaboration in 2003  \cite{Choi:2003ue}.
The $X$ has a remarkably narrow width, and it has other properties consistent with a hidden-charm tetraquark meson.
A new front in the revolution was recently opened up by the discovery 
of the first double-charm tetraquark meson $T_{cc}^+(3875)$ (or, more concisely, $T_{cc}^+$)
by the LHCb collaboration \cite{LHCb:2021vvq}.
The width of $T_{cc}^+$ may be even narrower than that of the $J/\psi$ \cite{LHCb:2021auc},
whose discovery in 1974 launched the first revolution in hadron spectroscopy \cite{E598:1974sol,SLAC-SP-017:1974ind}.
The quark model introduced in 1964 provides a simple explanation 
for the patterns of most of the hadrons discovered in the 20$^\mathrm{th}$ century, 
both light  hadrons and heavy hadrons  that contain  charm and bottom quarks \cite{Gell-Mann:1964ewy,Zweig:1964}.
The development of quantum chromodynamics (QCD) provided a fundamental explanation for these patterns.
The patterns of exotic heavy hadrons discovered in the second revolution are not yet understood.
They present a major challenge to our understanding of QCD.

Until the discovery of $T_{cc}^+$, $X$ was unique among the exotic heavy hadrons not only in its narrow width 
but also in how close it is to the threshold for a pair of hadrons to which it can couple.
The mass of $X$  is extremely close to the $D^{*0} \bar D^0$ scattering threshold.
Recent precise measurements of its energy $\varepsilon_X$ relative the $D^{*0} \bar D^0$ threshold
by the LHCb collaboration give $\varepsilon_X = -0.07 \pm 0.12$~MeV \cite{Aaij:2020qga,Aaij:2020xjx}.
The $J^{PC}$ quantum numbers of $X$ were determined by the LHCb collaboration in 2013 to be $1^{++}$ \cite{Aaij:2013zoa}.
They imply that $X$ has an S-wave coupling to $D^{*0} \bar D^0$.
The universality of near-threshold S-wave resonances 
for particles with short-range interactions is therefore applicable \cite{Braaten:2004rn}.
This remarkable aspect of quantum mechanics guarantees that $X$ has universal properties 
determined by $\varepsilon_X$  \cite{Braaten:2003he}.
The dominant component of the wavefunction of $X$ is a loosely bound charm-meson molecule
with flavor $( D^{*0} \bar D^0 +   D^0 \bar D^{*0} )/\sqrt2$.
If $\varepsilon_X<0$ so that  $X$ is  a bound state, the mean separation of the charm mesons is 
$\langle r \rangle = 1/\sqrt{8 \mu |\varepsilon_X|}$, where $\mu$ is the reduced mass of $D^{*0} \bar D^0$.
The measured value of $\varepsilon_X$ implies $\langle r \rangle > 4.7$~fm at the 90\% confidence level.
Thus the radius of $X$ is probably an order of magnitude larger than that of most hadrons.

The $T_{cc}^+$ is a second exotic heavy hadron to which the
universality of near-threshold S-wave resonances is applicable.
The mass of  $T_{cc}^+$  is extremely close to the $D^{*+} D^0$ scattering threshold.
The energy $\varepsilon_T$ relative to the $D^{*+} D^0$ threshold 
measured by the LHCb collaboration assuming a Breit-Wigner line shape is $-273 \pm 63$~keV \cite{LHCb:2021vvq}. 
The real part $\varepsilon_T$ of the pole energy assuming a line shape that takes into account 
the nearby $D^{*+} D^0$ threshold is \cite{LHCb:2021auc}
\begin{equation}
\varepsilon_T  = -360 \pm 40~\mathrm{keV}.
\label{energyT}
\end{equation}
The analysis by the LHCb collaboration suggests that its $J^P$ quantum numbers are $1^+$.
This implies that $T_{cc}^+$ has an S-wave coupling to $D^{*+} D^0$.
Universality then implies that the dominant component of the wavefunction of  $T_{cc}^+$ 
is a loosely bound charm-meson molecule with flavor $D^{*+} D^0$.
The mean separation of the charm mesons is 
$\langle r \rangle = 1/\sqrt{8 \mu |\varepsilon_T|}$, where $\mu$ is the reduced mass of $D^{*+}D^0$.
The measured value of $\varepsilon_T$ implies $\langle r \rangle = 3.7 \pm 0.2$~fm,
which is almost an order of magnitude larger than  the radius of most hadrons.

Universality identifies the dominant components of the wavefunctions of $X$ and $T_{cc}^+$
to be a loosely bound charm-meson  molecule.
The universal wavefunction has the form $\psi(r) = (1/r) \exp(- \gamma r)$.
This universal wavefunction is applicable only at separations $r$ larger than the size of a charm meson.
Universality says nothing about the wavefunction at shorter distances.
The wavefunction of $X$ at shorter distances has a small $D^{*+} D^- + D^{*-} D^+$ component. 
It could also have a charmonium component ($c \bar c$) or a  compact tetraquark component ($c \bar c q \bar q$).
The wavefunction of $T_{cc}^+$ at shorter distances has a small $D^{*0} D^+$ component. 
It could also have a compact tetraquark component ($c c \bar q \bar q$) 
or a component with $\bar q \bar q$ bound to a heavy diquark ($cc$).

A physicist who is skeptical about the relevance of universality to the $X$ and $T_{cc}^+$ 
could ask for direct experimental evidence for the large size of the loosely bound charm-meson molecule.
One might hope to find evidence for the nature of $X$ and $T_{cc}^+$ from their decays.
However the only decays sensitive to the long-distance wavefunction are 
those with contributions from the decay of a constituent $D^*$ or $\bar D^*$.
In the case of the $X$, the only such decay modes are $D^0 \bar D^0 \pi^0$ and $D^0 \bar D^0 \gamma$.
In the case of the $T_{cc}^+$, the only such decay modes are $D^0 D^0 \pi^+$, $D^+ D^0 \pi^0$, and $D^+ D^0 \gamma$.
There have been several theoretical calculations of the partial decay rates into these three decay modes
\cite{Meng:2021jnw,Ling:2021bir,Fleming:2021wmk,Yan:2021wdl,Ren:2021dsi}.
If the decay rates can be calculated sufficiently precisely, measurements of the three branching fractions
could provide evidence that $T_{cc}^+$ is a loosely bound charm-meson molecule.
There have also been several theoretical calculations of the line shape in the $D^0 D^0 \pi^+$ channel
\cite{Feijoo:2021ppq,Dai:2021wxi,Albaladejo:2021vln,Du:2021zzh}
and the invariant mass distributions for  $D^0 D^0$ and $D^+ D^0$ \cite{Fleming:2021wmk,Du:2021zzh}.
Precise measurements of these distributions
could also provide evidence that $T_{cc}^+$ is a loosely bound charm-meson molecule.

One might also hope to find evidence for the nature of $X$ and $T_{cc}^+$ from their production.
The production of the $T_{cc}^+$ at the LHC has been studied under the assumption that it proceeds by the fragmentation of a $cc$ diquark jet \cite{Qin:2020zlg,Jin:2021cxj} or by the coalescence of $D^*$ and $D$ charm mesons \cite{Jin:2021cxj}.
The production of $X$ and $T_{cc}^+$ in heavy ion collisions may also provide information about their nature
\cite{Cho:2013rpa,MartinezTorres:2014son,Abreu:2016qci,Zhang:2020dwn,Wu:2020zbx,Chen:2021akx,Hong:2018mpk,Fontoura:2019opw,Hu:2021gdg,Abreu:2021jwm}.

One way in which the production  of a hadron can reveal its nature is through {\it triangle singularities}.
A triangle singularity is a kinematic singularity that arises if three virtual particles that form a triangle 
in a Feynman diagram can all be on their mass shells simultaneously \cite{Karplus:1958zz,Landau:1959}.
A triangle singularity can produce a double-log divergence in a reaction rate.
The effects of triangle singularities on the production of exotic heavy mesons
has been studied in Refs.~\cite{Szczepaniak:2015eza,Liu:2015taa}.
A Feynman diagram for the production of a charm-meson molecule can have a triangle in which the vertices are
(a) the creation of two charm mesons at short distances,
(b) a transition between two charm mesons in which a pion or photon is emitted, 
and (c) the coalesence of two charm mesons into the molecule.
The two charm mesons at vertex (c) can both be on shell in the limit as the binding energy goes to 0.
The other charm meson in the triangle can brought on shell by tuning the momentum of the pion or photon
that is emitted at vertex (b).
The log$^2$ divergence in the invariant mass distribution of the molecule and the recoiling  pion or photon 
is smoothed out into a narrow peak by the binding energy of the molecule
and by the decay widths of the charm mesons in the triangle.

The effects of triangle singularities on the production of  $X$ were first studied in Ref.~\cite{Braaten:2019yua,Braaten:2019sxh,Guo:2019qcn}.
Ref.~\cite{Braaten:2019yua} showed that in the exclusive decays of a $B$ meson into $K X \pi$, 
there is a narrow peak in the $X \pi$ invariant mass from a triangle singularity. 
A later study of that reaction  in Ref.~\cite{Sakai:2020ucu}
took into account only one of the three possible Lorentz structures in the short-distance amplitude.
Ref.~\cite{Braaten:2019sxh} showed that in the inclusive prompt production of $X \pi$
at a high-energy hadron collider, there is a narrow peak in the $X \pi$ invariant mass from a triangle singularity.
In both $B$ decay and prompt production, 
the peak in the $X \pi^+$ invariant mass  is predicted to be about 
6.1~MeV above the $X \pi^+$ threshold with a width of about 1~MeV.
The peak in the $X \pi^0$ invariant mass is predicted to be about 7.3~MeV above the $X \pi^0$ threshold.
In Ref.~\cite{Guo:2019qcn}, Guo emphasized that the triangle singularity makes the line shape 
in $X \gamma$ strongly sensitive to the mass of $X$.
In Ref.~\cite{Sakai:2020crh}, the effect of the triangle singularity was studied
 in $e^+ e^-$ annihilation into $X \gamma + \pi^0$ and in $p \bar p$ annihilation into $X \gamma$.
Back in 2006, Dubynskiy and Voloshin pointed out that in $e^+ e^-$ annihilation into $X \gamma$,
there should be a narrow peak at a center-of-mass energy near the $D^{*0} \bar D^{*0}$ threshold \cite{Dubynskiy:2006cj}.
The narrow peak comes from a charm-meson triangle singularity \cite{Braaten:2019gfj}.
The peak is predicted to be at a center-of-mass energy near 4016~MeV with a width of about 5~MeV \cite{Braaten:2019gfj,Braaten:2019gwc}. 
Other studies involving triangle singularities and $X$
have appeared in Refs.~\cite{Molina:2020kyu,Nakamura:2019nwd,Braaten:2020iye}.

In this paper, we study the effects of a charm-meson triangle singularity on the inclusive production
of $T_{cc}^+ \pi$ from the rescattering of $D^* D^*$ created at short distances 
in high-energy hadron collisions, such as proton-proton collisions at the Large Hadron Collider (LHC).
In Section~\ref{sec:Molecule}, we summarize some universal aspects of loosely bound S-wave molecules.
In Section~\ref{sec:XEFT},  we describe the effective field theory XEFT for charm mesons and pions
that is applicable to loosely bound charm-meson molecules.
We give Feynman rules for the double-charm sector relevant to $T_{cc}^+$.
In Section~\ref{sec:Production}, we discuss production of loosely bound charm-meson molecules at a hadron collider.
In the subsequent sections, we apply XEFT to various cross sections at a high energy hadron collider.
In Section~\ref{sec:D*D*}, we consider the production of two charm mesons with small relative momentum
in channels without a resonance near the threshold.
In Section~\ref{sec:D*D,T}, we consider the production of $D^{*+} D^0$ with small relative momentum
and the production of $T_{cc}^+$ without an accompanying soft pion.
In  Section~\ref{sec:Xsoftpi}, we calculate the cross sections for producing $T_{cc}^+\pi^+$ and $T_{cc}^+\pi^0$.
We show that a charm-meson triangle singularity produces
narrow peaks in their invariant mass distributions about 6.1~MeV and 7.3~MeV above their thresholds, respectively.
We summarize our results and discuss their implications in  Section~\ref{sec:Summary}.
In Appendix~\ref{sec:TriAmpLimit}, we determine the charm-meson triangle amplitudes in various limits.
In Appendix~\ref{sec:TriAmpCC}, we give expressions for the triangle amplitudes 
in a coupled-channel model that takes into account the $D^{*0} D^+$ component of $T_{cc}^+$.


\section{Loosely Bound S-wave  Molecules}
\label{sec:Molecule}

If two particles with short-range interactions have an S-wave resonance
extremely close to their scattering threshold,
the few-body physics of those particles has universal aspects that are determined by 
their {\it scattering length} $1/\gamma$ \cite{Braaten:2004rn}. 
In this Section, we describe the universal wavefunction for a bound state extremely close to the scattering threshold,
and we present a coupled-channel model for the wavefunctions at shorter distances.

\subsection{Universal wavefunction}
\label{sec:UniWavefunction}

If the resonance is a bound state with a negative energy $\varepsilon$ relative to 
the scattering threshold,
the inverse scattering length or {\it binding momentum} is 
$\gamma = \sqrt{2 \mu |\varepsilon|}$, where $\mu$ is the reduced mass of the two particles.
The normalized universal  wavefunction of the bound state is
\begin{equation}
\psi(r) = \frac{ \sqrt{\gamma/2\pi}}{r}\exp(- \gamma r).
\label{psi-r}
\end{equation}
This wavefunction diverges at the origin.
The corresponding normalized  momentum-space wavefunction is
\begin{equation}
\psi(k) = \frac{ \sqrt{8 \pi \gamma}}{k^2 + \gamma^2}.
\label{psi-k}
\end{equation}
The spatial wavefunction  at the origin can be expressed as an integral of the momentum-space wavefunction:
$\psi(r\!=\!0) = \int\! d^3k\, \psi(k)/(2\pi)^3 $.
This integral is ultraviolet divergent: 
it can be regularized by imposing a sharp momentum cutoff $|\bm{k}| < (\pi/2)\Lambda$ with $\Lambda \gg \gamma$.
The resulting expression for the wavefunction at the origin, up to corrections that go to 0 as $\Lambda \to \infty$,  is
\begin{equation}
\psi(r\!=\!0) =   \big(\Lambda - \gamma \big)\sqrt{\gamma/2\pi}.
\label{psi-0}
\end{equation}
The ultraviolet cutoff $\Lambda$ can be interpreted as the momentum scale beyond which $\psi(k)$
decreases more rapidly than the prediction $1/k^2$ from the universal wavefunction in Eq.~\eqref{psi-k}.  
The wavefunction at the origin can be used to take into account short-distance components of  the
bound state that are not described explicitly.

The universal aspects of the low-energy scattering of the two particles  can be described by
a simple function of the complex energy $E$ relative to the scattering threshold:
\begin{equation}
f(E) = \frac{1}{-\gamma+\sqrt{-2\mu E }}.
\label{f-E}
\end{equation}
The universal elastic scattering amplitude at relative momentum $k$ 
is obtained by evaluating $f(E)$ at $E=k^2/(2\mu) + i \epsilon$.
By the optical theorem, the  inclusive production rate from the creation of the  two particles at short distances  
 is proportional to the imaginary part of $f(E)$: 
\begin{equation}
\mathrm{Im}[f(E+ i \epsilon)] = \frac{\pi \gamma}{\mu} \delta(E + \gamma^2/2\mu)
+ \frac{\sqrt{2\mu E}}{\gamma^2+2\mu E} \theta(E).
\label{Imf-E}
\end{equation}
The delta function comes from the production of the bound state
and the theta function comes from the  production of the two particles above the threshold. 
Given an ultraviolet cutoff $\Lambda$,
the corresponding energy $\Lambda^2/(2\mu)$ can be interpreted as the energy scale beyond which
the inclusive production rate no longer decreases as $E^{-1/2}$ as predicted by Eq.~\eqref{Imf-E}.

In the case of $T_{cc}^+$, the resonant S-wave channel consists of the charm mesons $D^{*+}D^0$.
The energy $\varepsilon_T$ of $T_{cc}^+$ relative to the $D^{*+}D^0$ threshold is given in Eq.~\eqref{energyT}.
Its binding momentum is $\gamma_T=26.4\pm 1.5$~MeV.
An order-of-magnitude estimate for the ultraviolet cutoff $\Lambda$ is the pion mass $m_\pi$.

\subsection{Model wavefunction at shorter distances}
\label{sec:WavefunctionSD}

A sharp ultraviolet cutoff on the momentum $\bm{k}$ gives unphysical results for some observables.
A simple model that is equivalent to a smooth ultraviolet cutoff can be defined by the 
normalized momentum-space wavefunction
\begin{equation}
\psi^{(\Lambda)}(k) = \frac{ \sqrt{8 \pi (\Lambda+\gamma)\Lambda\gamma}}{\Lambda-\gamma}
\left(  \frac{1}{k^2 + \gamma^2} -  \frac{1}{k^2 + \Lambda^2} \right) .
\label{psi-k:reg}
\end{equation}
This regularized wavefunction was first applied to $X(3872)$ by Suzuki \cite{Suzuki:2005ha}.
Its leading behavior at large $k$ is
\begin{equation}
 \psi^{(\Lambda)}(k)\longrightarrow \sqrt{8 \pi (\Lambda+\gamma)^3\Lambda\gamma}/k^4.
\label{psireg-largek}
\end{equation}
The spatial wavefunction at the origin  is
\begin{equation}
 \psi^{(\Lambda)}(r\!=\!0) =  \sqrt{(\Lambda+\gamma)\Lambda\gamma/2\pi}.
\label{psi-0:reg}
\end{equation}
The sharp cutoff $|\bm{k}| < (\pi/2)\Lambda$ used to calculate $\psi(r\!=\!0)$ in Eq.~\eqref{psi-0}
was chosen so it would have the same limit  for $\Lambda \gg \gamma$ as 
$\psi^{(\Lambda)}(r\!=\!0)$ in Eq.~\eqref{psi-0:reg}.

The regularized wavefunction in Eq.~\eqref{psi-k:reg}  is at best a model with the same momentum dependence 
as the universal wavefunction $\psi(k)$ at small $k$ and more physical qualitative behavior at large $k$.
In this model, the coefficient of $1/k^4$ in Eq.~\eqref{psireg-largek} 
and the wavefunction at the origin in Eq.~\eqref{psi-0:reg} are both determined by the same parameter $\Lambda$.
In general, there is no simple relation between these two quantities.
Sensitivity to $\Lambda$ in this model can reveal
aspects of a problem that are sensitive to momenta much larger than $\gamma$.
If the momentum scale where the EFT breaks down is identified, 
the model can be used to estimate the order of magnitude of short-distance effects by replacing $\Lambda$ by that momentum scale.

\subsection{Model wavefunction for coupled channel}
\label{sec:WavefunctionCC}

There could be another S-wave channel coupled to the resonant channel 
that has a scattering threshold higher by an energy $\delta$.
In this case, the bound state will also have a component in the coupled channel with a smaller probability.
For simplicity, we consider the case of a coupled channel consisting of particles with the same masses
and a symmetry relating the two channels that is broken by the energy difference $\delta$. 
We assume the symmetry requires the wavefunctions in the two channels to be  equal at short distances.
Note that this condition is not identical to requiring the wavefunctions in the two channels 
to be equal at large momenta.

The binding momentum for the coupled channel is $\gamma_\mathrm{cc} = \sqrt{2 \mu(\delta + |\varepsilon|)}$.
A simple model for the coupled-channel wavefunction  is 
\begin{equation}
\psi_\mathrm{cc}(k) =
\frac{\Lambda - \gamma}{\Lambda - \gamma_\mathrm{cc}} \,  \frac{ \sqrt{8 \pi \gamma}}{k^2 + \gamma_\mathrm{cc}^2}.
\label{psicc-k}
\end{equation}
We have chosen its normalization 
so that the wavefunction at the origin defined by a sharp ultraviolet cutoff $|\bm{k}| < (\pi/2)\Lambda$
is equal to that for the resonant channel in Eq.~\eqref{psi-0}: $\psi_\mathrm{cc}(r\!=\!0) = \psi(r\!=\!0)$.
Note that this symmetry condition at short distances is not equivalent to requiring 
$\psi_\mathrm{cc}(k)$ to approach $\psi(k)$ at large $k$ except in the limit $\Lambda \to \infty$.

An alternative model for the coupled-channel  wavefunction that corresponds to a smooth ultraviolet cutoff is
\begin{equation}
\psi^{(\Lambda)}_\mathrm{cc}(k) = 
\frac{ \sqrt{8 \pi(\Lambda+\gamma)\Lambda\gamma}}{\Lambda-\gamma_{cc}}
\left( \frac{1}{k^2 +  \gamma_\mathrm{cc}^2}- \frac{1}{k^2 + \Lambda^2} \right).
\label{psicc-k:reg}
\end{equation}
We have chosen its normalization 
so that the wavefunction at the origin is equal to that for the resonance channel in Eq.~\eqref{psi-0:reg}:
\begin{equation}
\psi^{(\Lambda)}_\mathrm{cc}(r\!=\!0) = \psi^{(\Lambda)}(r\!=\!0).
\label{psicc:r=0}
\end{equation}
This condition could be required by a symmetry between the two channels at short distances.
The relative probability for the coupled-channel  wavefunction  is
\begin{equation}
Z_\mathrm{cc} \equiv \int \frac{d^3k}{(2\pi)^3} \big|\psi^{(\Lambda)}_\mathrm{cc}(k)\big|^2  = 
\frac{(\Lambda+\gamma)\gamma}{(\Lambda+\gamma_\mathrm{cc})\gamma_\mathrm{cc}}.
\label{prob0+:reg}
\end{equation}
This is less than 1 provided $\gamma < \gamma_\mathrm{cc}$.

The coupled-channel  wavefunction $\psi^{(\Lambda)}_\mathrm{cc}(k)$ in Eq.~\eqref{psicc-k:reg}
can be used in conjunction with the regularized wavefunction $\psi^{(\Lambda)}(k)$ in Eq.~\eqref{psi-k:reg}
as a qualitative model for the bound state in which these two components are described explicitly
and all others are taken into account through the wavefunction at the origin.
The total probability in the two channels can be normalized to 1
by multiplying both $\psi^{(\Lambda)}(k)$ in Eq.~\eqref{psi-k:reg} 
and $\psi^{(\Lambda)}_{cc}(k)$ in Eq.~\eqref{psicc-k:reg} by $1/\sqrt{1+Z_\mathrm{cc}}$.

In the case of $T_{cc}^+$, the coupled S-wave channel consists of the charm mesons $D^{*0}D^+$.
We sometimes denote this coupled channel simply by 0+.
We will assume that at short distances the resonance  is in the isospin-0 combination  $(D^{*+} D^0 - D^{*0} D^+)/\sqrt2$ of the two coupled channels.
This is consistent with the observation by the LHCb collaboration 
of a peak near threshold in the $D^0 D^+$ invariant mass distribution, 
which can come from the $D^{*0} D^+$ component of $T_{cc}^+$.
The possibility that the resonance has isospin 1 is disfavored
by the nonobservation of peaks in the $D^+ D^+$ and $D^+ D^0 \pi^+ $ invariant mass distributions, 
which could come from $D^{*+} D^+$.
In many of the analyses of the decays of $T_{cc}^+$,
the resonance was assumed to be a linear combination of isospin 0 and isospin 1
\cite{Meng:2021jnw,Feijoo:2021ppq,Yan:2021wdl,Fleming:2021wmk}.
In Ref.~\cite{Feijoo:2021ppq}, a fit to the $D^0 D^0 \pi^+$ energy distribution 
was used to infer that the $T_{cc}^+$ resonance is mostly isospin 0.

The energy difference between the $D^{*0} D^+$ and $D^{*+} D^0$  scattering  thresholds
 is $\delta = 1.41 \pm 0.03$~MeV. 
The two channels are related by isospin symmetry, which is broken by the energy difference $\delta$.
The binding energy of $T_{cc}^+$ for the $D^{*0} D^+$ channel  
is $\delta +|\varepsilon_T| = 1.77\pm 0.05$~MeV.
The binding momentum for that channel is $\gamma_{0+} =58.5 \pm 0.8$~MeV.
A simple coupled-channel model defined by wavefunctions analogous to Eqs.~\eqref{psi-k} and \eqref{psicc-k}
was used in Refs.~\cite{Meng:2021jnw,Yan:2021wdl}.
We introduce a  coupled-channel model defined by wavefunctions analogous to  Eqs.~\eqref{psi-k:reg} and \eqref{psicc-k:reg},
which have more physical behavior at large momentum.
We assume that isospin symmetry requires the wavefunctions at the origin in the two channels to be equal,
as in Eq.~\eqref{psicc:r=0}.
If $\Lambda$ is varied from $m_\pi/2$ to $m_\pi$ to $2m_\pi$,
the ratio $Z_{0+}$  of the probabilities for the $D^{*0} D^+$ and $D^{*+} D^0$ components 
from Eq.~\eqref{prob0+:reg} ranges from 0.34 to 0.38 to 0.41.


\section{XEFT for the Double-Charm Sector}
\label{sec:XEFT}

In this Section, we describe the effective field theory XEFT for low-energy charm mesons and pions,
and we give the Feynman rules for XEFT relevant to $T_{cc}^+$.

\subsection{Effective field theories for charm mesons and pions}
\label{sec:EFTcharmpi}

The universal  wavefunction in Eq.~\eqref{psi-r} 
and the  scattering amplitude  in Eq.~\eqref{f-E} can be derived from 
a zero-range effective field theory (ZREFT) with a single scattering channel  \cite{Braaten:2004rn}.
The simplest single-channel ZREFT has been applied previously to the $X(3872)$ 
and its constituents $D^{*0}\bar D^0$ and $D^0\bar D^{*0}$ \cite{Braaten:2004rn}.
Its region of validity extends at most up to the $D^{*+} D^-$ scattering threshold,
which is 8.2~MeV above the  $D^{*0} \bar D^0$ threshold.
ZREFT cannot describe accurately the effects of $D^0 \bar D^0\pi^0$ states,
which can be reached by the decay of a constituent $D^{*0}$ or $\bar D^{*0}$.
An analogous ZREFT can describe $T_{cc}^+(3875)$ and  its constituents $D^{*+} D^0$.
Its region of validity  extends at most up to the $D^{*0} D^+$ scattering threshold,
which is 1.4~MeV above the  $D^{*+} D^0$ threshold.
ZREFT cannot describe accurately the effects of  $D^0D^0\pi^+$ or $D^+D^0\pi^0$ states,
which can be reached by the decay of a constituent $D^{*+}$.

Fleming {\it et al.}\ developed an effective field theory called XEFT that describes $X$ 
and its meson constituents with a much larger region of validity \cite{Fleming:2007rp}.
XEFT is a nonrelativistic effective field theory for charm mesons $D^{(*)}$ and $\bar D^{(*)}$ and pions $\pi$.
The states described explicitly by XEFT are $D^* \bar D$,  $D \bar D^*$, $D \bar D \pi$, and $X$
with total energy in the region near the $D^* \bar D$ thresholds.
XEFT can equally well be applied to $T_{cc}^+$ and its meson constituents.
The states described explicitly by XEFT are $D^*D$, $DD\pi$, and $T_{cc}^+$ with total energy 
in the region near the $D^*D$ thresholds. 

The region of validity of XEFT is limited by the nonrelativistic approximation for the pion
to momenta less than   the pion mass $m_\pi$.
The natural scale for the ultraviolet momentum cutoff $\Lambda$ of XEFT is therefore $m_\pi$.
The corresponding scale for the kinetic energy of a pion is $m_\pi$.
The corresponding scale for the kinetic energy of two charm mesons is $m_\pi^2/M$, 
where $M$ is the charm-meson mass, which is about 10~MeV.

A Galilean-invariant formulation of XEFT that exploits the 
approximate conservation of mass in the transitions $D^* \leftrightarrow D \pi$
was developed in Ref.~\cite{Braaten:2015tga}.
In  Galilean-invariant XEFT, the spin-0 charm mesons $D^0$ and $D^+$ have the same kinetic mass $M$ and
 the pions $\pi^0$ and $\pi^+$  have the same kinetic mass $m$.
Conservation of kinetic mass requires
the spin-1 charm mesons $D^{*0}$ and $D^{*+}$ to have the same kinetic mass $M+m$.
The difference between the physical  mass and the kinetic mass of a particle
is taken into account through its rest energy.
In XEFT, the number of charm mesons with a charm quark 
and the number of charm mesons with a charm antiquark are both conserved.
In Galilean-invariant XEFT, the pion number defined by the sum of the numbers of $D^*$, $\bar D^*$, and $\pi$ mesons
is also conserved.
The conservation of pion number simplifies calculations in XEFT by reducing the number of diagrams.
Galilean invariance also  simplifies the analytic expressions for loop diagrams.
Furthermore, it simplifies the renormalization of XEFT by constraining ultraviolet divergences.
An improved formulation of Galilean-invariant XEFT 
that is particularly convenient for  calculations beyond leading order was developed in Ref.~\cite{Braaten:2020nmc}.

In Ref.~\cite{Braaten:2010mg}, Braaten, Hammer, and Mehen pointed out that XEFT 
could also be applied to sectors with pion number larger than 1.
It was applied specifically to the sector with pion number 2
consisting of $D^* \bar D^*$, $D^*  \bar D \pi$, $D \bar D^*  \pi$, $D \bar D \pi \pi$, and $X\pi$
with total energy in the region near the  $D^* \bar D^*$ thresholds  \cite{Braaten:2010mg}.
The states in the pion-number 2 sector with double charm
described explicitly by XEFT are $D^*D^*$, $D^*D\pi$, $DD\pi\pi$, and $T_{cc}^+\pi$
with total energy in the region near the  $D^* D^*$ thresholds.

\subsection{Feynman rules}
\label{sec:XEFTrules}

We denote the masses of the spin-0 charm mesons  $D^0$ and $D^+$ by $M_0$ and $M_+$,
the masses of the spin-1 charm mesons $D^{*0}$ and $D^{*+}$ by $M_{*0}$ and $M_{*+}$,
and the masses of the pions  $\pi^0$ and $\pi^+$ by $m_0$ and $m_+$ (or collectively by $m_\pi$).
We choose the kinetic mass $M$ of the spin-0 charm mesons to be $M_0$
and the kinetic mass $m$ of the pions to be $m_+$.
Galilean invariance then requires the kinetic mass of the spin-1 charm mesons to be $M_*=M+m$
and the kinetic mass of $T_{cc}^+$ to be $M_T=2M+m$.
The Galilean-invariant reduced masses of $D^* D$ and $D \pi$ are $\mu = MM_*/M_T$ 
and $\mu_\pi  = Mm/M_*$.
The reduced mass of $T_{cc}^+$ and a pion is $\mu_{\pi T} = M_Tm/(2M_*)$.

We proceed to give the Feynman rules for Galilean-invariant XEFT at leading order (LO)
applied to the $DD\pi$ and $DD\pi\pi$ sectors of QCD.
Our Feynman rules are essentially those in Ref.~\cite{Braaten:2020nmc},
in which the geometric series of $D^{*+} D^0$ bubble diagrams have been summed up into a  $T_{cc}^+$ propagator.
The Feynman rule for the propagator of $D^{*+}$ with energy $E$ relative to the $D^0\pi^+$ threshold,
momentum $\bm{p}$, and vector indices $i$ and $j$ is
\beq
\boxed{
\frac{i \, \delta^{ij}}{E -  p^2/(2(M+m))  - \delta_{0+} + i \Gamma_{*+}/2}, }
\label{propD*}
\eeq
where $\delta_{0+}=M_{*+} \!-\! M_0 \!-\! m_+ = 5.9$~MeV 
and $\Gamma_{*+} = 83.4 \pm 1.8$~keV is the measured decay width of  $D^{*+}$.
The Feynman rule for the complete propagator of $T_{cc}^+$ with  energy $E$ relative to the $D^0D^0\pi^+$ threshold,
momentum $\bm{P}$, and vector indices $i$ and $j$ is
\beq
\boxed{
\frac{-i\, \delta^{ij}}{- \gamma_T + \sqrt{- 2 \mu (E_\mathrm{cm} - \delta_{0+} + i \Gamma_{*+}/2)}},}
\label{pairproprule}
\eeq
where $E_\mathrm{cm} = E - P^2/[2(2M+m)]$ is the Galilean-invariant combination of $E$ and $P$.
The real binding momentum $\gamma_T$ is a significant simplification over XEFT applied to $X$,
whose binding momentum $\gamma_X$ must be complex to take into account
short-distance decay channels such as $X \to J/\psi\, \pi^+\pi^-$. 
The external-line factor for an outgoing $T_{cc}^+$ with polarization vector $\bm{\varepsilon}$ and vector index $i$ is
\beq
\boxed{
\sqrt{\gamma_T/\mu} \, \varepsilon^{i*}.}
\label{Texternal}
\eeq
The vertex connecting $D^{*+} D^0$  lines to the $T_{cc}^+$ propagator is
\beq
\boxed{
-i  \sqrt{2\pi/\mu}\, \delta^{ij},}
\label{vertex:TtoD*D}
\eeq
where $i$ and $j$ are the vector indices of $T_{cc}^+$ and $D^{*+}$. 
(In Ref.~\cite{Braaten:2020nmc},
 the factor $\sqrt{2\pi/\mu}$  was removed from this vertex 
in favor of multiplying the propagator in Eq.~\eqref{pairproprule} by $2\pi/\mu$
and multiplying the external line factor in Eq.~\eqref{Texternal} by  $\sqrt{2\pi/\mu}$.)
Because the complete  $T_{cc}^+$ propagator is obtained by summing a geometric series of $D^{*+} D^0$ 
bubble diagrams, the $D^{*+} D^0$ lines emerging from the vertex  in Eq.~\eqref{vertex:TtoD*D} 
are not allowed to close into a bubble before some other interaction, such as the emission of a pion.

The Feynman rule for the $D^{*+} \leftrightarrow D^0\pi^+$ vertex
in Galilean-invariant XEFT  is \cite{Braaten:2015tga,Braaten:2020nmc}
\beq	
 \boxed{\pm \, \frac{g}{\sqrt{2m} f_\pi} \frac{(M \bm{q} - m\bm{p}_0)^i}{M+m},}
\label{vertex:D*toDpi}
\eeq
where $i$ is the vector index for $D^{*+}$ and $\bm{q}$ and  $\bm{p}_0$ are the momenta of $\pi^+$ and $D^0$. 
The overall sign is $+$ if the $D^0\pi^+$  lines are outgoing and $-$ if they are incoming.
The Feynman rules for the $D^{*0} \leftrightarrow D^0\pi^0$,  
$D^{*+} \leftrightarrow D^+\pi^0$, and $D^{*0} \leftrightarrow D^+\pi^-$ vertices 
differ by the Clebsch-Gordan factors $+1/\sqrt{2}$, $-1/\sqrt2$ and +1, respectively. 
In the prefactor in Eq.~\eqref{vertex:D*toDpi}, $f_\pi = 130.5$~MeV  is the pion decay constant
and $g$ is a dimensionless coupling constant that can be determined from the decay width of $D^{*+}$ 
and its branching fraction into $D^0 \pi^+$.
Having chosen $m=m_+$, the value of $g$ is given by $g^2 = 0.329 \pm 0.008$. 
In the center-of-momentum (CM) frame defined by $\bm{p}_0 + \bm{q} = 0$, 
the momentum-dependent factor in Eq.~\eqref{vertex:D*toDpi} reduces to $q^i$.
In original XEFT, the momentum-dependent factor is $q^i$ in all frames.

A coupled-channel model for a loosely bound molecule with two coupled channels 
related by a symmetry at short distances was introduced in Section~\ref{sec:Molecule}.
The wavefunctions for the two coupled channels  in  Eqs.~\eqref{psi-k:reg} and \eqref{psicc-k:reg}
satisfy the symmetry condition  in Eq.~\eqref{psicc:r=0}.
If an amplitude in XEFT for producing $T_{cc}^+$ is expressed in a form with a factor of 
$1/(k^2+\gamma^2)$ from a $D^0$ propagator,
where $k$ is the relative momentum of the constituents $D^{*+}$ and $D^0$,
then the corresponding amplitude in the coupled-channel model can be obtained by making the substitution
\begin{equation}
 \frac{1}{k^2 + \gamma^2} \longrightarrow \frac{1}{\sqrt{1+Z_{0+}}}\,
 \frac{ \sqrt{(\Lambda+\gamma)\Lambda}}{\Lambda-\gamma}
\left( \frac{1}{k^2 + \gamma^2} - \frac{1}{k^2 + \Lambda^2} \right),
\label{psi-sub}
\end{equation}
where $Z_{0+} = (\Lambda+\gamma)\gamma/[(\Lambda+\gamma_{0+})\gamma_{0+}]$
is the relative probability of the $D^{*0}D^+$ channel.
This is equivalent to replacing the universal wavefunction $\psi_T(k)$ by $\psi^{(\Lambda)}_T(k)/\sqrt{1+Z_{0+}}$.
If an amplitude for producing $T_{cc}^+$ through the $D^{*0}D^+$ channel
is expressed in a form with  Eq.~\eqref{vertex:TtoD*D} as the $D^{*0} D^+$-to-$T_{cc}^+$ vertex 
and with a factor of $1/(k^2+\gamma_{0+}^2)$ from a $D^+$ propagator, 
where $k$ is the relative momentum of $D^{*0}$ and $D^+$,
then the amplitude in the coupled-channel model can be obtained by making the substitution
\begin{equation}
 \frac{1}{k^2 + \gamma_{0+}^2} \longrightarrow - \frac{1}{\sqrt{1+Z_{0+}}}\,
 \frac{ \sqrt{(\Lambda+\gamma)\Lambda}}{\Lambda-\gamma_{0+}}
\left( \frac{1}{k^2 + \gamma_{0+}^2} - \frac{1}{k^2 + \Lambda^2} \right).
\label{psi0+-sub}
\end{equation}
The relative minus sign compared to Eq.~\eqref{psi-sub}
comes from the isospin-0 combination $(D^{*+}D^0 - D^{*0}D^+)/\sqrt2$.
The integrals over $\bm{k}$ of the right sides of Eqs.~\eqref{psi-sub} and \eqref{psi0+-sub} differ only by a minus sign.
This is consistent with our assumption that isospin symmetry at short distances requires the
wavefunctions at the origin for the channels $D^{*+}D^0$ and $D^{*0}D^+$  to be equal, as in Eq.~\eqref{psicc:r=0}.
The ratio of the integrals over $\bm{k}$ of the squares of the right sides of Eqs.~\eqref{psi0+-sub} and \eqref{psi-sub} 
is equal to the relative probability $Z_{0+}$ of the $D^{*0}D^+$ channel.


\section{Production at a Hadron Collider}
\label{sec:Production}

In this Section, we consider the production of $T_{cc}^+(3875)$  at a high-energy hadron collider such as the LHC.  
We compare various aspects of its production with that of $X(3872)$.

\subsection{Production mechanisms}
\label{sec:ProductionMech}

The production of $X$ at a hadron collider has two contributions that can be resolved experimentally:
{\it bottom hadron decay} and {\it prompt production}.
In bottom hadron decay, a $b$ or $\bar b$ is created at the primary vertex for the colliding hadrons.  
It hadronizes into a bottom hadron, which travels a measurable distance before decaying through the weak interaction
at a secondary vertex into a final state that includes $X$.
The decay products of $X$, such as $J/\psi\, \pi^+\pi^-$, emerge from that secondary vertex.
In the prompt production of $X$, the $c \bar c$ constituents of $X$ are created at the primary vertex
by QCD interactions and the decay products of $X$ emerge from the primary vertex.
Bottom hadron decay can be distinguished from prompt production by the distribution of the measured distance
between the $X$ decay vertex and the primary collision vertex.
In the production of $T_{cc}^+$  at a hadron collider, there is no significant production mechanism 
analogous to bottom hadron decay. The production of $T_{cc}^+$ is entirely prompt.
Its decay products, such as $D^0D^0\pi^+$, emerge from the primary vertex.

\begin{figure}[t]
\includegraphics*[width=0.9\linewidth]{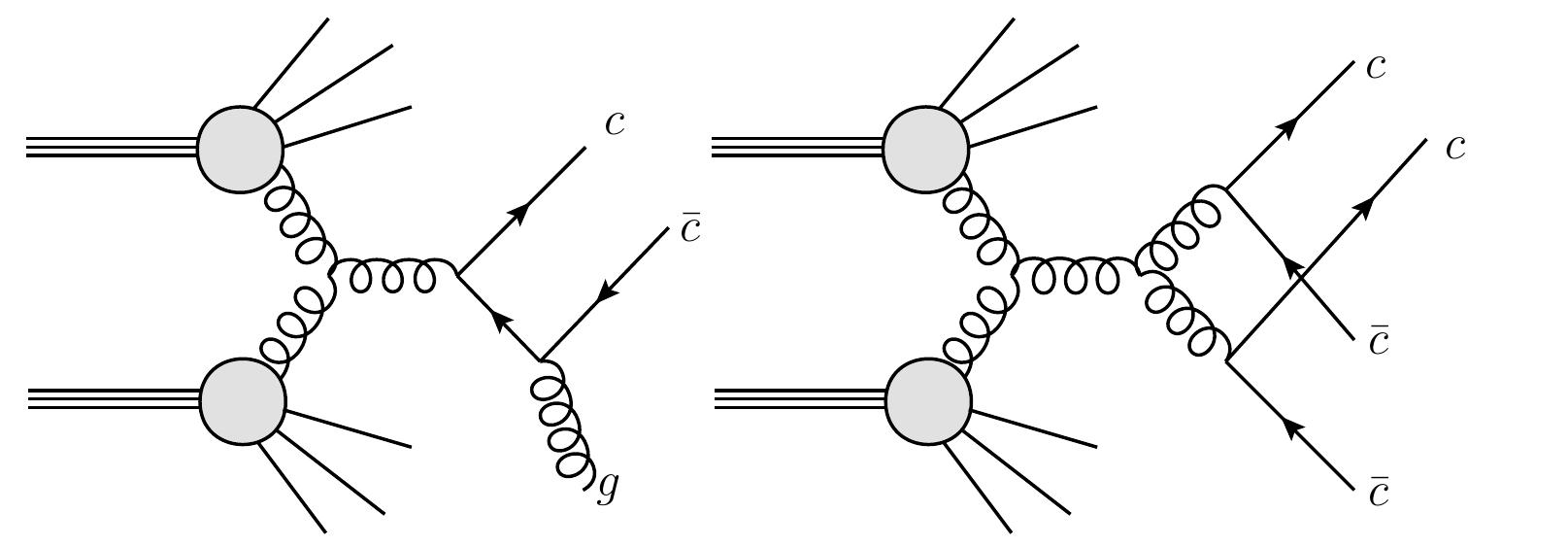} 
\caption{
Feynman diagrams for the prompt production of X (left) and $T_{cc}^+$ (right) through SPS. }
\label{fig:SPS}
\end{figure}
\begin{figure}[t]
\includegraphics*[width=0.9\linewidth]{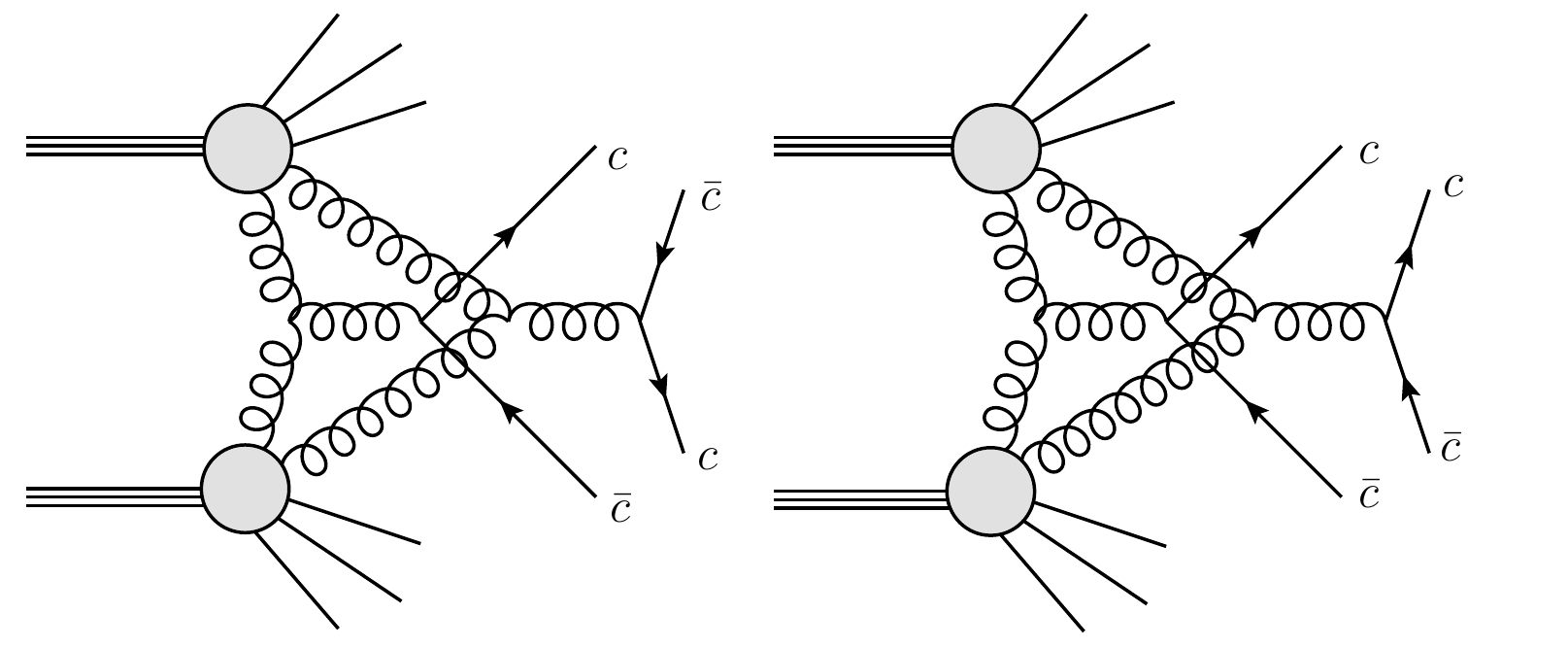} 
\caption{Feynman diagrams for the prompt production of X (left) and $T_{cc}^+$ (right) through DPS.}
\label{fig:DPS}
\end{figure}
At a hadron collider,
there are two distinct mechanisms for the prompt production of $X$ or the production of $T_{cc}^+$:
{\it single-parton scattering} (SPS) and {\it double-parton scattering} (DPS).
In SPS, the $c \bar c$  constituents of $X$ and the $cc$ constituents of $T_{cc}^+$ are created
with small relative momentum by a single gluon-gluon collision.
At leading order in the QCD coupling constant $\alpha_s$, 
the parton reaction that produces the $c \bar c$ constituents of $X$ is $g g \to c \bar c +g$, with diagrams like that on the left side of Fig.~\ref{fig:SPS}. 
This reaction, whose cross section is order $\alpha_s^3$,
also produces a gluon jet recoiling  against the collinear $c\bar c$ pair.
At leading order in $\alpha_s$, 
the parton reaction that produces the $c c$ constituents of $T_{cc}^+$ is  $g g \to cc \bar c \bar c$,
with diagrams like that on the right side of Fig.~\ref{fig:SPS}.
This reaction, whose cross section is order $\alpha_s^4$,
also produces two charm antiquark jets recoiling against the collinear $cc$.
In DPS, the $c \bar c$  constituents of $X$ and the $cc$ constituents of $T_{cc}^+$ are created 
with small relative momentum by two separate gluon-gluon collisions,
such as $g g \to  c \bar c$ whose cross section is order $\alpha_s^2$. 
The Feynman diagrams for $X$ and $T_{cc}^+$ include those on the left and right side of Fig.~\ref{fig:DPS}.
There is a small probability that the $c$ from one gluon-gluon collision
and the $\bar c$ or $c$ from the other have small relative momentum,
 in which case they can become constituents of $X$ or $T_{cc}^+$.

An intermediate step between the creation of a charm quark or charm antiquark 
and its becoming a constituent of $X$ or $T_{cc}^+$ is the hadronization of $c$ or $\bar c$ into a charm meson.
A $c \bar c$ pair created with small relative momentum
can hadronize into a pair of charm mesons $D^{(*)} \bar D^{(*)}$ with small relative momentum.
If the charm mesons are $D^{*0} \bar D^0$ or $D^0 \bar D^{*0}$, they may bind to form $X$.
Two charm quarks created with small relative momentum
can hadronize into two charm mesons $D^{(*)}  D^{(*)}$ with small relative momentum.
If the charm mesons are $D^{*+} D^0$, they may bind to form $T_{cc}^+$.

An alternative intermediate step between the creation of a $c \bar c$ pair with small relative momentum
and the formation of $X$ is the hadronization of $c \bar c$ into a more compact meson 
that is a component of the wavefunction of $X$ at short distances.
The $X$ is likely to have a short-distance component  that is the $\chi_{c1}(2P)$ charmonium state,
whose mass was expected to be less than 100~MeV above the $D^{*0} \bar D^0$ threshold.
Quantitative calculations of the prompt production rate of $X$ at the LHC
through the $\chi_{c1}(2P)$ component of its wavefunction starting from SPS reactions 
have been carried out using NRQCD factorization at next-to-leading order \cite{Butenschoen:2013pxa,Meng:2013gga,Butenschoen:2019npa}.
The probability for the $\chi_{c1}(2P)$ component of $X$ is a multiplicative factor in the cross section.
It can be adjusted to bring the calculated prompt production rate of $X$ into agreement with measurements at the LHC.
One cannot, a priori, exclude the possibility that $X$ also has a short-distance component that is a compact $c \bar c q \bar q$ tetraquark meson, although the production rate of $X$ through such a component is much more difficult to quantify.
The formulation of the production rate of $X$ entirely  in terms of the production of charm mesons
does not require its production through $\chi_{c1}(2P)$ or a compact tetraquark state to be ignored.
That contribution can be taken into account through the wavefunction at the origin $\psi_X(r\!=\!0)$ of $X$.

An alternative intermediate step between the creation of $c c$ with small relative momentum
and the formation of $T_{cc}^+$ is the hadronization of $c c$ into a tetraquark meson 
that is a component of the wavefunction of $T_{cc}^+$ at short distances.
Such a meson could be a compact $cc \bar q \bar q$ meson
or it could consist of $\bar q \bar q$ bound to a $cc$ diquark core.
The production rate of $T_{cc}^+$ through such a component would be difficult to quantify.
The formulation of the production rate of $T_{cc}^+$ entirely  in terms of the production of charm mesons
does not require its production through a more compact tetraquark meson to be ignored.
That contribution can be taken into account through the wavefunction at the origin $\psi_T(r\!=\!0)$ of $T_{cc}^+$.

\subsection{Short-distance production}
\label{sec:Shortdistance}

A charm-meson triangle singularity can be relevant to the production of $X$ or $T_{cc}^+$ only if
the process involves the creation of two charm mesons at points whose separation is much smaller than
the radius $\langle r \rangle$ of the loosely bound molecule.
There is a significant difference between the SPS and DPS mechanisms in the distance between
the points where the charm mesons are created.
With the SPS mechanism, the points where the collinear $c \bar c$ or $cc$ are created can be localized to within
the reciprocals of their transverse momenta to a single point where the gluon-gluon collision occurs.
Their subsequent hadronization can produce two charm mesons emerging from that point.
With the DPS mechanism, the points where the collinear $c \bar c$ or $cc$ are created 
can be localized to within the reciprocals of their transverse momenta to two separate points 
where the gluon-gluon collisions occur.
Their subsequent hadronization  can produce two charm mesons emerging from points separated 
by a distance comparable to the radius of the proton.
Thus the two charm mesons from the SPS mechanism are created at significantly shorter distances 
than those from the DPS mechanism.
However the DPS mechanism may still create charm mesons at short enough distances for a charm-meson triangle singularity 
to be relevant.

The charm meson $D^{(*)}$ has 4 spin states (1 for $D$ and 3 for $D^*$)
and 3 light-flavor states ($\bar u$, $\bar d$, and $\bar s$).
Because the available energy in $pp$ collisions at the LHC is so large, 
a charm quark  has approximately equal probabilities to hadronize 
at short distances into each of the 12 $D^{(*)}$ flavor/spin states.
At longer distances, the $D^*$’s all decay into $D \pi$ or $D \gamma$,
and the resulting probabilities for $D^0$, $D^+$ and $D_s^+$  are roughly  in the proportions 6:2:4. 
A charm quark and antiquark created with small relative momentum have  approximately equal probabilities 
to hadronize at short distances into each of the 144 $D^{(*)} \bar D^{(*)}$  flavor/spin states.
Because of the effects of identical bosons,
two charm quarks created with small relative momentum have  approximately equal probabilities 
to hadronize at short distances into each of the 78 $D^{(*)}D^{(*)}$  flavor/spin states.
For example, the short-distance hadronization probabilities for each of the 6 spin states of $D^{*+} D^{*+}$
and each of the 9 spin states of $D^{*+} D^{*0}$
are approximately equal to those for each of the 3 light-flavor states $D^+ D^+$, $D^+ D^0$, and $D^0 D^0$.
The production rates of two charm mesons with small relative momentum may be modified at longer distances 
in channels with a resonance near the threshold.
The existence of the $T_{cc}^+$ implies that there is an S-wave  resonance near the threshold in the $D^{*+}D^0$ channel.

At a high energy proton-proton collider like the LHC,  
the reactions that produce $T_{cc}^+$ or two charm mesons $D^{(*)} D^{(*)}$
also produce hundreds or even thousands  of additional particles.
It is convenient to consider the reaction in the CM frame of $D^{(*)} D^{(*)}$.
In this frame, the colliding protons and most of the additional particles have very large momenta.
If all the additional particles have momenta larger than $q_\mathrm{max}$ in that frame,
the two charm mesons $D^{(*)} D^{(*)}$ are  guaranteed to be created at points separated by less than 
 $1/q_\mathrm{max}$.
For reactions involving the charm mesons that involve momenta less than $q_\mathrm{max}$,
they might as well be created at a point.
An effective field theory for charm mesons and pions, such as XEFT,
 can be applied to the short-distance production of $D^{(*)} D^{(*)}$ by introducing local operators 
that create two charm mesons at a point.
The amplitude for producing a given set of final-state particles from the creation of $D^{(*)} D^{(*)}$
at a point can be expressed as a sum of Feynman diagrams
whose initial state is the creation of the charm mesons at a point with a vertex $\mathcal{A}_{D^{(*)} D^{(*)}}$
determined by the local operator.
The vertices $\mathcal{A}_{D^{(*)} D^{(*)}}$  must be such that 
the short-distance production rates for each of the 78 $D^{(*)}D^{(*)}$  flavor/spin states are approximately equal.

A local operator can create particles with arbitrarily large energies.
In an effective field theory, the inclusive production rate from a local operator that creates two charm mesons 
is ultraviolet divergent and it therefore requires regularization.
A possible ultraviolet cutoff is an upper limit $q_\mathrm{max}$ on the momenta of  particles 
in the CM frame of the two charm mesons.
Renormalization may require the vertices $\mathcal{A}_{D^{(*)} D^{(*)}}$ to depend on $q_\mathrm{max}$.
If the effective field theory is XEFT,  its results for all production amplitudes 
can be accurate only if $q_\mathrm{max}$ is at most of order $m_\pi$.
If the effective field theory is Galilean-invariant XEFT,  the particles that can be produced by the local operator
are strongly constrained by the conservation of charm-quark number and pion number.
A vertex of the form $\mathcal{A}_{DD}$, $\mathcal{A}_{D^* D}$, or $\mathcal{A}_{D^* D^*}$
produces final states  with pion number 0, 1, or 2, respectively.
A final state with pion number 1 or 2 can consist of a pion  with large relative momentum
and a recoiling  system with pion number lower by 1.
The recoiling system can be created by a vertex of the form $\mathcal{A}_{DD}$ or $\mathcal{A}_{D^* D}$.
Final states that include a pion with relative momentum $q < q_\mathrm{max}$
are described explicitly in the  effective field theory. 
The effects of pions with $q > q_\mathrm{max}$ can be taken into account 
through the dependence  on $q_\mathrm{max}$ of vertices of the form $\mathcal{A}_{DD}$ and $\mathcal{A}_{D^* D}$.
Vertices of the form $\mathcal{A}_{D^* D^*}$ do not acquire any dependence on $q_\mathrm{max}$ 
from the interactions of Galilean invariant XEFT.

It has been argued that the prompt production rates of $X$ at the Tevatron and the LHC are orders of magnitude too large
for a charm-meson molecule \cite{Bignamini:2009sk}.
The argument is based on the assumption that an order-of-magnitude estimate 
of the production rate of a molecule is the production rate of its constituents 
with relative momentum less than its binding momentum $\gamma$. 
The production rate of a molecule whose constituents are produced at short distances
is actually proportional to the square $|\psi(r\!=\!0)|^2$ of its wavefunction  at the origin \cite{Artoisenet:2009wk}.
For a generic molecule, $|\psi(r\!=\!0)|^2$ can be expressed as $\Lambda^3$ 
for some momentum scale $\Lambda$ of order $\gamma$.
The production rate of the molecule can therefore be approximated by the production rate of its constituents 
with relative momentum less than $\Lambda$.
Since the production rate of the constituents scales as $\Lambda^3$, 
this gives at best an order-of-magnitude estimate of the production rate of the molecule.
This estimate does not apply to a loosely bound S-wave molecule, because
the universal wavefunction in Eq.~\eqref{psi-r} is ultraviolet divergent  at the origin.
In this case, $|\psi(r\!=\!0)|^2$ can be expressed as $\Lambda^2 \gamma$ 
for some momentum scale $\Lambda$ much larger than $\gamma$.
The production rate of a loosely bound S-wave molecule can therefore be approximated 
by the production rate of its constituents with relative momentum less than $(\Lambda^2 \gamma)^{1/3}$.
 If $\Lambda$ is taken to be of order $m_\pi$,  the resulting order-of-magnitude estimates 
 for the prompt production rates of $X$ are compatible with the observed production rates
 at the Tevatron and the LHC \cite{Artoisenet:2009wk,Albaladejo:2017blx,Braaten:2018eov}.

\subsection{Multiplicity dependence}
\label{sec:Multiplicity}

The total number of light hadrons in the final state is the multiplicity.
At the LHC, the total multiplicity of an event is often in the thousands.
An additional hard-parton scattering can increase the multiplicity.
The multiplicities of $X$  or $T_{cc}^+$ events produced by DPS 
are therefore expected to be larger than those produced by SPS.
An additional jet produced by a hard scattering can increase the multiplicity.
The multiplicities of $T_{cc}^+$ events produced by SPS are therefore expected to  be larger than those 
for $X$ events produced by SPS.
However, the increase in the multiplicity from an additional hard scattering or from an additional jet
are probably small compared to the total multiplicity.  
The dominant effect of the multiplicity  on the production rate for $X$  or $T_{cc}^+$
could be through the environment  a charm meson or a loosely bound charm-meson molecule 
must propagate  through after it is produced.
One possible effect of a higher multiplicity is a higher probability for
particles produced by a hard scattering to interact with comoving partons or hadrons.
Esposito {\it et al.}\  have considered the effects on the production of $X$ from its breakup by collisions with comovers
and from its formation through recombination reactions involving comovers  \cite{Esposito:2020ywk}.
Once a loosely bound charm-meson molecule has formed, almost any interaction with a comover will break it up.
It is possible that the formation of the molecule from the two charm mesons created by DPS 
occur most often after they have traveled beyond the reach of comovers.
In this case, its production by DPS 
would be less suppressed by interactions with  comovers than its production  by SPS.

The LHCb collaboration has studied the multiplicity dependence of the prompt production of $X$ and $\psi(2S)$
and their production from  bottom hadron decays \cite{LHCb:2020sey}.
They measured the yields as functions of the number of charged tracks $N_\mathrm{tracks}$  in the vertex detector,
which has a range of several units of rapidity.
The prompt fraction of $X$ decreases with $N_\mathrm{tracks}$ 
from about 94\% in the lowest bin near $N_\mathrm{tracks}=30$
to about 71\% in the highest bin near 120.
The decreasing prompt fraction suggests that the prompt production of $X$ may be dominated by SPS.
A theoretical analysis by Esposito {\it et al.}\  showed that the prediction of the comover interaction model 
for the multiplicity dependence of the $X$-to-$\psi(2S)$ ratio is in good agreement with the LHCb data 
if $X$ has a size consistent with a compact tetraquark \cite{Esposito:2020ywk}.
They used a coalescence model that takes into account the recombination of charm-meson pairs 
to calculate the multiplicity dependence of the production of $X$ if it is a molecule.
Their result  for the $X$-to-$\psi(2S)$ ratio is a rapidly increasing function of $N_\mathrm{tracks}$.
The analysis in Ref.~\cite{Braaten:2020iqw} showed that a good fit to all the LHCb data 
on the multiplicity dependence of the production of $X$ and $\psi(2S)$ can be obtained 
if the break-up cross section of $X$ with comoving pions is roughly 3~mb.  This value is plausible,
given that the break-up cross section for a loosely bound charm-meson molecule
should be approximately equal to the cross section for scattering from a charm-meson constituent.

The LHCb collaboration has also studied the multiplicity dependence of the inclusive production of $T_{cc}^+$ \cite{LHCb:2021auc}.
The ratio of the yield of $T_{cc}^+$ in the decay channel $D^0D^0\pi^+$ to the yield of $D^0 \bar D^0$
seems to be about 2 or 3 times larger at values of  $N_\mathrm{tracks}$ greater than about 80 
than at smaller values of $N_\mathrm{tracks}$.
It is possible that the increased  yield  at larger $N_\mathrm{tracks}$ arises from the DPS mechanism.  
In this case, the restriction to $N_\mathrm{tracks}< 80$ could produce a sample of $T_{cc}^+$ events 
 in which a larger fraction is produced by the SPS mechanism.


\section{Production of Two Charm Mesons}
\label{sec:D*D*}

In this section, we consider the production of two spin-1 charm mesons $D^*D^*$
without any accompanying soft pions.
Our treatment is similar to that for the prompt production of $D^* \bar D^*$ in Ref.~\cite{Braaten:2019sxh},
but there are additional complications associated with identical bosons.

\subsection{Distinguishable charm mesons}
\label{sec:AmpD*D*}

\begin{figure}[t]
\includegraphics*[width=0.3\linewidth]{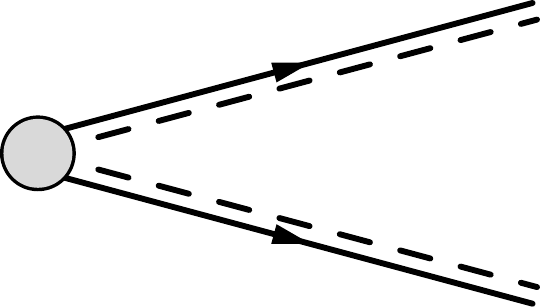} 
\caption{
Feynman diagram in XEFT for the production of $D^*D^*$ from their creation at a point.
A $D^*$ is represented by a double (solid+dashed) line with an arrow.}
\label{fig:D*D*}
\end{figure}

We consider the production of two charm mesons $D^{(*)} D^{(*)}$
plus additional particles $y$ that all have large momenta in the $D^{(*)} D^{(*)}$ CM frame.
The short-distance amplitude for creating $D^* D^*$  is represented in XEFT by  
the vertex in Fig.~\ref{fig:D*D*},  in which  the solid+dashed lines of the $D^*$’s emerge from a point.
The amplitudes for creating $D^* D$ and $DD$ can be represented  by  
analogous Feynman diagrams with a solid line for a $D$. 
A spin-0 charm meson is described by a scalar field $D$.
A spin-1 charm mesons is described by a vector field $D^{*i}$ with vector index $i=1,2,3$.
We denote the short-distance vertex factor for creating $DD$ at a point  with small relative momentum
while producing additional particles $y$ with large momentum in the $DD$ CM frame by $i \mathcal{A}_{DD+y}$.
We denote the analogous short-distance vertex factor for creating $D^*D$  at a point by $i \mathcal{A}^i_{D^*D+y}$.
We denote the analogous short-distance vertex factor for creating $D^* D^*$ at a point 
by $i \mathcal{A}^{ij}_{D^* D^*+y}$.
If the two $D^{*}$’s have the same light flavor, 
then $\mathcal{A}^{ij}_{D^* D^*+y}$ is symmetric in the indices $i$ and $j$.
We take the relative momentum $\bm{k}$ of the charm mesons in their CM frame to be
smaller than some ultraviolet cutoff $q_\mathrm{max}$ of order $m_\pi$.
Since the momenta of all the additional particles $y$ in that frame are larger than $q_\mathrm{max}$,
we take the limit  $\bm{k} \to 0$  in the short-distance vertex factors. 

We first consider the short-distance production of the two distinguishable spin-1 charm mesons  $D^{*+} D^{*0}$. 
Under the assumption that there is no resonance near threshold in that channel,
the matrix element for producing the final state $D^{*+}D^{*0}+y$
is obtained by contracting the short-distance amplitude $\mathcal{A}^{ij}_{D^{*+}D^{*0}+y}$
with the polarization vectors $\varepsilon^{i*}$ and  $\varepsilon^{\prime j*}$ for the $D^*$’s.
The inclusive differential cross section for producing $D^{*+} D^{*0}$ 
with relative momentum $\bm{k}$ in their CM frame can be expressed as
\begin{equation}
d\sigma[D^{*+}   D^{*0}] = \sum_{D^*~\mathrm{spins}}
\Big\langle\mathcal{A}^{ij}_{D^{*+}   D^{*0}} \big(\mathcal{A}^{kl}_{D^{*+}   D^{*0}}\big)^*\Big\rangle
(\varepsilon^i   \varepsilon^{\prime j})^* (\varepsilon^k   \varepsilon^{\prime l} )  \frac{d^3k}{(2 \pi)^3M_{*}}.
\label{dsigmaD*+D*0}
\end{equation}
The factor in angular brackets involves only short distances:
\begin{equation}
\Big\langle\mathcal{A}^{ij}_{D^{*+}   D^{*0}} \big(\mathcal{A}^{kl}_{D^{*+}   D^{*0}}\big)^*\Big\rangle \equiv
\frac{1}{\mathrm{flux}} \sum_y \int d\Phi_{(D^*  D^*)+y}
\mathcal{A}^{ij}_{D^{*+}   D^{*0}+y} \big(\mathcal{A}^{kl}_{D^{*+}   D^{*0}+y} \big)^*.
\label{<AA-D*+D*0>}
\end{equation}
The product of the short-distance amplitude and its complex conjugate 
is integrated over the relativistic differential phase space $d\Phi_{(D^*  D^*)+y}$ for the additional particles $y$
plus a composite particle  with mass $2M_{*}$ denoted by $(D^*  D^*)$,
summed over the additional particles $y$ (including their spins), 
and multiplied by the flux factor 1/flux for the colliding protons.
The differential phase space for $D^* D^*$  has been expressed as the product of $d^3P/[(2 \pi)^3 2P_0]$ 
for the composite particle $(D^*  D^*)$, which is included in $d\Phi_{(D^*  D^*)+y}$,
and $d^3k/[(2 \pi)^3M_{*}]$, where $M_*$ is twice the $D^*D^*$ reduced mass.

We next  consider the short-distance production of the two distinguishable spin-0 charm mesons  $D^+ D^0$. 
Under the assumption that there is no resonance near threshold in that channel,
the matrix element for producing $D^+ D^0$ plus additional particles $y$
is just the short-distance amplitude $\mathcal{A}_{D^+D^0+y}$.
The inclusive differential cross section for producing $D^+ D^0$ 
with relative momentum $\bm{k}$ in  their CM frame can be expressed as
\begin{equation}
d\sigma[D^+   D^0] = 
\Big\langle\mathcal{A}_{D^+   D^0} \big(\mathcal{A}_{D^+  D^0}\big)^*\Big\rangle
 \frac{d^3k}{(2 \pi)^3M}.
\label{dsigmaD+D0}
\end{equation}
The factor  in angular brackets involves only short distances:
\begin{equation}
\Big\langle\mathcal{A}_{D^+   D^0} \big(\mathcal{A}_{D^+   D^0}\big)^*\Big\rangle \equiv
\frac{1}{\mathrm{flux}} \sum_y \int d\Phi_{(D  D)+y}
\mathcal{A}_{D^+   D^0+y} \big(\mathcal{A}_{D^+   D^0+y} \big)^*.
\label{<AA-D+D0>}
\end{equation}
The product of the short-distance amplitude and its complex conjugate 
is integrated over the relativistic differential phase space $d\Phi_{(DD)+y}$ for the additional particles $y$
plus a composite particle  with mass $2M$ denoted by $(DD)$.
The cross section in Eq.~\eqref{dsigmaD+D0} is for the production of $D^+   D^0$ at short distances.
It does not include the feeddown from the production of $D^*D^*$ or $D^*D$ at short distances
followed by decays $D^* \to D\pi,D\gamma$.

In the CM frame of $D^{*+}D^{*0}$, the  short-distance amplitude  $\mathcal{A}^{ij}_{D^{*+}D^{*0}+y}$
is a Cartesian tensor with vector indices $ij$.  
The indices $ij$ can be carried by the metric tensor $\delta^{ij}$,
by momentum vectors of  additional particles $y$ or the colliding protons,
or by polarization vectors, spinors, or tensors associated with  their spins.  
The indices  $ij$ cannot be carried by the relative momentum vector $\bm{k}$ of the two $D^{*}$’s,
because the limit $\bm{k} \to 0$ has been taken in the short-distance amplitude.
The weighted average $\langle \mathcal{A}^{ij}_{D^{*+}D^{*0}}(\mathcal{A}^{kl}_{D^{*+}D^{*0}})^*\rangle$ 
of the product of short-distance amplitudes in Eq.~\eqref{<AA-D*+D*0>} is a Cartesian tensor
with vector indices $i jkl$. The indices cannot be carried by the momentum vector of any of the additional particles $y$, 
because they have been integrated over.
They cannot be carried by the  polarization vector, spinor, or tensor associated with one of their spins, 
because the spins have been summed over.
The indices can however be carried by the momentum vector of one of the colliding protons or by its polarization spinor.
That possibility can be removed by averaging over the spins of the colliding protons
and by averaging over the directions of their  momenta in the CM frame of the two charm mesons.
Averaging over the directions of the proton momenta in the charm-meson CM frame 
has the same effect as averaging over the directions of the total momentum $\bm{P}$ of the charm mesons
in the $pp$ CM frame.
From now  on, it will be understood that the weighted average of an amplitude and its complex conjugate, 
such as that in Eq.~\eqref{<AA-D*+D*0>} or Eq.~\eqref{<AA-D+D0>},
is also averaged over the spins of the colliding protons 
and averaged over the directions of their momenta in the CM frame of the two charm mesons. 
The weighted average in Eq.~\eqref{<AA-D*+D*0>} must then be 
a linear combination of $\delta^{ik} \delta^{jl}$, $\delta^{il} \delta^{jk}$, and $\delta^{ij} \delta^{kl}$.
The condition that the 78 flavor/spin states of $D^{(*)} D^{(*)}$ 
are produced equally often at short distances can be implemented by keeping only the $\delta^{ik} \delta^{jl}$ term.
The weighted average in Eq.~\eqref{<AA-D*+D*0>}
can be related to the weighted average in Eq.~\eqref{<AA-D+D0>}.
Since the difference between the masses  $2M_*$ and $2M$ of the composite particles $(D^*D^*)$ and $(D D)$
in the phase-space integrals in Eq.~\eqref{<AA-D*+D*0>} and Eq.~\eqref{<AA-D+D0>}
 is tiny compared to the collision energy, it  can be ignored.
The resulting relation between the weighted averages has the form
\begin{equation}
\Big\langle  \mathcal{A}^{ij}_{D^{*+} D^{*0}}\big(\mathcal{A}^{kl}_{D^{*+}   D^{*0}}\big)^*   \Big\rangle
= \Big\langle\mathcal{A}_{D^+   D^0} \big(\mathcal{A}_{D^+  D^0}\big)^*\Big\rangle \,  \delta^{ik} \, \delta^{jl}.
\label{<AA-D*+D*0>tensor}
\end{equation}
After multiplying by the polarization vectors in Eq.~\eqref{dsigmaD*+D*0}
and summing over the spin states of $D^{*+} D^{*0}$, we obtain
\begin{equation}
\sum_{D^*~\mathrm{spins}}
\Big\langle\mathcal{A}^{ij}_{D^{*+}   D^{*0}} \big(\mathcal{A}^{kl}_{D^{*+}   D^{*0}}\big)^*\Big\rangle
(\varepsilon^i   \varepsilon^{\prime j})^* (\varepsilon^k   \varepsilon^{\prime l} )
= 9 \, \Big\langle \mathcal{A}_{D^+   D^0} \big(\mathcal{A}_{D^+  D^0}\big)^*\ \Big\rangle.
\label{<AA-D*+D*0>spinsum}
\end{equation}
The prefactor in Eq.~\eqref{<AA-D*+D*0>tensor} was chosen 
so the prefactor in Eq.~\eqref{<AA-D*+D*0>spinsum} is the number of $D^{*+} D^{*0}$ spin states.

Our final result for the $D^{*+} D^{*0}$  cross section  is
obtained by inserting Eq.~\eqref{<AA-D*+D*0>spinsum} into Eq.~\eqref{dsigmaD*+D*0}:
\begin{equation}
d\sigma[D^{*+}   D^{*0}] =9\,
\Big\langle \mathcal{A}_{D^+   D^0} \big(\mathcal{A}_{D^+  D^0}\big)^* \Big\rangle \frac{d^3k}{(2 \pi)^3M_{*}}
\label{sigmaD*+D*0}
\end{equation}
The differential cross section $d\sigma/d^3k$ for $D^{*+}   D^{*0}$
differs from that for $D^+D^0$ from Eq.~\eqref{dsigmaD+D0} by the spin factor 9 and the mass ratio $M/M_*$.

\subsection{Identical charm mesons}
\label{sec:sigmaD*D*}

We  next consider the short-distance production of the two identical spin-1 charm mesons $D^{*+} D^{*+}$. 
Under the assumption that there is no resonance near threshold in that channel,
the matrix element for producing the final state $D^{*+}D^{*+}+y$
is obtained by contracting the short-distance amplitude $\mathcal{A}^{ij}_{D^{*+}D^{*+}+y}$
with the polarization vectors $\varepsilon^{i*}$ and  $\varepsilon^{\prime j*}$ for the $D^*$’s.
The inclusive differential cross section for producing $D^{*+} D^{*+}$ 
with relative momentum $\bm{k}$ in  their CM frame can be expressed as
\begin{equation}
d\sigma[D^{*+}   D^{*+}] = \frac12 \sum_{D^*~\mathrm{spins}}
\Big\langle\mathcal{A}^{ij}_{D^{*+}   D^{*+}} \big(\mathcal{A}^{kl}_{D^{*+}   D^{*+}}\big)^*\Big\rangle
(\varepsilon^i   \varepsilon^{\prime j})^* (\varepsilon^k   \varepsilon^{\prime l} )  \frac{d^3k}{(2 \pi)^3M_{*}}.
\label{dsigmaD*+D*+}
\end{equation}
The short-distance factor is defined by a weighted average 
analogous to that in Eq.~\eqref{<AA-D*+D*0>}.
The prefactor of 1/2 in Eq.~\eqref{dsigmaD*+D*+}
compensates for overcounting the states of the identical bosons $D^{*+}   D^{*+}$. 

We next  consider the short-distance production of the two identical spin-0 charm mesons  $D^+ D^+$. 
Under the assumption that there is no resonance near threshold in that channel,
the matrix element for producing $D^+ D^+$ plus additional particles $y$
is just the short-distance amplitude $\mathcal{A}_{D^+D^++y}$.
The inclusive differential cross section for producing $D^+ D^+$
with relative momentum $\bm{k}$ in  their CM frame can be expressed as
\begin{equation}
d\sigma[D^+   D^+] = \frac12
\Big\langle\mathcal{A}_{D^+   D^+} \big(\mathcal{A}_{D^+  D^+}\big)^*\Big\rangle
 \frac{d^3k}{(2 \pi)^3M}.
\label{dsigmaD+D+}
\end{equation}
The short-distance factor is defined by a weighted average analogous to that in Eq.~\eqref{<AA-D+D0>}.
The prefactor of 1/2 in Eq.~\eqref{dsigmaD+D+}
compensates for overcounting the states of the identical bosons $D^+D^+$. 
The cross section in Eq.~\eqref{dsigmaD+D+} is  for the production of $D^+ D^+$ at short distances.
It does not include the feeddown from the production of $D^*D^*$ or $D^*D$ at short distances
followed by decays $D^* \to D\pi,D\gamma$.

In the CM frame of $D^{*+}D^{*+}$, the short-distance amplitude $\mathcal{A}^{ij}_{D^{*+}D^{*+}+y}$
is a Cartesian tensor with  vector indices $ij$.  
The weighted average $\langle \mathcal{A}^{ij}_{D^{*+}D^{*+}}(\mathcal{A}^{kl}_{D^{*+}D^{*+}})^*\rangle$ 
analogous to that in Eq.~\eqref{<AA-D*+D*0>}  is a Cartesian tensor with vector indices $i jkl$ that
can only be carried by the metric tensor. 
Because the $D^{*+}D^{*+}$ are identical bosons,
the indices must be symmetric under interchange of $i$ and $j$ and under interchange of $k$ and $l$.
It must therefore be a linear combination of 
$\delta^{ik} \delta^{jl} +\delta^{il} \delta^{jk}$ and $\delta^{ij} \delta^{kl}$.
The condition that the 78 flavor/spin states of $D^{(*)} D^{(*)}$ 
are produced equally often at short distances can be implemented by keeping only the 
$\delta^{ik} \delta^{jl} +\delta^{il} \delta^{jk}$ term.
The weighted average in Eq.~\eqref{dsigmaD*+D*+}
can be related to the weighted average in Eq.~\eqref{dsigmaD+D+}:
\begin{equation}
\Big\langle  \mathcal{A}^{ij}_{D^{*+} D^{*+}}\big(\mathcal{A}^{kl}_{D^{*+}   D^{*+}}\big)^*   \Big\rangle
= \frac12  \Big\langle \mathcal{A}_{D^+   D^+} \big(\mathcal{A}_{D^+  D^+}\big)^*\ \Big\rangle
\left( \delta^{ik} \delta^{jl}  +  \delta^{il} \delta^{jk}\right).
\label{<AA-D*+D*+>tensor}
\end{equation}
After multiplying by the polarization vectors in Eq.~\eqref{dsigmaD*+D*+}
and summing over the spin states of $D^{*+} D^{*+}$, we obtain
\begin{equation}
\sum_{D^*~\mathrm{spins}}
\Big\langle\mathcal{A}^{ij}_{D^{*+}   D^{*+}} \big(\mathcal{A}^{kl}_{D^{*+}   D^{*+}}\big)^*\Big\rangle
(\varepsilon^i   \varepsilon^{\prime j})^* (\varepsilon^k   \varepsilon^{\prime l} )
= 6 \,  \Big\langle \mathcal{A}_{D^+   D^+} \big(\mathcal{A}_{D^+  D^+}\big)^*\ \Big\rangle.
\label{<AA-D*+D*+>spinsum}
\end{equation}
The prefactor in Eq.~\eqref{<AA-D*+D*+>tensor} was chosen 
so the prefactor in Eq.~\eqref{<AA-D*+D*+>spinsum} is the number of $D^{*+} D^{*+}$ spin states.

Our final result for the $D^{*+} D^{*+}$ cross section  is
obtained by inserting Eq.~\eqref{<AA-D*+D*+>spinsum} into Eq.~\eqref{dsigmaD*+D*+}:
\begin{equation}
d\sigma[D^{*+}   D^{*+}] = 6\,
\Big\langle  \mathcal{A}_{D^+  D^+} \big(  \mathcal{A}_{D^+  D^+} \big)^* \Big\rangle \frac{d^3k}{2(2 \pi)^3M_{*}}.
\label{sigmaD*+D*+}
\end{equation}
The differential cross section $d\sigma/d^3k$ for $D^{*+}   D^{*+}$
differs from that for $D^+D^+$ from  Eq.~\eqref{dsigmaD+D+}
by the spin factor 6 and the mass ratio $M/M_*$.

The short-distance cross sections for the two identical bosons $D^+   D^+$ in Eq.~\eqref{dsigmaD+D+} 
and the two distinguishable bosons $D^+   D^0$ in  Eq.~\eqref{dsigmaD+D0} must be equal.
The short-distance factors in  those cross sections must therefore differ by a factor of 2:
\begin{eqnarray}
\Big\langle\mathcal{A}_{D^+   D^+} \big(\mathcal{A}_{D^+  D^+}\big)^*\Big\rangle =
2\, \Big\langle\mathcal{A}_{D^+   D^0} \big(\mathcal{A}_{D^+  D^0}\big)^*\Big\rangle.
\label{factor-D+D+/D+D0}
\end{eqnarray}
The cross sections for the two identical bosons $D^{*+} D^{*+}$ in  Eq.~\eqref{sigmaD*+D*+}
and the two distinguishable bosons $D^{*+}   D^{*0}$  in Eq.~\eqref{sigmaD*+D*0}
therefore differ by the ratio 2/3 of the numbers of their spin states.


\section{Production of $\bm{D^{*+}D^0}$ and $\bm{T_{cc}^+}$}
\label{sec:D*D,T}

In this section, we consider the production of $D^{*+}D^0$ and $T_{cc}^+$
without any accompanying soft pions.
Our treatment is similar to that for the prompt production of $D^{*0} \bar D^0$, $D^0 \bar D^{*0}$, and $X$
in Ref.~\cite{Braaten:2019sxh}.
The results are consistent with factorization formulas first derived in Ref.~\cite{Braaten:2005jj}.

\subsection{Production of $\bm{D^{*+}D^0}$}
\label{sec:prodD*D}

\begin{figure}[t]
\includegraphics*[width=0.8\linewidth]{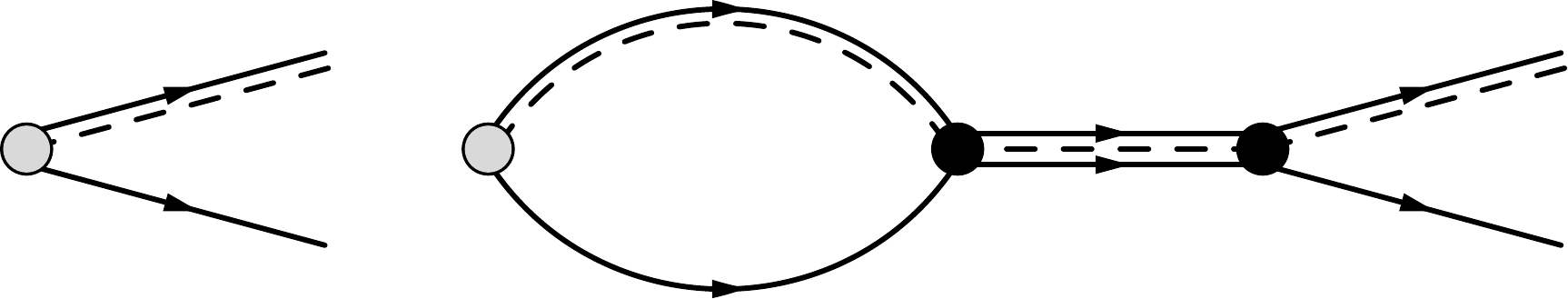} 
\caption{
Feynman diagrams in XEFT for the production of $D^{*+}D^0$ from the creation of the charm mesons at a point.
A $D^0$ is represented by a solid line with an arrow.
The $T_{cc}^+$ is represented by a triple (solid+dashed+solid) line.
}
\label{fig:D*+D0}
\end{figure}

The existence of $T_{cc}^+$ implies that there is an S-wave resonance near threshold 
in the $D^{*+} D^0$ channel.
The resonance must be taken into account in the production of $D^{*+} D^0$  with small relative momentum
as well as in the production of $T_{cc}^+$.
We first consider the production of $D^{*+} D^0$.
The two Feynman diagrams in XEFT for the production of $D^{*+} D^0$ by their creation at a point 
are shown  in Fig.~\ref{fig:D*+D0}.
The blob on the left side of each diagram is the vertex $i \mathcal{A}^i_{D^{*+} D^0 + y}$
for creating the charm mesons at a point while producing additional particles $y$
with large momenta in the $D^{*+} D^0$ CM frame.
The first diagram in Fig.~\ref{fig:D*+D0} is the tree amplitude for producing $D^{*+} D^0$ 
without any subsequent interaction between the charm mesons.
The second diagram in Fig.~\ref{fig:D*+D0} is the loop amplitude for producing $D^{*+} D^0$ 
with one or more subsequent rescatterings of the charm mesons.
These rescattering amplitudes form a geometric series that can be summed up in terms of the
complete propagator for $T_{cc}^+$ in Eq.~\eqref{pairproprule}.
We take the relative momenta of the charm mesons in their CM frame to be $\bm{\ell}$ for the $D^{*+} D^0$ 
that are created and $\bm{k}$ for the final-state $D^{*+} D^0$.
The loop  integral over $\bm{\ell}$ should be evaluated
at a total energy $E= \delta_{0+} - i \Gamma_{*+}/2 +k^2/(2\mu)$
given by the sum of the complex energy of the $D^{*+}$ and the real energy of the $D^0$.
The resulting expression for the sum of the two diagrams is
\begin{equation}
 \mathcal{A}_{D^{*+} D^0+y}(\bm{k})  = 
\mathcal{A}^i_{D^{*+}  D^0+y}
\left( 1 + \frac{4\pi}{-\gamma_T-ik} 
\int \frac{d^3 \ell}{(2\pi)^3} \frac{1}{\ell^2 - (k^2 + i\epsilon)} \right)
 \varepsilon^{i*},
\label{amp-DstarDint}
\end{equation}
where $\gamma_T = \sqrt{2\mu|\varepsilon_T|}$.
The ultraviolet-divergent loop integral can be evaluated analytically 
after imposing a sharp ultraviolet cutoff $|\bm{\ell}| < (\pi/2)\Lambda$ on the loop momentum.
The amplitude for producing $D^{*+} D^0$ with polarization vector $\bm{\varepsilon}$ for the $D^{*+}$ is
\begin{equation}
 \mathcal{A}_{D^{*+} D^0+y}(\bm{k})  = 
\mathcal{A}^i_{D^{*+}  D^0+y}
\,\frac{\Lambda - \gamma_T}{-\gamma_T-ik}  \varepsilon^{i*}.
\label{amp-DstarD}
\end{equation}
The factor $\Lambda - \gamma_T$ in the numerator 
can be expressed as $\sqrt{2\pi/\gamma_T}\, \psi_T(r\!=\!0)$, 
where $\psi_T(r\!=\!0)$ is the universal wavefunction at the origin for $T_{cc}^+$ given by Eq.~\eqref{psi-0}
with $\gamma = \gamma_T$.

The inclusive differential cross section for producing $D^{*+} D^0$ with
small relative momentum $\bm{k}$ in their CM frame can be expressed as
\begin{eqnarray}
d\sigma[D^{*+}   D^0] =
\frac{1}{\mathrm{flux}} \sum_{D^*\, \mathrm{spins}} \sum_y  \int d\Phi_{(D^*  D)+y}
\Big| \mathcal{A}_{D^{*+}   D^0 +y}(\bm{k})  \Big|^2  \frac{d^3k}{(2 \pi)^3 2\mu},
\label{factor-DstarD}
\end{eqnarray}
where $d\Phi_{(D^*  D)+y}$ is the relativistic differential phase space for all the additional particles $y$
plus a composite particle  denoted by $(D^*  D)$ with mass $M_{*} \!+\! M$.
The relativistic differential phase space for $D^{*+}   D^0$ has been expressed as the product of
the differential phase space $d^3P/[(2 \pi)^3 2P_0]$ for the composite particle  $(D^*  D)$
and $d^3k/[(2\pi)^3 2\mu]$, where $\mu$ is  the $D^*D$ reduced mass.
The cross section in Eq.~\eqref{factor-DstarD} does not include the feeddown
 from the production of $D^*D^*$ at short distances followed by decays $D^* \to D \pi,D \gamma$.

In the CM frame of $D^{*+}D^0$,
the amplitude $\mathcal{A}^i_{D^{*+}D^0+y}$ is a Cartesian vector with index $i$.  
The weighted average $\langle \mathcal{A}^i_{D^{*+}D^0} (\mathcal{A}^j_{D^{*+}D^0})^* \rangle$ 
of the product of amplitudes can be defined as in Eq.~\eqref{<AA-D*+D*0>},
except that the composite particle is $(D^*D)$ with mass $M_* \!+\! M$.  
The weighted average is a Cartesian tensor whose vector indices $i j$ 
can only be carried by the metric tensor $\delta^{ij}$.
This weighted average can be related to the corresponding weighted average 
$\langle \mathcal{A}_{D^+D^0}(\mathcal{A}_{D^+D^0})^*\rangle$ for two spin-0 charm mesons. 
Since the difference between the masses $M_* \!+\! M$  and $2M$ of the composite particles $(D^*D)$ and $(DD)$ 
in the phase-space integrals in Eqs.~\eqref{dsigmaD+D0} and \eqref{factor-DstarD}
is  tiny compared to the collision energy, it can be ignored. 
After multiplying the weighted average $\langle \mathcal{A}^i_{D^{*+}D^0} (\mathcal{A}^j_{D^{*+}D^0})^* \rangle$
by the polarization vectors for $D^{*+}$ and summing over its spin states, we obtain
\begin{equation}
\sum_{D^{*+}~\mathrm{spins}}
\Big\langle\mathcal{A}^i_{D^{*+}   D^0} \big(\mathcal{A}^j_{D^{*+}   D^0}\big)^*\Big\rangle
\varepsilon^{i*}    \varepsilon^j   
= 3 \, \Big\langle \mathcal{A}_{D^+   D^0} (\mathcal{A}_{D^+   D^0})^* \Big\rangle.
\label{sumAASS-D*+D0}
\end{equation}
The prefactor is the number of $D^{*+}$ spin states.

After inserting  the amplitude in Eq.~\eqref{amp-DstarD} into Eq.~\eqref{factor-DstarD}
and then using Eq.~\eqref{sumAASS-D*+D0}, 
the inclusive differential cross section for producing $D^{*+} D^0$ 
with small relative momentum $\bm{k}$ in  their CM frame reduces to
\begin{eqnarray}
d\sigma[D^{*+}   D^0] = 3\,
\Big\langle \mathcal{A}_{D^+   D^0} (\mathcal{A}_{D^+   D^0})^* \Big\rangle 
\big| \psi_T(r\!=\!0) \big|^2\, \frac{2\pi/ \gamma_T}{k^2+\gamma_T^2}\frac{d^3k}{(2 \pi)^32\mu}.
\label{factor-DstarD-2}
\end{eqnarray}

\subsection{Production of $\bm{T_{cc}^+}$}
\label{sec:prodT}

\begin{figure}[t]
\includegraphics*[width=0.4\linewidth]{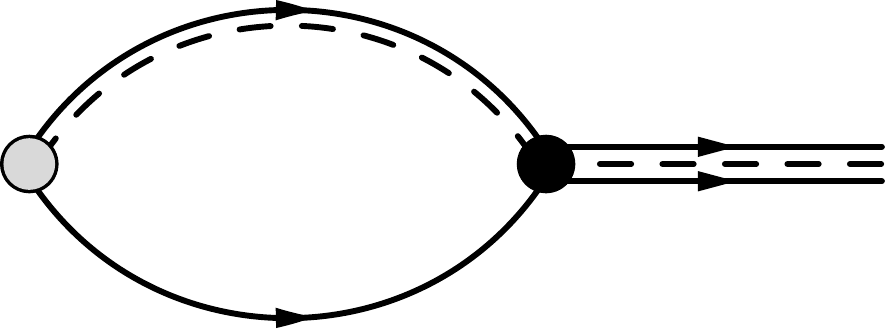} 
\caption{
Feynman diagram in XEFT for the production of $T_{cc}^+$ from the creation of $D^{*+}D^0$ at a point.}
\label{fig:Tcc+}
\end{figure}

We now turn to the production of $T_{cc}^+$.  
The Feynman diagram in XEFT for the production of $T_{cc}^+$ from the creation of $D^{*+}   D^0$ at a point
is shown in Fig.~\ref{fig:Tcc+}.
The blob on the left side of the diagram is the vertex $i \mathcal{A}^i_{D^{*+} D^0+y}$ 
for creating the charm mesons at a point 
while producing additional particles $y$ with large momenta in the $D^{*+} D^0$ CM frame.
The loop integral should be evaluated at the complex pole energy of $T_{cc}^+$.
In XEFT at LO, the imaginary part of its pole energy is $-\Gamma_{*+}/2$. 
The complex pole energy of $T_{cc}^+$ in its rest frame relative to the $D^0D^0\pi^+$ threshold
is therefore $\delta_{0+} + \varepsilon_T - i \Gamma_{*+}/2$.
Upon evaluating the loop  integral at this complex energy,
the amplitude for producing $T_{cc}^+$ with polarization vector $\bm{\varepsilon}$  is
\begin{equation}
 \mathcal{A}_{T_{cc}^++y}  = 
- \left(\mathcal{A}^i_{D^{*+}  D^0+y} \sqrt{M_T/2M_* M} \, \right) \sqrt{8\pi \gamma_T} \,  \varepsilon^{i*}
\int \frac{d^3\ell}{(2\pi)^3} \frac{1}{\ell^2 - 2\mu(\varepsilon_T + i \epsilon)} .
\label{amp-Xint}
\end{equation}
The factor $\sqrt{M_T/2M_* M}$ takes into account the difference between relativistic and nonrelativistic
 normalizations of states.
The ultraviolet-divergent loop integral can be evaluated analytically 
after imposing a sharp ultraviolet cutoff $|\bm{\ell}| < (\pi/2)\Lambda$:
\begin{equation}
 \mathcal{A}_{T_{cc}^++y} = 
- \left(\mathcal{A}^i_{D^{*+}  D^0+y} \sqrt{M_T/2M_* M} \, \right) \psi_T(r\!=\!0) \,  \varepsilon^{i*},
\label{amp-Tcc+}
\end{equation}
where $\psi_T(r\!=\!0)$ is the wavefunction at the origin in Eq.~\eqref{psi-0} with $\gamma= \gamma_T$.

The inclusive cross section for producing $T_{cc}^+$ 
from the creation of $D^{*+} D^0$ at short distances can be expressed as
\begin{eqnarray}
\sigma[T_{cc}^+,\mathrm{no}\,\pi] =
\frac{1}{\mathrm{flux}} \sum_{T_{cc}^+\, \mathrm{spins}} \sum_y \int d\Phi_{(D^*  D)+y}
\big| \mathcal{A}_{T_{cc}^+ +y}  \big|^2 .
\label{factor-Tcc}
\end{eqnarray}
We have denoted this cross section by $\sigma[T_{cc}^+,\mathrm{no}\,\pi]$ to emphasize that 
it is the cross section for producing $T_{cc}^+$ without any pion
with momentum less than the ultraviolet cutoff $q_\mathrm{max}$ used to define the short-distance vertices.
After inserting the expression for the amplitude in Eq.~\eqref{amp-Tcc+}
and then using Eq.~\eqref{sumAASS-D*+D0}, the cross section for $T_{cc}^+$ reduces to
\begin{eqnarray}
\sigma[T_{cc}^+,\mathrm{no}\,\pi] =
\Big\langle \mathcal{A}_{D^+   D^0} (\mathcal{A}_{D^+   D^0})^* \Big\rangle 
  \frac{3}{2\mu} \big| \psi_T(r\!=\!0) \big|^2.
\label{factor-DstarD-3}
\end{eqnarray}

The short-distance factor in angular brackets in the cross section for $T_{cc}^+$ in Eq.~\eqref{factor-DstarD-3} is
the same as that in the differential cross section for producing $D^{*+} D^0$ in Eq.~\eqref{factor-DstarD-2}.
Those short-distance factors
can therefore be eliminated to get a relation between the cross sections. 
The factors of $| \psi_T(r\!=\!0)|^2$ are also eliminated in that relation.
The invariant kinetic energy of $D^{*+} D^0$ is its total kinetic energy $E=k^2/(2\mu)$ in the $D^{*+} D^0$  CM frame.
The cross section for $D^{*+} D^0$ differential in  $E$  is
\begin{eqnarray}
\frac{d\sigma}{dE}[D^{*+}   D^0] =
\sigma[T_{cc}^+,\mathrm{no}\,\pi] \, \frac{\mu/(\pi \gamma_T)}{2 \mu E+\gamma_T^2}\, (2 \mu E)^{1/2}  .
\label{DstarD-T}
\end{eqnarray}
Note that $\sigma[T_{cc}^+,\mathrm{no}\,\pi]$ has a factor of $\gamma_T$ that cancels the explicit factor of $1/\gamma_T$.
This relation between the cross sections  
is consistent with the imaginary part of the universal scattering amplitude in Eq.~\eqref{Imf-E}.

The $T_{cc}^+$ can also be produced from the creation of $D^{*0} D^+$ at short distances,
with the subsequent formation of $T_{cc}^+$ proceeding through the $D^{*0} D^+$ component of its wavefunction.
This contribution can be taken into account in the coupled-channel model introduced in Section~\ref{sec:Molecule},
which can be implemented by using the prescriptions in Eqs.~\eqref{psi-sub} and \eqref{psi0+-sub}.
The amplitudes for producing $T_{cc}^+$  through its $D^{*+} D^0$ and $D^{*0} D^+$ components
have different  short-distance factors $\mathcal{A}^i_{D^{*+}  D^0+y}$ and $\mathcal{A}^i_{D^{*0}  D^++y}$.
The cross section therefore has interference terms with the short-distance factors
$\langle \mathcal{A}^i_{D^{*+}  D^0} (\mathcal{A}^j_{D^{*0}  D^+})^* \rangle$
and $\langle \mathcal{A}^i_{D^{*0}  D^+} (\mathcal{A}^j_{D^{*+}  D^0})^* \rangle$.
They are suppressed by the random phases in the sum over the many additional particles $y$.
The terms with short-distance factors $\langle \mathcal{A}^i_{D^{*+}  D^0} (\mathcal{A}^j_{D^{*+}  D^0})^* \rangle$
and  $\langle \mathcal{A}^i_{D^{*0}  D^+} (\mathcal{A}^j_{D^{*0}  D^+})^* \rangle$
are not suppressed, because they are sums of positive quantities.
The contribution to the  $T_{cc}^+$ cross section from its $D^{*+} D^0$ and $D^{*0} D^+$ components 
can be obtained by replacing $|\psi_T(r\!=\!0)|^2$ in Eq.~\eqref{factor-DstarD-3}
by $|\psi_T^{(\Lambda)}(r\!=\!0)|^2/(1+Z_{0+})$ and $|\psi_{0+}^{(\Lambda)}(r\!=\!0)|^2/(1+Z_{0+})$, respectively.
Isospin symmetry at short distances implies that these two contributions are equal.
The total cross section in the coupled-channel model for producing $T_{cc}^+$  
without an accompanying soft pion is therefore
\begin{eqnarray}
 \sigma^{(\Lambda)}[T_{cc}^+,\mathrm{no}\,\pi] =
\Big\langle \mathcal{A}_{D^+   D^0} (\mathcal{A}_{D^+   D^0})^* \Big\rangle 
  \frac{3}{(1+Z_{0+})\mu}   \big| \psi_T^{(\Lambda)}(r\!=\!0) \big|^2.
\label{sigmaT-CC}
\end{eqnarray}
The coefficient of $| \psi_T^{(\Lambda)}(r\!=\!0) |^2$ is larger than 
the coefficient of $| \psi_T(r\!=\!0) |^2$ in Eq.~\eqref{factor-DstarD-3}
by the factor $2/(1+Z_{0+})$, which is 1.45 if $\Lambda = m_\pi$.


\section{Production of  $\bm{T_{cc}^+}$ and a Soft Pion}
\label{sec:Xsoftpi}

In this section, we consider the production of $T_{cc}^+$ and a  soft pion.
Our treatment of the triangle-singularity peaks
is similar to that for the prompt production of $X$ and a soft pion in Ref.~\cite{Braaten:2019sxh}.
We also consider the production of $T_{cc}^+$ and a pion with larger relative momentum.

\subsection{Triangle-Singularity Peaks}
\label{sec:XsoftpiAmp}

\subsubsection{Amplitude for $T_{cc}^+\pi^+$}
\label{sec:AmpT+pi+}

\begin{figure}[t]
\includegraphics*[width=0.4\linewidth]{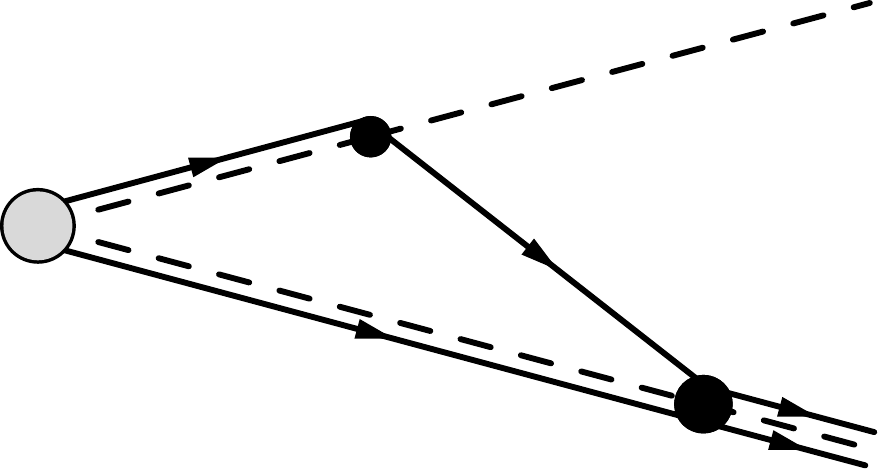} 
\caption{
Feynman diagram in XEFT for $D^* D^*$ created at a point to rescatter into $T_{cc}^+ \pi$.
The pion is represented by a dashed line.
}
\label{fig:DDtoXpi}
\end{figure}

Charm mesons $D^{*+} D^{*+}$ created at short distances can rescatter into $T_{cc}^+\pi^+$. 
The Feynman diagram in XEFT for the production of $T_{cc}^+\pi^+$ 
from the creation of $D^{*+}   D^{*+}$ at a point is shown in Fig.~\ref{fig:DDtoXpi}.
The blob on the left side is the vertex factor $i\, \mathcal{A}^{ij}_{D^{*+}   D^{*+} + y}$
for creating $D^{*+} D^{*+}$ at a point while producing additional particles $y$
with large momenta in the $D^{*+} D^{*+}$ CM frame.
The vertex factor is symmetric in the vector indices $ij$.
The $D^{*+} $-to-$D^0\pi^+$ vertex  is given in Eq.~\eqref{vertex:D*toDpi}.
The $D^{*+} D^0$-to-$T_{cc}^+$ vertex  is given in Eq.~\eqref{vertex:TtoD*D}.

We take the relative  momentum of $T_{cc}^+\pi^+$ in their CM frame to be $\bm{q}$.
The integral over the loop energy in the diagram in Fig.~\ref{fig:DDtoXpi}
is conveniently evaluated by contours using the pole of the 
propagator for the  $D^{*+}$ line attached to the $T_{cc}^+$.
The resulting amplitude for producing $T_{cc}^+\pi^+$ is
\begin{eqnarray}
\mathcal{A}_{T_{cc}^+\,\pi^+ +y} (\bm{q}) 
&=& i \left(\mathcal{A}^{ij}_{D^{*+}   D^{*+}+y} \sqrt{M_T m/M_*^2} \right)
 4\pi G_\pi M_* \sqrt{\gamma_T}  \, \varepsilon^{i*}
\nonumber\\
 && \hspace{1cm}
 \times
  \!\!\int\!\! \frac{d^3k}{(2\pi)^3} 
\frac{1}{\big( \bm{k} + (\mu/M)  \bm{q}\big)^2 +\gamma^2}\, 
 \frac{q^j  + (m/M_*) k^j }{\bm k^2  - (\mu/\mu_{\pi}) \bm q^2 + M_* E_+ },
\label{amplitudeXpi0intk}
\end{eqnarray}
where  $\bm{\varepsilon}$ is the polarization vector for $T_{cc}^+$ and $G_\pi = \sqrt{g^2/4 \pi m f_\pi^2}$.
The factor $\sqrt{M_T m/M_*^2}$ takes into account the difference between relativistic 
and nonrelativistic normalizations of states.
The integral is over the loop momentum $\bm k$ of the $D^{*+}$  that becomes a constituent of $T_{cc}^+$.
In the first denominator in the integrand,
$\gamma$ is the complex binding momentum:  $\gamma^2 =  - 2 \mu(\varepsilon_T+ i \,\Gamma_{*+}/2)$,
where $\Gamma_{*+} = 83$~keV is the $D^{*+}$ decay width.
In the second denominator, $E_+$ is the complex energy
\begin{equation}
E_+ =  \delta_{0+} -\varepsilon_T - i \, \Gamma_{*+},
\label{E+}
\end{equation}
where $\delta_{0+}=M_{*+} \!-\! M_0 \!-\! m_+ = 5.9$~MeV. 
A necessary (but not necessarily sufficient) condition for the validity of the amplitude  in Eq.~\eqref{amplitudeXpi0intk} 
is that the integral should be dominated by regions in which the relative momentum between the two charm mesons connected to $T_{cc}^+$ is less than order $m_\pi$.
If we require the relative momentum to be less than $m_\pi/2$, $m_\pi$, or $2 m_\pi$, the total kinetic energy $E = q^2 /(2\mu_{\pi T})$ of $T_{cc}^+ \pi^+$ is required to be less than 7, 12, or 32~MeV.

The two denominators in Eq.~\eqref{amplitudeXpi0intk}
can be combined into a single denominator by introducing an integral over a Feynman parameter. 
After evaluating the integral over the loop momentum, 
the amplitude for producing $T_{cc}^+ \pi^+$ can be reduced to the form
\begin{eqnarray}
\mathcal{A}_{T_{cc}^+\,\pi^+ +y} (\bm{q})
= - G_\pi \sqrt{M_T m\gamma_T/4} \, \mathcal{A}^{ij}_{D^{*+}   D^{*+}+y} \varepsilon^{i*}q^j \,  T_+(q^2,\gamma^2).
\label{amplitude-Tpi+}
\end{eqnarray}
The triangle amplitude $T_+(q^2,\gamma^2)$ depends on its two explicit arguments and also on
the complex energy $E_+$ in Eq.~\eqref{E+}.
It can be expressed as a Feynman parameter integral of the form
\begin{eqnarray}
T_+(q^2,\gamma^2) = \int_0^1 dx \frac{1 - (m/M_T) x}{\sqrt{a+bx+cx^2}}.
\label{G+integral}
\end{eqnarray}
The integral over $x$ in Eq.~\eqref{G+integral} can be evaluated analytically:
\begin{eqnarray}
T_+(q^2,\gamma^2) = 
    \left(1 +\frac{m b}{2M_T c} \right)\frac{1}{\sqrt{c}}
    \log\frac{\sqrt{a}+\sqrt{c} + \sqrt{a+b+c}}{\sqrt{a}-\sqrt{c} + \sqrt{a+b+c}} 
+ \frac{m}{M_T c} \left(\sqrt{a} -  \sqrt{a+b+c}  \right).~~~
\label{Gqfunction}
\end{eqnarray}
The coefficients  in these equations are
\begin{subequations}
\begin{eqnarray}
a &=&  (\mu/\mu_{\pi})  q^2 -  M_*E_+ ,
\label{parametera}
 \\ 
b &=&- 2 (\mu/\mu_{\pi}) (\mu / M)q^2 + M_*E_+
 - \gamma^2, 
\label{parameterb}
\\ 
c &=& (\mu / M)^{2} q^{2}.
\label{parameterc}
\end{eqnarray}
\label{parameterabc}%
\end{subequations}
Note that $a+b+c$ does not depend on  $q^2$, and it approaches 0 in the limits $\varepsilon_T \to 0$,  $\Gamma_{*+}\to 0$.
Its square root is $\sqrt{a+b+c} = i\,  \gamma$.

The denominator of the argument of the logarithm in Eq.~\eqref{Gqfunction}
has a zero at a complex value of $q^2$ that approaches the real axis in the limit where
the binding energy $|\varepsilon_T|$ and the width $\Gamma_{*+}$ both go to zero.
This is the triangle singularity.  It is convenient to express the singularity in terms of the
total kinetic energy $E=q^2/(2\mu_{\pi T})$ of $T_{cc}^+\pi^+$.
The triangle amplitude $T_+(q^2,\gamma^2)$ has a logarithmic branch point 
at the triangle-singularity energy
\begin{eqnarray}
E_{\triangle +} = \frac{M_*}{4\mu^2}\left( \sqrt{2 \mu E_+ - \gamma^2} - i \sqrt{m/M_T} \, \gamma \right)^2.
\label{Etriangle+}
\end{eqnarray}
This is the complex energy where the three charm-meson lines that form a triangle 
in the Feynman diagram in Fig.~\ref{fig:DDtoXpi} are all simultaneously on shell.
The limit of the  triangle-singularity energy as $\varepsilon_T \to 0$, $\Gamma_{*+}\to 0$ is
\begin{eqnarray}
E_{\triangle +} \longrightarrow (M_T/2M) \delta_{0+} = 6.1~\mathrm{MeV}.
\label{Etriangle+limit}
\end{eqnarray}
The triangle amplitude $T_+(q^2,\gamma^2)$ also has a square-root branch point at $E=E_+$
from the $\sqrt{a}$ terms in Eq.~\eqref{Gqfunction}.
The limiting behavior of $T_+(q^2,\gamma^2)$ near the triangle singularity is 
determined by the interplay between the singularities at 
$E_{\triangle +}$ and $E_+$, as discussed in Appendix~\ref{sec:TriAmpLimit}.

\subsubsection{Amplitude for $T_{cc}^+\pi^0$}
\label{sec:AmpT+pi0}

Charm mesons $D^{*+} D^{*0}$ created at short distances can rescatter into $T_{cc}^+ \pi^0$.  
In the Feynman diagram in Fig.~\ref{fig:DDtoXpi},
the vertex factor for the creation of $D^{*+} D^{*0}$ at a point is $i\mathcal{A}^{ij}_{D^{*+}   D^{*0}+y}$.
 The amplitude for producing $T_{cc}^+\pi^0$ with small relative momentum $\bm{q}$  in their CM frame is
\begin{eqnarray}
\mathcal{A}_{T_{cc}^+\,\pi^0 +y} (\bm{q}) = 
- G_\pi \sqrt{M_T m\gamma_T/8}  \, \mathcal{A}^{ij}_{D^{*+}   D^{*0}+y}  \varepsilon^{i*} q^j \,   T_0(q^2,\gamma^2).
\label{amplitude-Tpi0}
\end{eqnarray}
The triangle amplitude $T_0(q^2,\gamma^2)$ is given by the right side of Eq.~\eqref{Gqfunction} with 
 $E_+$ in the coefficients $a$ and $b$ replaced by the complex energy
\begin{equation}
E_0 =\delta_{00}  -  \varepsilon_T - i \, (\Gamma_{*0} + \Gamma_{*+})/2,
\label{E0}
\end{equation}
where $\delta_{00}=M_{*0} \!-\! M_0 \!-\! m_0= 7.0$~MeV
and $\Gamma_{*0} \approx 55$~keV is the predicted decay width of $D^{*0}$.

The amplitude $T_0(q^2,\gamma^2)$ has a  triangle singularity from the logarithm in Eq.~\eqref{Gqfunction}.
The logarithmic branch point is at the complex triangle-singularity energy
\begin{eqnarray}
E_{\triangle 0} = \frac{M_*}{4\mu^2}\left( \sqrt{2 \mu E_0 - \gamma^2} - i \sqrt{m/M_T} \, \gamma \right)^2.
\label{Etriangle0}
\end{eqnarray}
The limit of the  triangle-singularity energy as $\varepsilon_T \to 0$,  $\Gamma_{*+}\to 0$, $\Gamma_{*0}\to 0$  is
\begin{eqnarray}
E_{\triangle 0}  \longrightarrow (M_T/2M) \delta_{00} = 7.3~\mathrm{MeV}.
\label{Etriangle0limit}
\end{eqnarray}
The triangle amplitude $T_0(q^2,\gamma^2)$ also has a square-root branch point at $E=E_0$.

\subsubsection{Cross sections}
\label{sec:Tripeaks+}

The inclusive differential cross section for producing $T_{cc}^+ \pi^+$ with 
 small relative momentum $\bm{q}$ in their CM frame can be expressed as
\begin{eqnarray}
d\sigma[T_{cc}^+ \, \pi^+] =
\frac{1}{\mathrm{flux}} \sum_{T_{cc}^+\, \mathrm{spins}}  \sum_y  \int d\Phi_{(D^*  D^*)+y}
\Big| \mathcal{A}_{T_{cc}^+\pi^+ +y}(\bm{q})  \Big|^2  \frac{d^3q}{(2 \pi)^3 2\mu_{\pi T}},
\label{sigma-Tpi}
\end{eqnarray}
where $d\Phi_{(D^*  D^*)+y}$ is defined after Eq.~\eqref{<AA-D*+D*0>}.
The relativistic differential phase space for $T_{cc}^+ \pi^+$ has been expressed as the product of
the differential phase space $d^3P/[(2 \pi)^3 2P_0]$ for the composite particle  $(D^*  D^*)$
and $d^3q/[(2\pi)^3 2\mu_{\pi T}]$, where $\mu_{\pi T}$ is  the $T_{cc}^+ \pi^+$ reduced mass.
The differential cross section for producing $T_{cc}^+ \pi^0$ is obtained by replacing
$\mathcal{A}_{T_{cc}^+\pi^+ +y}(\bm{q})$ in Eq.~\eqref{sigma-Tpi} by $\mathcal{A}_{T_{cc}^+\pi^0 +y}(\bm{q}) $.
The weighted average $\langle \mathcal{A}^{ij}_{D^{*} D^{*}}(\mathcal{A}^{kl}_{D^{*} D^{*}})^*\rangle$
of the product of short-distance amplitudes is defined by Eq.~\eqref{<AA-D*+D*0>}
followed by the average over the spin states of the colliding protons 
and over the directions of their momenta  in the $T_{cc}^+ \pi$ rest frame.
After multiplying by $(\varepsilon^i q^j )^* (\varepsilon^k q^l )$,
the weighted averages can be simplified using Eqs.~\eqref{<AA-D*+D*0>tensor} and \eqref{<AA-D*+D*+>tensor}.
The sum over the spin states of $T_{cc}^+$ results in a factor $q^2$:
\begin{subequations}
\begin{eqnarray}
\sum_{T_{cc}^+~\mathrm{spins}} \left\langle\mathcal{A}^{ij}_{D^{*+} D^{*+}} 
\big(\mathcal{A}^{kl}_{D^{*+} D^{*+}}\big)^*\right\rangle
(\varepsilon^i q^j )^* (\varepsilon^k q^l )
&=&  2 q^2\left\langle  \mathcal{A}_{D^{+} D^{+}}  \big(\mathcal{A}_{D^{+} D^{+}}  \big)^*\right\rangle ,
\label{<AA>epseps++}%
\\
\sum_{T_{cc}^+~\mathrm{spins}} \left\langle\mathcal{A}^{ij}_{D^{*+} D^{*0}} 
\big(\mathcal{A}^{kl}_{D^{*+} D^{*0}}\big)^*\right\rangle
(\varepsilon^i q^j )^* (\varepsilon^k q^l )
&=&  3 q^2\left\langle   \mathcal{A}_{D^{+} D^{0}}\big( \mathcal{A}_{D^{+} D^{0}} \big)^*  \right\rangle .
\label{<AA>epseps+0}%
\end{eqnarray}
\end{subequations}

The total kinetic energy $E = q^2/(2 \mu_{\pi T})$ of $T_{cc}^+ \pi$ in their CM  frame
is called the invariant kinetic energy, because it is invariant under Galilean boosts.
The differential cross sections for $T_{cc}^+ \pi^+$ and $T_{cc}^+ \pi^0$ as functions of $E$  are
\begin{subequations}
\begin{eqnarray}
\frac{d\sigma}{dE} [T_{cc}^+ \pi^+] &=&
\Big\langle  \mathcal{A}_{D^{+} D^{0}}  \big(\mathcal{A}_{D^{+} D^{0}}  \big)^*\Big\rangle 
\frac{G_\pi^2 M_T  m \gamma_T}{4\pi^2} (2 \mu_{\pi T} E)^{3/2}
 \big| T_+(2 \mu_{\pi T} E,\gamma^2) \big|^2 ,
\label{dsigmaTpi+tri}
\\
\frac{d\sigma}{dE} [T_{cc}^+ \pi^0] &=&
\Big\langle  \mathcal{A}_{D^{+} D^{0}}  \big(\mathcal{A}_{D^{+} D^{0}}  \big)^*\Big\rangle 
\frac{3G_\pi^2 M_T m \gamma_T}{32\pi^2}(2 \mu_{\pi T} E)^{3/2}
 \big| T_0 (2 \mu_{\pi T} E,\gamma^2) \big|^2 ,
\label{dsigmaTpi0tri}
\end{eqnarray}
\label{dsigmaTpitri}%
\end{subequations}
where $G_\pi^2 = g^2/(4 \pi m f_\pi^2)$.
We have used Eq.~\eqref{factor-D+D+/D+D0} to express both cross sections in terms of the same
short-distance factor $\langle \mathcal{A}_{D^{+}   D^{0}} (\mathcal{A}_{D^{+}   D^{0}})^* \rangle$
that appears in the cross section for $T_{cc}^+$ in Eq.~\eqref{factor-DstarD-3}.
These cross sections depend on $\varepsilon_T$ through the explicit factor of $\gamma_T$ 
and through the triangle amplitudes $T_+$ and $T_0$.

\begin{figure}[t]
\includegraphics*[width=0.8\linewidth]{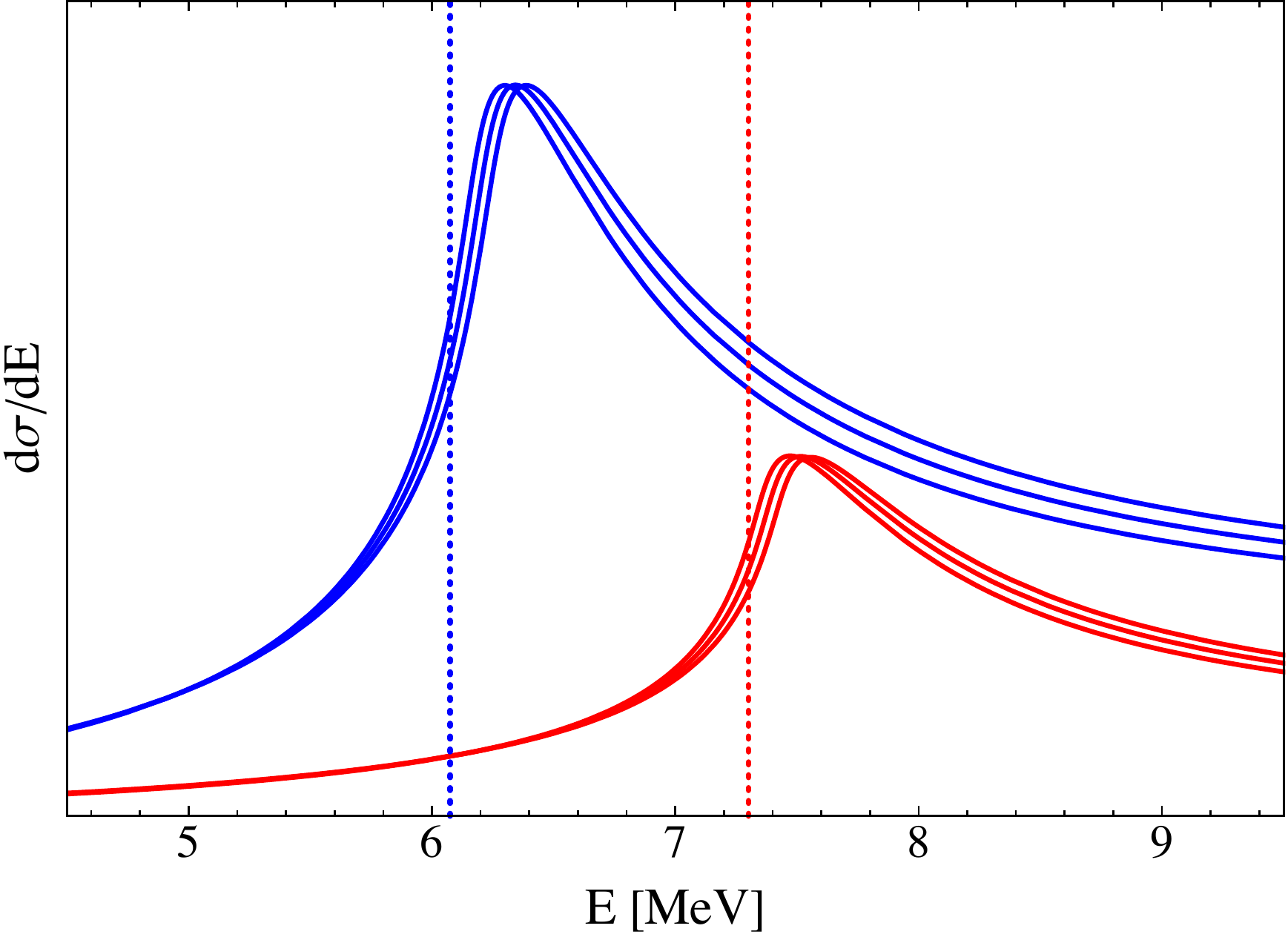} 
\caption{
Differential cross sections $d\sigma/dE$ from Eqs.~\eqref{dsigmaTpitri} as  functions of the invariant kinetic energy $E$
for  $T_{cc}^+\pi^+$ (left blue curves) and for $T_{cc}^+\pi^0$ (right red curves).
The binding energies of $T_{cc}^+$ are 320, 360, and 400~keV 
in order of increasing energy at the peak.
The vertical dotted lines are at the limiting  triangle-singularity energies 
in Eqs.~\eqref{Etriangle+limit} and \eqref{Etriangle0limit}.
The scale on the vertical axis is arbitrary.
}
\label{fig:Xpi-q}
\end{figure}

The dependence of the differential cross sections $d\sigma/dE$ for $T_{cc}^+ \pi^+$ and  $T_{cc}^+ \pi^0$
on the invariant kinetic energy $E$ is illustrated in Fig.~\ref{fig:Xpi-q} 
for three values of the $T_{cc}^+$ binding energy: $|\varepsilon_T| = 320$, 360, and 400~keV.
The differential cross sections for $T_{cc}^+ \pi^+$ and $T_{cc}^+ \pi^0$ each has  a narrow peak
near the limiting triangle-singularity energy $E_{\triangle +}$ in Eq.~\eqref{Etriangle+limit} and 
$E_{\triangle 0}$ in Eq.~\eqref{Etriangle0limit}, respectively.
The full width at half maximum of the peak is about 1~MeV.
As $|\varepsilon_T|$ decreases, the energy at the peak approaches the limiting triangle-singularity energy.
It decreases through that energy when $|\varepsilon_T|$ decreases below about 0.1~MeV.
The shape of $d \sigma/dE$ near the peak is determined by the interplay 
between the  logarithmic singularity and the square-root singularity  in the triangle amplitude. 
Beyond the triangle-singularity peaks, the cross sections predicted by Eqs.~\eqref{dsigmaTpitri} 
decrease to a local minimum and then begin to increase.
The energy  at the local minimum is insensitive to $\varepsilon_T$: $E_{\mathrm{min},+}=17.5$~MeV
for $T_{cc}^+ \pi^+$ and $E_{\mathrm{min},0} =21.2$~MeV for $T_{cc}^+ \pi^0$.

\begin{figure}[t]
\includegraphics*[width=0.8\linewidth]{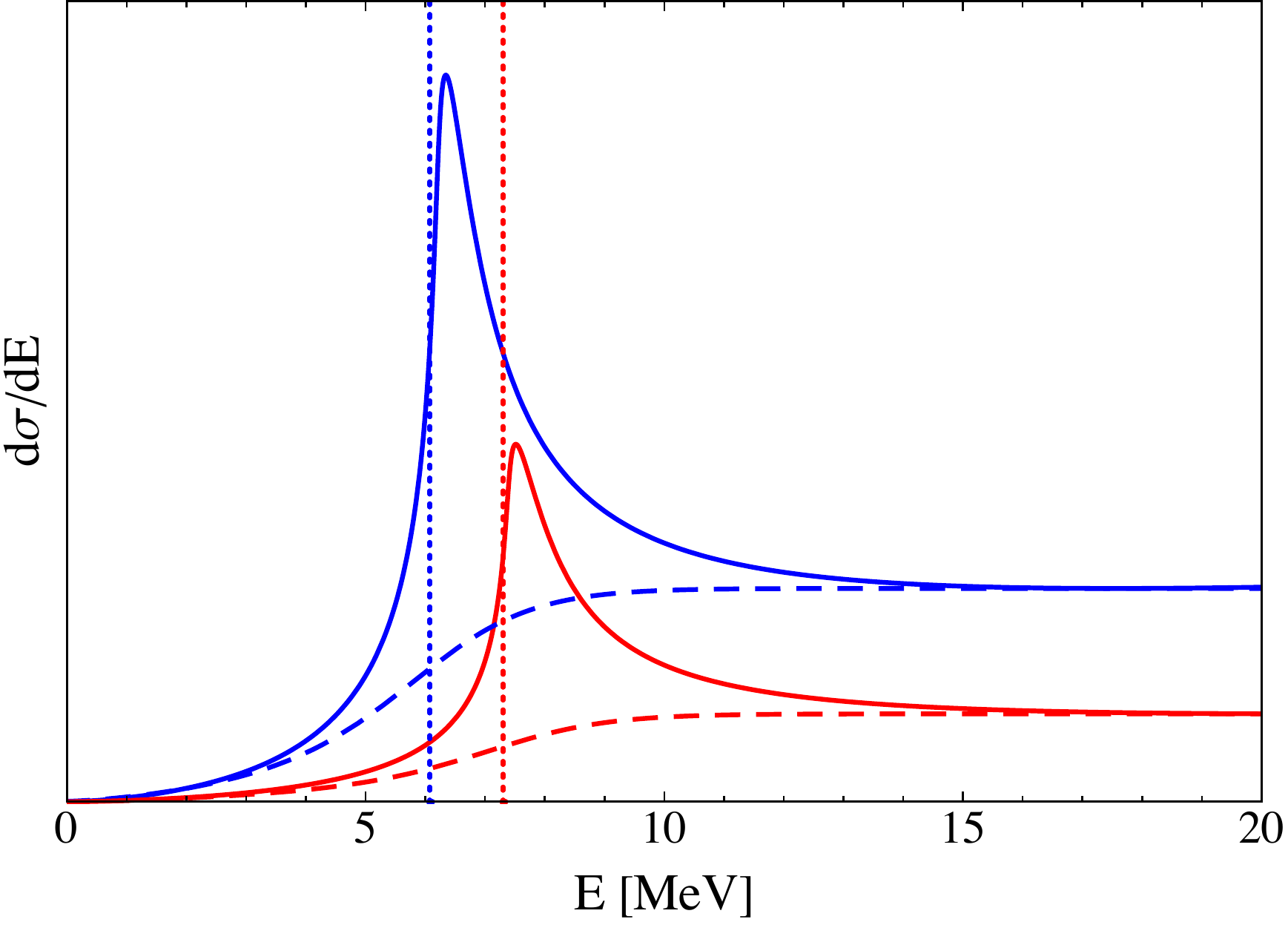} 
\caption{
Differential cross sections $d\sigma/dE$ from Eqs.~\eqref{dsigmaTpitri} as functions of the invariant kinetic energy $E$ 
for $T_{cc}^+\pi^+$ (left blue curves) and for $T_{cc}^+\pi^0$  (right red curves). 
The  binding energy of $T_{cc}^+$ is $|\varepsilon_T| =360$~keV.
The dashed curves are simple models for the backgrounds.
The vertical dotted lines are at the limiting triangle-singularity  energies in Eqs.~\eqref{Etriangle+limit} and \eqref{Etriangle0limit}.
The scale on the vertical axis is arbitrary.
}
\label{fig:Xpi-EX}
\end{figure}

We would like quantitative estimates of the contributions to the cross sections 
for $T_{cc}^+ \pi^+$ and $T_{cc}^+ \pi^0$ from the triangle-singularity peaks.
To quantify such a cross section, it is necessary to make a model for the background under the peak.
A simple model for the background for $d\sigma/dE$ can be obtained by 
interpolating between the leading power of $E$ at small $E$, which is  $E^{3/2}$,
and a constant at large $E$ equal to the value of $d\sigma/dE$ at the local minimum.
The factor $q^3 \, |T_+(q^2,\gamma^2)|^2$
in the differential cross section $d\sigma/dE$ for the production of $T_{cc}^+ \pi^+$ 
has a local minimum at a momentum $q_{\mathrm{min},+}\!=\!\sqrt{2\mu_{\pi T} E_{\mathrm{min},+}}$
well above the triangle-singularity peak.
Our model for the background function for $T_{cc}^+ \pi^+$ is
\begin{eqnarray}
q^3\, \big|T_+^\mathrm{(bg)}(q^2,\gamma^2) \big|^2 &=&
n\big( (E-M_T\delta_{0+}/2M)/\Gamma_\times \big) \,  q^3\, \big| T_+(0,\gamma^2) \big|^2
\nonumber\\
&&+ \big[ 1 - n\big( (E-M_T\delta_{0+}/2M)/\Gamma_\times\big) \big] \, 
q^3_{\mathrm{min},+}\, \big| T_+(q^2_{\mathrm{min},+},\gamma^2) \big|^2,
\label{Gbackground}
\end{eqnarray}
where $n(x) = 1/(e^x+1)$ and $E= q^2/(2 \mu_{\pi T})$.  
Our model for the background function  for $T_{cc}^+ \pi^0$ can be obtained from Eq.~\eqref{Gbackground}
by replacing $T_+(q^2,\gamma^2)$ by  $T_0(q^2,\gamma^2)$,
$\delta_{0+}$ by $\delta_{00}$, and $q_{\mathrm{min},+}$ by $q_{\mathrm{min},0}$.
The adjustable parameter $\Gamma_\times$ in Eq.~\eqref{Gbackground} controls the width of the crossover 
from $q^3| T_+(0,\gamma^2)|^2$ to the constant $q^3_{\mathrm{min},+}| T_+(q^2_{\mathrm{min},+},\gamma^2) |^2$.
We choose $\Gamma_\times = 1$~MeV.  The resulting background curves 
for $T_{cc}^+  \pi^+$ and $T_{cc}^+  \pi^0$ are shown in Fig.~\ref{fig:Xpi-EX}.

We denote the peaks in the cross sections above the background curves 
by $(T_{cc}^+  \pi^+)_\triangle$ and $(T_{cc}^+  \pi^0)_\triangle$.
The cross sections for $(T_{cc}^+  \pi^+)_\triangle$ and $(T_{cc}^+  \pi^0)_\triangle$
can be estimated by integrating over the regions below the curves given by Eqs.~\eqref{dsigmaTpitri}
and above the corresponding backgrounds from the threshold to the local minimum.
The integrated cross sections for producing $(T_{cc}^+ \pi)_\triangle$ can be 
expressed in terms of the cross section for producing $T_{cc}^+$ without an accompanying soft pion
by eliminating the short-distance factor using Eq.~\eqref{factor-DstarD-3}:
\begin{subequations}
\begin{eqnarray}
\sigma\big[  (T_{cc}^+\,  \pi^+)_\triangle \big] &\approx& 
\left(  8.6\pm 0.5 \right) \times 10^{-3}\, \frac{m_\pi^2 \gamma_T/{2\pi}}{|\psi_T(r\!=\!0)|^2} 
\sigma[T_{cc}^+,\mathrm{no}\,\pi]\, ,
\label{sigmaTpi+tri}
\\
\sigma\big[  (T_{cc}^+\,  \pi^0)_\triangle \big] &\approx& 
\left(  4.8 \pm 0.2 \right) \times10^{-3}\, \frac{m_\pi^2 \gamma_T/{2\pi}}{|\psi_T(r\!=\!0)|^2} 
\sigma[T_{cc}^+,\mathrm{no}\,\pi]\,  .
\label{sigmaTpi0tri}
\end{eqnarray}
\label{sigmaTpitri}%
\end{subequations}
The errors in the numerical prefactors come from the uncertainty in the binding energy $|\varepsilon_T| = 360\pm 40$~keV.
The largest uncertainty comes from
the factor $(m_\pi^2 \gamma_T/2\pi)/|\psi_T(r\!=\!0)|^2$.  
Using the expression for the universal wavefunction at the origin in Eq.~\eqref{psi-0}, this factor
can be approximated by $(m_\pi/\Lambda)^2$,
where $\Lambda$ is the ultraviolet cutoff.
If $\Lambda$ is larger or smaller than $m_\pi$ by a factor of 2, that factor is smaller or larger than 1 by a factor of 4.

\subsection{Coupled-channel model}
\label{sec:Xsoftpisigma}

\subsubsection{Cross sections }
\label{sec:sigmaCC}

The limiting behavior of the triangle amplitude $T_+(q^2,\gamma^2)$  
at large $q^2$ is determined  in Eq.~\eqref{Ginfinity} of Appendix~\ref{sec:TriAmpLimit}: 
$T_+(q^2,\gamma^2) \to 0.724/q$.
The triangle amplitude $T_0(q^2,\gamma^2)$  has the same  limiting behavior.
The differential cross sections $d\sigma/dE$ for $T_{cc}^+  \pi^+$ and $T_{cc}^+  \pi^0$ in Eqs.~\eqref{dsigmaTpitri}
therefore increase asymptotically as $E^{1/2}$ at large $E$.
This unphysical behavior is an artifact of using the universal approximation for $T_{cc}^+$ beyond its range of applicability.

The coupled-channel model introduced in Section~\ref{sec:Molecule}
is a simple model with universal behavior at long distances and more physical qualitative behavior at short distances.
The model is specified by wavefunctions for both the $D^{*+} D^0$ and $D^{*0}D^+$ components of $T_{cc}^+$,
 whose parameters are the binding momenta $\gamma_T$ and $\gamma_{0+}$ 
and a larger momentum scale $\Lambda$ that we assume to be order $m_\pi$.
The model can be implemented by making the substitutions in Eqs.~\eqref{psi-sub} and \eqref{psi0+-sub}
in amplitudes from XEFT.
The model gives predictions for the production of not only  
$T_{cc}^+  \pi^+$ and $T_{cc}^+  \pi^0$ but $T_{cc}^+  \pi^-$ as well.
The amplitudes in the coupled-channel model are given in Appendix~\ref{sec:TriAmpCC}.
They are expressed in terms of  simple triangle amplitudes $T_+$,  $T_0$, and $T_-$
that do not depend on $\Lambda$.
The amplitude for $T_{cc}^+  \pi^+$ has a contribution only from the  $D^{*+} D^0$ component of $T_{cc}^+$.
The amplitude for $T_{cc}^+  \pi^-$ has a contribution only from the  $D^{*0} D^+$ component of $T_{cc}^+$.
The amplitude for $T_{cc}^+  \pi^0$ has contributions from both 
the  $D^{*+} D^0$ and $D^{*0} D^+$ components of $T_{cc}^+$
with short-distance factors $\mathcal{A}^{ij}_{D^{*+}  D^{*0}+y}$
and $\mathcal{A}^{ij}_{D^{*0}  D^{*+}+y}$, respectively.
The cross section for $T_{cc}^+ \pi^0$ has interference terms with the short-distance factors
$\langle \mathcal{A}^{ij}_{D^{*0}  D^{*+}} (\mathcal{A}^{kl}_{D^{*+}  D^{*0}})^* \rangle$ 
and $\langle \mathcal{A}^{ij}_{D^{*+}  D^{*0}} (\mathcal{A}^{kl}_{D^{*0}  D^{*+}})^* \rangle$.
They are suppressed by the random phases in the sum over the many additional particles $y$.
The terms with the short-distance factors
$\langle \mathcal{A}^{ij}_{D^{*+}  D^{*0}} (\mathcal{A}^{kl}_{D^{*+}  D^{*0}})^* \rangle$
and $\langle \mathcal{A}^{ij}_{D^{*0}  D^{*+}} (\mathcal{A}^{kl}_{D^{*0}  D^{*+}})^* \rangle$ are not suppressed,
because they are sums of positive quantities.
The short-distance factors in the cross sections for  $T_{cc}^+  \pi^+$, $T_{cc}^+  \pi^0$, and $T_{cc}^+  \pi^-$
can be reduced to the same factor $\langle \mathcal{A}_{D^+   D^0} (\mathcal{A}_{D^+   D^0})^*\rangle$ 
as in the cross section for $T_{cc}^+$ in  Eq.~\eqref{sigmaT-CC}.

\begin{figure}[t]
\includegraphics*[width=0.8\linewidth]{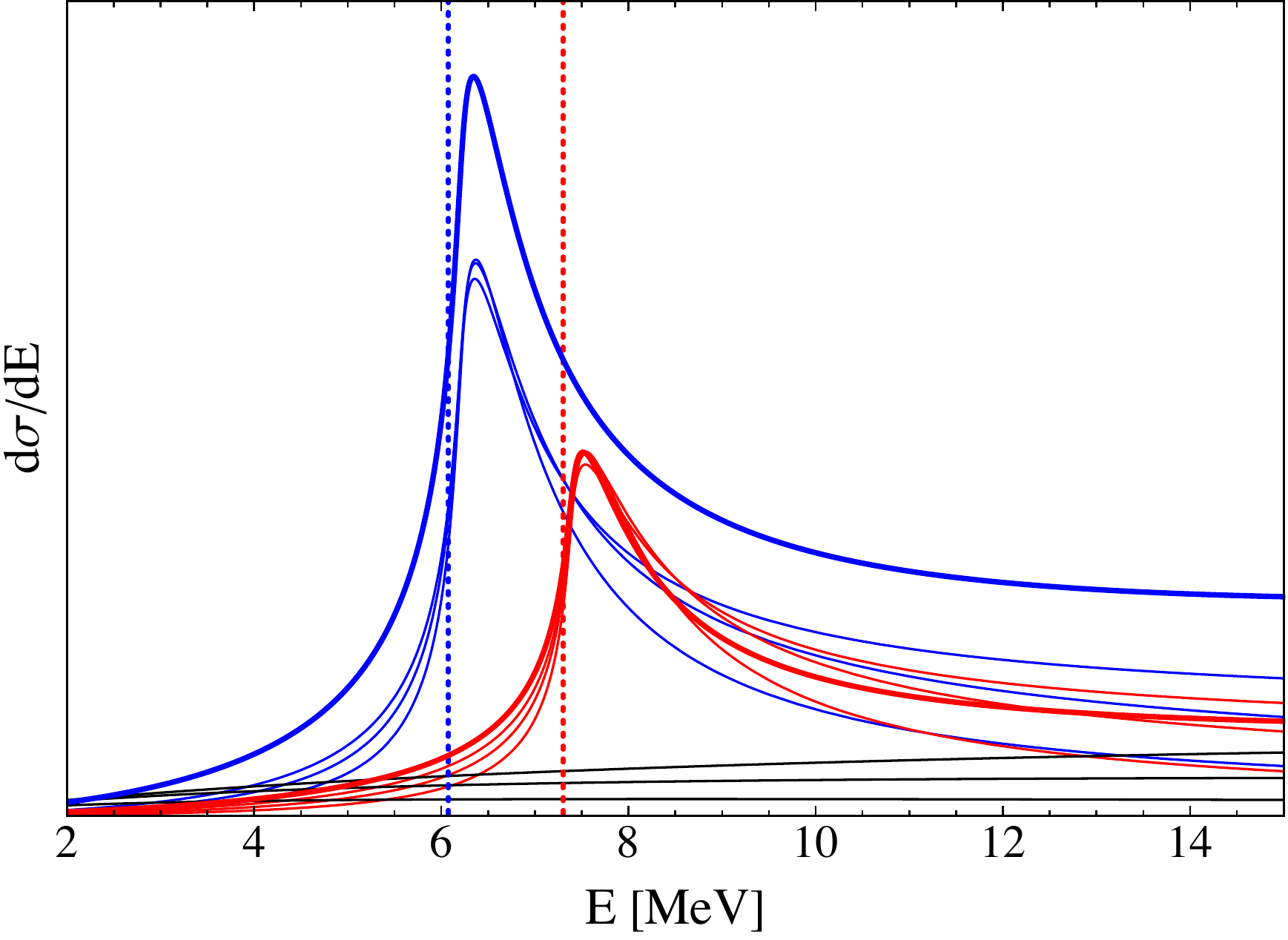} 
\caption{
Differential cross sections $d\sigma/dE$ as functions of the invariant kinetic energy $E$ 
for $T_{cc}^+\pi^+$ (left blue curves), $T_{cc}^+\pi^0$  (right red curves), and $T_{cc}^+\pi^-$ (lower black curves).
The  binding energy of $T_{cc}^+$ is $|\varepsilon_T| =360$~keV.
The thicker curves for $T_{cc}^+\pi^+$ and $T_{cc}^+\pi^0$ were calculated using Eqs.~\eqref{dsigmaTpitri}.
The thinner curves for the coupled-channel model were calculated using Eqs.~\eqref{dsigmaTpiLambda},
with $\Lambda/m_\pi = 1/2$, 1, and 2 in order of increasing cross sections at small $E$ and at large $E$.
The vertical dotted lines are at the limiting triangle-singularity energies in Eqs.~\eqref{Etriangle+limit} and \eqref{Etriangle0limit}.
The scale on the vertical axis is arbitrary.
}
\label{fig:Xpi-ELambda}
\end{figure}

The differential cross sections for  $T_{cc}^+  \pi^+$, $T_{cc}^+  \pi^0$, and $T_{cc}^+  \pi^-$
in the  coupled-channel model as functions of the invariant kinetic energy $E$ are
\begin{subequations}
\begin{eqnarray}
\frac{d\sigma}{dE} [T_{cc}^+ \pi^+] &=& \Big\langle \mathcal{A}_{D^+   D^0} (\mathcal{A}_{D^+   D^0})^* \Big\rangle
\frac{G_\pi^2 M_T m \gamma_T}{4\pi^2} (2 \mu_{\pi T} E)^{3/2}
 \big| T^{(\Lambda)}_+(2 \mu_{\pi T} E,\gamma^2) \big|^2,
\label{dsigmaTpi+Lambda}
\\
\frac{d\sigma}{dE}[T_{cc}^+ \pi^0] &=& \Big\langle \mathcal{A}_{D^+   D^0} (\mathcal{A}_{D^+   D^0})^* \Big\rangle
\frac{3 G_\pi^2 M_T m \gamma_T}{32\pi^2}(2 \mu_{\pi T} E)^{3/2}
\nonumber\\
&&\hspace{3cm}
\times \left( \big| T^{(\Lambda)}_0 (2 \mu_{\pi T} E,\gamma^2) \big|^2
+ \big| T^{\prime(\Lambda)}_0 (2 \mu_{\pi T} E,\gamma_{0+}^2) \big|^2 \right),
\label{dsigmaTpi0Lambda}
\\
\frac{d\sigma}{dE}[T_{cc}^+ \pi^-] &=& \Big\langle \mathcal{A}_{D^+   D^0} (\mathcal{A}_{D^+   D^0})^* \Big\rangle
\frac{G_\pi^2 M_T m \gamma_T}{4\pi^2}(2 \mu_{\pi T} E)^{3/2}
 \big| T^{(\Lambda)}_- (2 \mu_{\pi T} E,\gamma_{0+}^2) \big|^2.
\label{dsigmaTpi-Lambda}
\end{eqnarray}
\label{dsigmaTpiLambda}%
\end{subequations}
The triangle amplitudes $T_+^{(\Lambda)}$,  $T_0^{(\Lambda)}$, $T_0^{\prime(\Lambda)}$, and $T_-^{(\Lambda)}$
are given in Appendix~\ref{sec:TriAmpCC}
in Eqs.~\eqref{G+Lambda-q}, \eqref{G0Lambda-q}, \eqref{G0’Lambda-q}, and \eqref{G-Lambda-q}.
The differential cross sections $d\sigma/dE$ are shown in Fig.~\ref{fig:Xpi-ELambda}
for $|\varepsilon_T| =360$~keV
and three values of the momentum scale $\Lambda$: $\Lambda/m_\pi = 1/2$, 1, and 2.
The cross sections for $T_{cc}^+  \pi^+$ and $T_{cc}^+  \pi^0$ with the universal triangle amplitudes are also shown.
The triangle-singularity peaks for $T_{cc}^+  \pi^+$ and $T_{cc}^+  \pi^0$
in the coupled-channel model have essentially the same shape as those with the universal triangle amplitudes.
The height of the peak for $T_{cc}^+  \pi^+$  in the coupled-channel model is smaller by
the multiplicative factor $1/(1+Z_{0+})$, which is 0.73 for $\Lambda = m_\pi$.
The height of the peak for $T_{cc}^+  \pi^0$ in the coupled-channel model is approximately equal to that with the universal triangle amplitude.  
This is the result of a fortuitous compensation between the multiplicative factor $1/(1+Z_{0+})$
and the additional contribution from the $D^{*0}D^+$ component of $T_{cc}^+$.
The limiting behaviors of $T_0(q^2,\gamma^2)$
and $T_0 (q^2,\gamma_{0+}^2)$
near the triangle singularity can be deduced from the limiting behavior of $T_+(q^2,\gamma^2)$ 
determined in Eq.~\eqref{Gtriangle:2} of Appendix~\ref{sec:TriAmpLimit}.  
The ratio of the contributions to the cross section at the triangle-singularity peak from the $D^{*0}D^+$ and $D^{*+}D^0$ 
components of $T_{cc}^+$ can be approximated by the absolute square of the ratio of 
the logarithms in $T^{(\mathrm{log})}_0(q^2,\gamma_{0+}^2)$ and $T^{(\mathrm{log})}_0(q^2,\gamma^2)$ at $q^2 = q^2_{\triangle 0}$, which is equal to 0.36  for $|\varepsilon_T| =360$~keV.
This ratio is close to the value $Z_{0+} = 0.38$ for $\Lambda = m_\pi$.
There is no triangle singularity in the production of $T_{cc}^+\pi^-$, 
because the mass of $D^{*0}$ is 2.4~MeV  below the threshold for decay into $D^+ \pi^-$.
This prevents the $D^{*0}$ and $D^+$ lines in the triangle diagram from being
simultaneously on shell.
The cross sections for $T_{cc}^+  \pi^-$ are therefore small and slowly increasing 
in the region where the cross sections for $T_{cc}^+  \pi^+$ and $T_{cc}^+  \pi^0$ have narrow peaks. 
At energies above the peaks, there is a significant decrease in all three cross sections as $\Lambda$ decreases.
The dependence on $\Lambda$ demonstrates that the cross sections 
above the triangle-singularity peaks are model dependent.

\subsubsection{High energy limits}
\label{sec:sigmaCChi}

\begin{figure}[t]
\includegraphics*[width=0.8\linewidth]{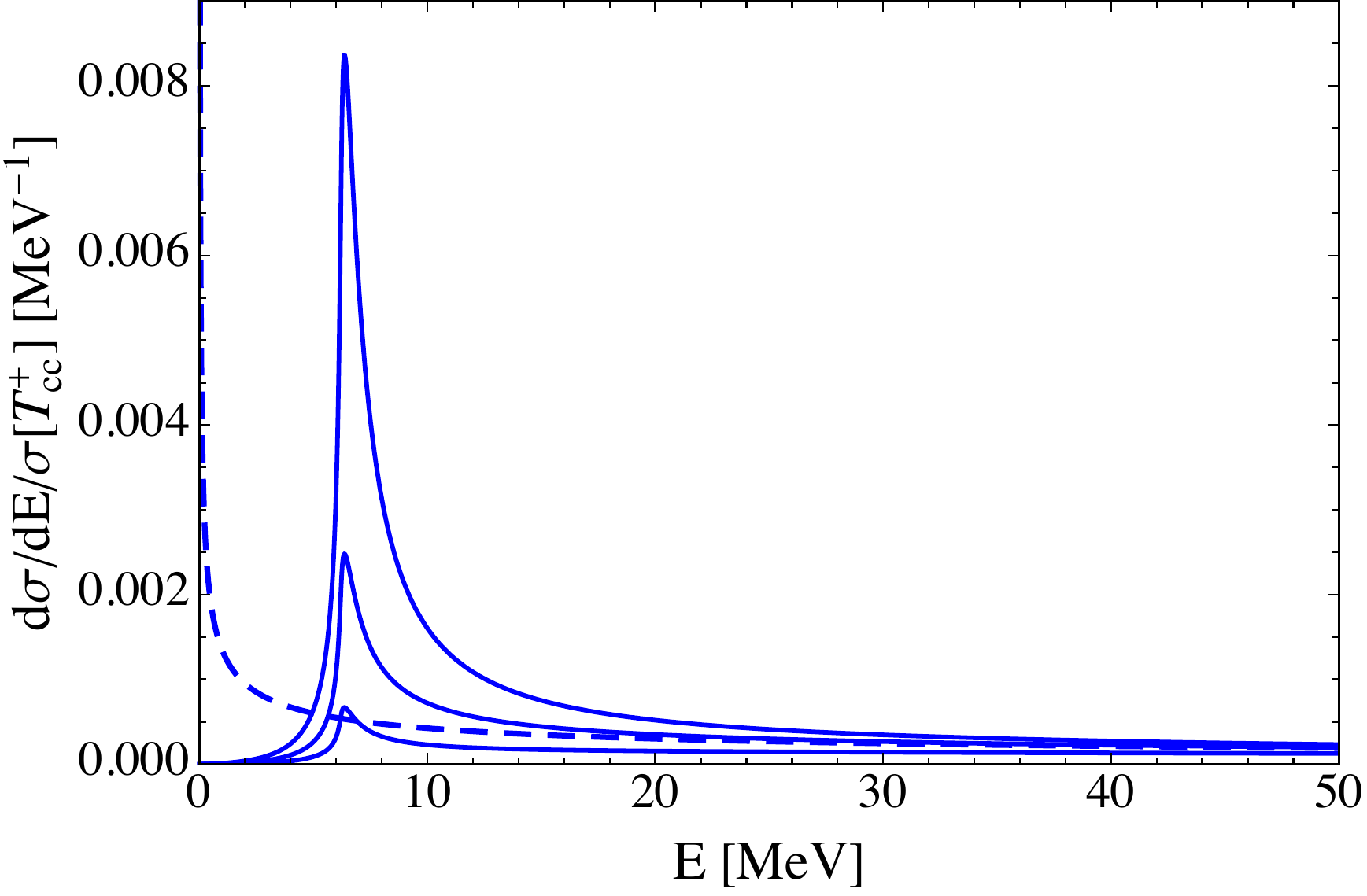} 
\caption{
Differential cross sections $d\sigma/dE$ divided by $\sigma^{(\Lambda)}[T_{cc}^+,\mathrm{no}\,\pi]$
as functions of the invariant kinetic energy $E$ for $T_{cc}^+\pi^+$. 
The  binding energy of $T_{cc}^+$ is $|\varepsilon_T| =360$~keV.
The solid curves are calculated using the triangle amplitude $T^{(\Lambda)}_+$
with $\Lambda/m_\pi = 1/2$, 1, and 2 in order of decreasing cross sections.
The dashed curve is the asymptotic result  from Eq.~\eqref{dsigmaTpi+:large}.
The scale on the vertical axis is in units of 1/MeV. }
\label{fig:Xpi-Elarge}
\end{figure}

The asymptotic behavior at large $q^2$ of the  triangle amplitude $T^{(\Lambda)}_+(q^2,\gamma^2)$ is determined in 
Eq.~\eqref{Greg-largepsi} of Appendix~\ref{sec:TriAmpCC}. 
It decreases asymptotically as $1/q^2$, with a coefficient that has a factor of $\psi_T^{(\Lambda)}(r\!=\!0)$.
The other triangle amplitudes  $T^{(\Lambda)}_0$, $T^{\prime(\Lambda)}_0$, and $T^{(\Lambda)}_-$
have the same asymptotic behavior up to a sign.
The differential cross sections for $T_{cc}^+  \pi^+$, $T_{cc}^+  \pi^0$, and $T_{cc}^+  \pi^-$
in Eqs.~\eqref{dsigmaTpiLambda} therefore all decrease asymptotically as $E^{-1/2}$ at large $E.$
The multiplicative short-distance factors in the cross sections for $T_{cc}^+ \pi$ in Eqs.~\eqref{dsigmaTpiLambda}
can be eliminated in favor of the cross section for $T_{cc}^+$ in Eq.~\eqref{sigmaT-CC}.
This also  eliminates the factors of $|\psi_T^{(\Lambda)}(r\!=\!0)|^2$.
The resulting expressions for the asymptotic behaviors of the differential cross sections 
for $T_{cc}^+ \pi^+$, $T_{cc}^+ \pi^0$, and $T_{cc}^+ \pi^-$ are
\begin{subequations}
\begin{eqnarray}
\frac{d\sigma}{dE}  [T_{cc}^+ \pi^+] &\longrightarrow& \sigma^{(\Lambda)}[T_{cc}^+,\mathrm{no}\,\pi]
\frac{8 G_\pi^2 \mu_{\pi T}^2 \mu_\pi}{3\pi} 
(2 \mu_{\pi T} E)^{-1/2} ,
\label{dsigmaTpi+:large}
\\
\frac{d\sigma}{dE}  [T_{cc}^+ \pi^0] &\longrightarrow&\sigma^{(\Lambda)}[T_{cc}^+,\mathrm{no}\,\pi]
\frac{2 G_\pi^2 \mu_{\pi T}^2 \mu_\pi}{\pi} 
 (2 \mu_{\pi T} E)^{-1/2} ,
\label{dsigmaTpi0:large}
\\
\frac{d\sigma}{dE} [T_{cc}^+ \pi^-] &\longrightarrow& \sigma^{(\Lambda)}[T_{cc}^+,\mathrm{no}\,\pi]
\frac{8 G_\pi^2 \mu_{\pi T}^2 \mu_\pi}{3\pi} 
 (2 \mu_{\pi T} E)^{-1/2} .
\label{dsigmaTpi-:large}
\end{eqnarray}
\label{dsigmaTpi:large}%
\end{subequations}
In Fig.~\ref{fig:Xpi-Elarge}, the differential cross section for $T_{cc}^+  \pi^+$ in Eq.~\eqref{dsigmaTpi+Lambda}
divided by $\sigma^{(\Lambda)}[T_{cc}^+,\mathrm{no}\,\pi]$ 
is compared with the asymptotic cross section in Eq.~\eqref{dsigmaTpi+:large} for $\Lambda/m_\pi = 1/2$, 1, and 2.
The height of the triangle-singularity peak depends dramatically on $\Lambda$,
but the curves all approach the asymptotic cross section as $E$ increases.

The cross section $\sigma^{(\Lambda)}[T_{cc}^+,\mathrm{no}\,\pi]$ on the right sides of Eqs.~\eqref{dsigmaTpi:large}
should not be interpreted literally as the cross section for $T_{cc}^+$ without any pion.  
It is actually the cross section for $T_{cc}^+$ without any pion
with relative momentum smaller than the ultraviolet cutoff $q_\mathrm{max}$
used to define the short-distance amplitudes. 
The momentum $q_\mathrm{max}$ is an arbitrary scale separating states described explicitly 
by the effective field theory from states described implicitly through the dependence of 
short-distance amplitudes on $q_\mathrm{max}$.
A pion with relative momentum less than $ q_\mathrm{max}$ is described explicitly.
The effects of pions with relative momentum larger than $q_\mathrm{max}$
must be taken into account through the short-distance amplitudes.
In the case of $T_{cc}^+ \pi$ states,
the corresponding invariant kinetic energy is $E_\mathrm{max} = q_\mathrm{max}^2/(2 \mu_{\pi T})$.
As $E_\mathrm{max}$ is increased, there are $T_{cc}^+$ events with no pion 
that are resolved into $T_{cc}^+ \pi^+$, $T_{cc}^+ \pi^0$, and $T_{cc}^+ \pi^-$ events,
so $\sigma^{(\Lambda)}[T_{cc}^+, \mathrm{no}\, \pi]$ must decrease accordingly.
The sum of $\sigma^{(\Lambda)}[T_{cc}^+, \mathrm{no}\, \pi]$ and the cross sections for  
$T_{cc}^+  \pi^+$, $T_{cc}^+  \pi^0$, and $T_{cc}^+  \pi^-$  
integrated over $E < E_\mathrm{max}$ should  not depend on $E_\mathrm{max}$.
This condition requires short-distance factors of the form $\langle \mathcal{A}_{D^* D} (\mathcal{A}_{D^* D})^*\rangle$
to be multiplied by a factor whose difference from 1  is order $G_\pi^2 \mu_{\pi T} \mu_\pi q_\mathrm{max}$.
Since $G_\pi^2 \mu_{\pi T} \mu_\pi m_\pi$ = 0.027,
the multiplicative factor is close to 1 if $q_\mathrm{max}< m_\pi$.
We have therefore not implemented the multiplicative factors that 
guarantee that cross sections are  independent of $q_\mathrm{max}$.

We would  like quantitative estimates of the integrated cross sections for $T_{cc}^+$ accompanied by a soft pion.
The differential cross sections for $T_{cc}^+  \pi^+$, $T_{cc}^+  \pi^0$, and  $T_{cc}^+  \pi^-$  
in the coupled-channel model are given in Eqs.~\eqref{dsigmaTpiLambda}.
Their high energy limits  in Eqs.~\eqref{dsigmaTpi:large} show that
the cross sections integrated over the energy $E$ up to some maximum $E_\mathrm{max}$
 increase asymptotically as $E_\mathrm{max}^{1/2}$.
As shown in Appendix~\ref{sec:AmpLargeq^2}, this is the correct asymptotic behavior for a general wavefunction.
The cross sections integrated up to an energy $E_\mathrm{max}$ much larger than the limiting triangle-singularity energies 
can be expressed as
\begin{subequations}
\begin{eqnarray}
\sigma\big[  T_{cc}^+\,  \pi^+ \big] &\approx& 
\left(  3.2 \sqrt{ \frac{E_\mathrm{max}}{m_\pi}} - 0.0^{+1.8}_{-1.3}  \right)\times
10^{-2}\,  \sigma^{(\Lambda)}\big[T_{cc}^+,\mathrm{no}\,\pi\big]\, ,
\label{sigmaTpi+}
\\
\sigma\big[  T_{cc}^+\,  \pi^0 \big] &\approx& 
\left(  2.4  \sqrt{ \frac{E_\mathrm{max}}{m_\pi}} - 0.0^{+1.3}_{-1.0}  \right)\times
10^{-2} \, \sigma^{(\Lambda)}\big[T_{cc}^+,\mathrm{no}\,\pi\big]\,  ,
\label{sigmaTpi0}
\\
\sigma\big[  T_{cc}^+\,  \pi^- \big] &\approx& 
\left(  3.2 \sqrt{ \frac{E_\mathrm{max}}{m_\pi}} -1.3^{+0.3}_{-0.5}  \right)\times
10^{-2}\,  \sigma^{(\Lambda)}\big[T_{cc}^+,\mathrm{no}\,\pi\big]\, .
\label{sigmaTpi-}
\end{eqnarray}
\label{sigmaTpi}%
\end{subequations}
The coefficients of $\sqrt{E_\mathrm{max}/m_\pi}$ were determined from the asymptotic behaviors 
of $d\sigma/dE$ in Eqs.~\eqref{dsigmaTpi:large}.
The numerical coefficients with error bars were deduced by fitting the subleading behavior at large $E_\mathrm{max}$
with $\Lambda = 2^{0\pm 1} m_\pi$.

\begin{figure}[t]
\includegraphics*[width=0.8\linewidth]{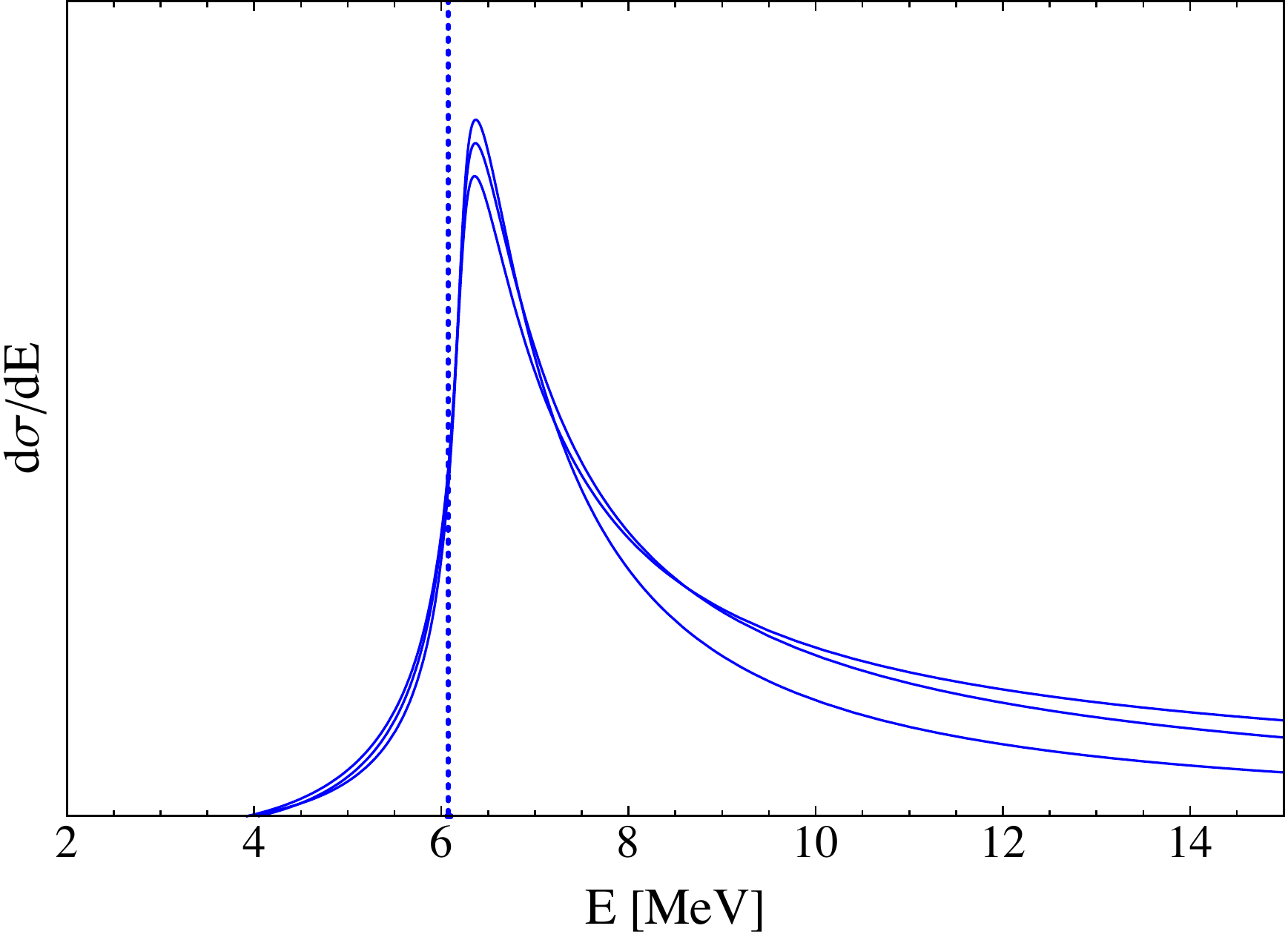} 
\caption{
Difference between the differential cross sections $d\sigma/dE$ 
for $T_{cc}^+\pi^+$ and $T_{cc}^+\pi^-$ in the coupled-channel model as functions of the invariant kinetic energy $E$.
The  binding energy of $T_{cc}^+$ is $|\varepsilon_T| =360$~keV.
The curves were calculated using  Eqs.~\eqref{dsigmaTpi+Lambda} and \eqref{dsigmaTpi-Lambda}
with $\Lambda/m_\pi = 1/2$, 1, and 2 in order of increasing cross sections at small $E$ and at large $E$.
The vertical dotted line is at the limiting triangle-singularity  energy in Eq.~\eqref{Etriangle+limit}.
The scale on the vertical axis is arbitrary.
}
\label{fig:Xpi-ELambda+-}
\end{figure}

Near the triangle-singularity peak in $d\sigma/dE$ for $T_{cc}^+  \pi^+$,
the differential cross section for  $T_{cc}^+  \pi^-$  is much smaller, as is evident in Fig.~\ref{fig:Xpi-ELambda}.
In experimental measurements of $d\sigma/dE$ for $T_{cc}^+  \pi^+$,
subtracting $d\sigma/dE$  for $T_{cc}^+  \pi^-$
would also remove the background from random pions from the $pp$ collision that have nothing to do 
with the creation of charm mesons.
The difference between the cross sections in the coupled-channel model 
in Eqs.~\eqref{dsigmaTpi+Lambda} and \eqref{dsigmaTpi-Lambda}
is shown as a function of $E$ in Fig.~\ref{fig:Xpi-ELambda+-}.
Since the differential cross sections for $T_{cc}^+  \pi^+$ and  $T_{cc}^+  \pi^-$ 
have the same limiting behavior at large $E$, 
the difference between their integrated cross sections is independent of $E_\mathrm{max}$:
\begin{equation}
\sigma\big[  T_{cc}^+\,  \pi^+ \big] - \sigma\big[  T_{cc}^+\,  \pi^- \big] \approx
\left(  1.3^{+1.5}_{-0.8}  \right)\times
10^{-2}\,  \sigma^{(\Lambda)}\big[T_{cc}^+,\mathrm{no}\,\pi\big]\, .
\label{dsigmaTpi+-}
\end{equation}
This difference is dominated by the triangle-singularity peak.
It is roughly compatible with the estimate of  the cross section for $(T_{cc}^+\,  \pi^+)_\triangle$ in Eq.~\eqref{sigmaTpi+tri}, but it has a smaller error bar from varying $\Lambda$.
We can use the difference in Eq.~\eqref{dsigmaTpi+-} as an estimate 
of the contribution to the integrated cross section  for $T_{cc}^+\,  \pi^+$ from the triangle-singularity peak.

\subsection{LHCb data}
\label{sec:sigmaLHC}

The production of $T_{cc}^+$ in $p p$ collisions at the LHC 
has been studied by the LHCb collaboration \cite{LHCb:2021vvq,LHCb:2021auc}.
The $T_{cc}^+$ was observed as a peak in the $D^0D^0\pi^+$ invariant mass distribution
below the $D^{*+} D^0$ threshold.
The number of events in the peak was $117 \pm 16$.
There is also evidence for the decay of $T_{cc}^+$ into $D^+ D^0 \pi^0$
in the form of a narrow peak in the invariant mass distribution for $D^+ D^0$ near its threshold.

At a hadron collider, it is much easier to detect a charged pion than a neutral pion.
Thus the production rates for $T_{cc}^+$ accompanied by a $\pi^+$ or $\pi^-$ can be measured.
The creation of $D^{*+} D^{*+}$ at short distances can produce $T_{cc}^+$ accompanied by a soft $\pi^+$.
The creation of $D^{*0} D^{*0}$ at short distances can produce  $T_{cc}^+$ accompanied by a soft $\pi^-$.
There may also be $T_{cc}^+ \pi^+$ and $T_{cc}^+ \pi^-$ events with random pions produced by the $pp$ collision
that have nothing to do with the creation of charm mesons at short distances.

The inclusive cross section for $T_{cc}^+$ is the sum of the cross section $\sigma^{(\Lambda)}[T_{cc}^+, \mathrm{no}\, \pi]$
for $T_{cc}^+$ without any pion with relative momentum less than $q_\mathrm{max}$
and the cross sections $\sigma[ T_{cc}^+  \pi^+]$, $\sigma[ T_{cc}^+  \pi^0]$, and $\sigma[ T_{cc}^+  \pi^-]$ 
integrated over the invariant kinetic energy up to $E_\mathrm{max} = q_\mathrm{max}^2/(2 \mu_{\pi T})$.
The fraction of $T_{cc}^+$ events accompanied by a soft $\pi^+$ or a soft $\pi^-$
can be estimated using the results in Eqs.~\eqref{sigmaTpi}. 
The fractions of  events having  $T_{cc}^+ \pi^+$ and $T_{cc}^+ \pi^-$ with
invariant kinetic energy less than $m_\pi$ are estimated to be
$(3.0^{+1.5}_{-1.2})$\% and $(1.8^{+0.2}_{-0.4})$\%, respectively.
Our estimates suggest that a few of the $T_{cc}^+$ events observed by the LHCb collaboration 
should be accompanied by a soft $\pi^+$ with relative momentum less than $m_\pi$,
and that there should be a smaller but comparable number accompanied by a soft $\pi^-$. 

The fraction of $T_{cc}^+$ events with a $\pi^+$ in the peak from the triangle singularity
can be estimated using the result in Eq.~\eqref{dsigmaTpi+-}.
Our estimate $(1.2^{+1.3}_{-0.7})$\%  suggests that a few of the $T_{cc}^+$ events 
observed by the LHCb collaboration could have a $\pi^+$ in the peak from the triangle singularity.
While the number of these events is small, they all have
invariant kinetic energy $E$ of $T_{cc}^+  \pi^+$ within 1~MeV of 6.1~MeV.
The creation of charm mesons at short distances should produce essentially no $T_{cc}^+ \pi^-$ events 
in that region of $E$. The production of $T_{cc}^+ \pi^-$ can therefore be used to  
measure the background from $T_{cc}^+ \pi^+$ events with a random $\pi^+$ from the $pp$ collision.


\section{Summary}
\label{sec:Summary}

We have studied the inclusive production of $T_{cc}^+$  at a high-energy hadron collider
through the creation of two charm mesons at short distances.
The formation of $T_{cc}^+$ was described by the effective field theory XEFT.
The $T_{cc}^+$  can be produced by the creation of its constituents $D^{*+} D^0$  at short distances
followed by the binding of the charm mesons into $T_{cc}^+$.
The $T_{cc}^+$  can also be produced by the creation of  spin-1 charm mesons $D^*  D^*$ at short distances
followed by the rescattering of the charm mesons into $T_{cc}^+ \pi$.
The universality of near-threshold S-wave resonances guarantees that there are aspects of the production 
that are determined by the binding momentum $\gamma_T$ of $T_{cc}^+$.
There are also aspects that involve larger momenta comparable to the ultraviolet cutoff $\Lambda$ of XEFT.
Those aspects were studied using a coupled-channel model for the $D^{*+}D^0$ and $D^{*0}D^+$
components of $T_{cc}^+$ with isospin symmetry at short distances.  The coupled-channel model can be
defined by the prescriptions in Eqs.~\eqref{psi-sub} and \eqref{psi0+-sub}.

The $T_{cc}^+$ can be produced without an accompanying soft pion by the creation of $D^{*+} D^0$ at short distances. 
The cross section $ \sigma[T_{cc}^+,\mathrm{no}\,\pi]$ is expressed in
Eq.~\eqref{factor-DstarD-3}
as the product of a short-distance factor and the square $| \psi_T(r\!=\!0)|^2$
of the universal wavefunction at the origin for $T_{cc}^+$.
The  factor $| \psi_T(r\!=\!0)|^2$
is sensitive to the binding energy of $T_{cc}^+$ through a multiplicative factor of $\gamma_T$.
It is more sensitive to the ultraviolet cutoff $\Lambda$, 
scaling approximately as $\Lambda^2$.
The cross section $\sigma^{(\Lambda)}[T_{cc}^+,\mathrm{no}\,\pi]$ in the coupled-channel model 
is given in Eq.~\eqref{sigmaT-CC}.
It differs from the cross section in Eq.~\eqref{factor-DstarD-3} by replacing $| \psi_T(r\!=\!0)|^2$
by $| \psi^{(\Lambda)}_T(r\!=\!0)|^2/(1+Z_{0+})$ 
and multiplying by 2 to take into account the $D^{*0}D^+$ component of $T_{cc}^+$.
The factor $| \psi^{(\Lambda)}_T(r\!=\!0)|^2$ could in principle be determined 
from other reactions involving $T_{cc}^+$,
such as the differential cross section for producing $T_{cc}^+\pi^+$ with large invariant kinetic energy.

The $T_{cc}^+$ can be produced with an accompanying soft $\pi^+$ or $\pi^0$
by the creation of $D^{*+} D^{*+}$ or $D^{*+} D^{*0}$ at short distances, respectively. 
In Eqs.~\eqref{dsigmaTpitri}, the differential cross sections for $T_{cc}^+\pi^+$ and $T_{cc}^+\pi^0$ 
are expressed in a form with the same short-distance factor as in $ \sigma[T_{cc}^+,\mathrm{no}\,\pi]$.
The differential cross sections $d\sigma/dE$ are shown in Fig.~\ref{fig:Xpi-q}
as functions of the invariant kinetic energy $E$ for  $T_{cc}^+\pi$.
They have a  narrow peak from a triangle singularity about 6.1~MeV above the threshold for $T_{cc}^+ \pi^+$
and about 7.3~MeV above the threshold for $T_{cc}^+ \pi^0$.
Since the peak is near the onset of a $D^*D^*$ threshold,
the calculation of the cross section integrated over the peak requires the construction of a smooth background,
such as that shown in Fig.~\ref{fig:Xpi-EX}.
Our results for the cross sections  for $T_{cc}^+\pi^+$ and $T_{cc}^+\pi^0$ 
integrated over the triangle-singularity peaks are given in Eqs.~\eqref{sigmaTpitri}.
The factor $| \psi_T(r\!=\!0)|^2$ in the denominator gives a very large uncertainty.

We used the coupled-channel model 
to calculate the cross sections for $T_{cc}^+\pi^+$, $T_{cc}^+\pi^0$, and  $T_{cc}^+\pi^-$
at energies near the triangle-singularity peaks and at higher energies.
The differential cross sections at $E$ above the triangle-singularity peaks
are sensitive to the momentum scale $\Lambda$, as illustrated in Fig.~\ref{fig:Xpi-ELambda}.
For $E$ well above the peak, the dependence of $d\sigma/dE$
on $\Lambda$ reduces to a multiplicative factor proportional to $|\psi^{(\Lambda)}_T(r\!=\!0)|^2$.
The short-distance factor and the factor  $|\psi^{(\Lambda)}_T(r\!=\!0)|^2$
can be eliminated from $d\sigma/dE$ in favor of  $\sigma^{(\Lambda)}[T_{cc}^+,\mathrm{no}\,\pi]$.
The resulting differential cross sections for $T_{cc}^+\pi^+$, $T_{cc}^+\pi^0$, and $T_{cc}^+\pi^-$  at large $E$
are given in Eqs.~\eqref{dsigmaTpi:large}.
Simple approximations for the cross section integrated over $E$ up to $E_\mathrm{max}$
are given in Eqs.~\eqref{sigmaTpi}.
In the case of  $T_{cc}^+\pi^+$ and $T_{cc}^+\pi^0$,
the subleading term  includes a contribution from the triangle-singularity peak.

The production of $T_{cc}^+$ accompanied by a soft $\pi^+$ can be studied at the LHC,
because the charged pion provides a clean signature.
Our estimate of the fraction of $T_{cc}^+$ events accompanied by a $\pi^+$ 
with invariant kinetic energy less than $m_\pi$ is $(3.0^{+1.5}_{-1.2})$\%.
The LHCb collaboration discovered $T_{cc}^+$ as a peak in the $D^0D^0\pi^+$ invariant mass distribution \cite{LHCb:2021vvq,LHCb:2021auc}.  The number of events in the peak was $117 \pm 16$.
Our estimate  suggests that several of those events 
should be accompanied by an additional $\pi^+$ with relative momentum less than $m_\pi$.
Our estimate for the fraction of $T_{cc}^+$ events 
with $T_{cc}^+ \pi^+$ in the narrow peak from the triangle singularity near $E = 6.1$~MeV is $(1.2^{+1.3}_{-0.7})$\%.
All of these events would be within 1~MeV of the triangle-singularity  energy.
There may be some $T_{cc}^+ \pi^+$ events near that peak
with a random pion from the proton-proton collision that is unrelated to the
creation of charm mesons.  The background from
this contribution can be determined experimentally by measuring $T_{cc}^+ \pi^-$ events.

Our calculation of the peak in the cross section for $T_{cc}^+\pi^+$ 
from a charm-meson triangle singularity is based on the assumption that the charm mesons 
are created at short distances much smaller than the mean radius $\langle r \rangle$ of $T_{cc}^+$.
This assumption is very well justified for the production of  $T_{cc}^+$ from single-parton scattering (SPS).
It is less well justified for the production of  $T_{cc}^+$ from double-parton scattering (DPS),
because the charm mesons may be created at distances comparable to the radius of a proton.
The triangle-singularity peak could stand out more clearly above the background in the contribution from SPS.
The LHCb collaboration has observed a larger yield of  $T_{cc}^+$ relative to $D^0 \bar D^0$
at larger values of the number $N_\mathrm{tracks}$ of tracks in the vertex detector \cite{LHCb:2021auc}.
If the increased yield at larger multiplicity arises from the DPS mechanism,
the restriction to $N_\mathrm{tracks}< 80$ could produce a sample of $T_{cc}^+$ events 
in which a larger fraction comes from the SPS mechanism.
Such a restriction could make the triangle-singularity peak stand out more clearly above the background.

We calculated the cross sections for $T_{cc}^+\pi^+$ and $T_{cc}^+\pi^0$ at low energies near the 
triangle-singularity peaks using XEFT at LO.
The coupled-channel model we used to calculate the cross sections 
for $T_{cc}^+\pi^+$, $T_{cc}^+\pi^0$, and  $T_{cc}^+\pi^-$ at higher energies
agrees with XEFT at LO at low energy
and at high energy it has power-law behavior  compatible with XEFT.
It smoothly connects the amplitudes in the intermediate energy region,
but in this region it is just a model.
It would be worthwhile to
extend the accuracy of our calculations to XEFT at NLO.
This effective field theory  has been applied to decays of the $T_{cc}^+$ at leading order in Refs.~\cite{Yan:2021wdl,Fleming:2021wmk}
and some next-to-leading order corrections were calculated in Ref.~\cite{Yan:2021wdl}.
The calculation of the cross section for $T_{cc}^+\pi^+$ 
near the peak from the triangle singularity in XEFT at 
NLO should be straightforward.
A systematically improvable calculation of the cross section at higher energies is a more challenging problem.

We have discussed the effect of the triangle singularity on the production of $T_{cc}^+$ accompanied by a pion.
The triangle singularity also affects the production of the constituents $D^{*+} D^0$ accompanied by a pion.
This reaction proceeds through the tree diagram in Fig.~\ref{fig:DDtoDDpi} 
and also through the loop diagram obtained from the triangle diagram in Fig.~\ref{fig:DDtoXpi}
by attaching $D^{*+}$ and $D^0$ lines to the outgoing $T_{cc}^+$ line.
The two diagrams produce an interesting interference effect called the {\it Schmid cancellation} \cite{Schmid:1967ojm}.
A convenient choice of Dalitz-plot variables is the invariant mass $s$  of $D^{*+} D^0$
and the  invariant mass $t$  of $D^0 \pi^+$.
The triangle singularity appears along the line $s = s_\triangle$, where $s_\triangle = (M_{*+} + M_0 + E_\triangle)^2$.
In the limit where the charm mesons in the triangle are all on shell,
the differential cross section as a function of $s$ and $t$ has a $\log^2|s-s_\triangle|$ divergence 
along that line for all values of $t$ inside the Dalitz plot.
The Schmid cancellation is that the differential cross section integrated over $t$ 
has only a single-log divergence $\log|s-s_\triangle|$. 
The cancellation can be most easily observed through a local minimum as  a function of $t$
in the differential cross section integrated over the region $s < s_\triangle$ \cite{Braaten:2020iye}.

At energies well above the  triangle singularity energy $E_\triangle$,
the differential cross section $d \sigma/dE$ for producing $T_{cc}^+\pi^+$ is predicted to decrease as $E^{-1/2}$.
This behavior provides a way of discriminating between a loosely bound charm-meson molecule and a compact tetraquark.
A compact tetraquark $T$ would have to have a suppressed coupling to $D^{*+} D^0$;
otherwise the resonant interactions of $D^{*+} D^0$ would transform $T$ into a large charm-meson molecule.
The Goldstone nature of the pion requires the production amplitude of $T\pi$ 
to be proportional to the relative momentum of the pion.
The differential cross section $d \sigma/dE$ should therefore  increase like $E^{3/2}$.
Measurements of the production rate of $T_{cc}^+\pi^+$ at energies well above $E_\triangle$
would therefore provide important clues to the nature of $T_{cc}^+$.

Loosely bound S-wave charm-meson molecules like $X$ and $T_{cc}^+$
have universal properties determined by their binding energies.
One of these properties is a narrow peak from a charm-meson triangle singularity
in the rate for their production accompanied by a pion.
Our estimate of the cross section for $T_{cc}^+ \pi^+$ from the triangle-singularity peak
is large enough to encourage the effort to observe the peak at the LHC.
The observation of such a  peak  would provide strong support for the identification 
of $T_{cc}^+$ as  a loosely bound charm-meson molecule.

\begin{acknowledgments}
This work was supported in part by the U.S.\  Department of Energy under grant DE-SC0011726, 
by the National Natural Science Foundation of China (NSFC) under grant 11905112,
by the Natural Science Foundation of Shandong Province of China under grant ZR2019QA012,
and by NSFC and the Deutsche Forschungsgemeinschaft (DFG)
through the Sino-German Collaborative Research Center TRR110 
“Symmetries and the Emergence of Structure in QCD” (NSFC Grant No.\ 12070131001, DFG Project-ID 196253076 - TRR110).
\end{acknowledgments}


\appendix

\section{Limiting Behavior of Triangle Amplitudes} 
\label{sec:TriAmpLimit}

An analytic expression for the triangle amplitude $T_+(q^2,\gamma^2)$ is given in Eq.~\eqref{Gqfunction},
where $a$, $b$, and $c$ are the coefficients in Eqs.~\eqref{parameterabc}. 
In this Appendix, we give limiting expressions for this triangle amplitude.

The triangle singularity in $T_+(q^2,\gamma^2)$ comes from the logarithm in Eq.~\eqref{Gqfunction}. 
In the simultaneous limits  $\varepsilon_T \to 0$, $\Gamma_{*+} \to 0$,
the triangle singularity is at the real value $q^2_{\triangle +} =(M_T/2 \mu) m \delta_{0+}$.
Near the triangle singularity, the square roots $\sqrt{a}$ and $\sqrt{c}$
in the argument of the logarithm are both comparable to $(\mu/M)q_{\triangle+}$.
This can be made more obvious by expressing the coefficient $a$ in Eq.~\eqref{parametera} in the form
\begin{equation}
a =  (\mu / M)^{2} q^2_{\triangle+} + (\mu/\mu_{\pi})  (q^2 - q^2_{\triangle+}) 
+  M_* (\varepsilon_T + i \, \Gamma_{* +}) .
\label{a-triangle}
\end{equation}
The difference $\sqrt{a}-\sqrt{c}$ can be comparable in magnitude to $\sqrt{a+b+c}=i \gamma$,
which is approximately $i \sqrt{2 \mu |\varepsilon_T|}$.
Near the triangle singularity, $T_+(q^2,\gamma^2)$ can be approximated by
simplifying the coefficient of the logarithm  in Eq.~\eqref{Gqfunction} and the additive term
by setting $q^2=q_{\triangle+}^2$ and taking the limits $\varepsilon_T \to 0$, $\Gamma_{*+} \to 0$:
\begin{equation}
T^{(\mathrm{log})}_+(q^2,\gamma^2)  = \sqrt{\frac{M/M_T}{\mu_{\pi T} \delta_{0+}} }
\left( \frac{2M}{M_*}  \log\frac{\sqrt{a}+ (\mu/M) q + i\, \gamma}{\sqrt{a}- (\mu/M) q  + i\, \gamma} 
+ \frac{m}{M_*}  \right).
\label{Gtriangle:2}
\end{equation}

\begin{figure}[t]
\includegraphics*[width=0.8\linewidth]{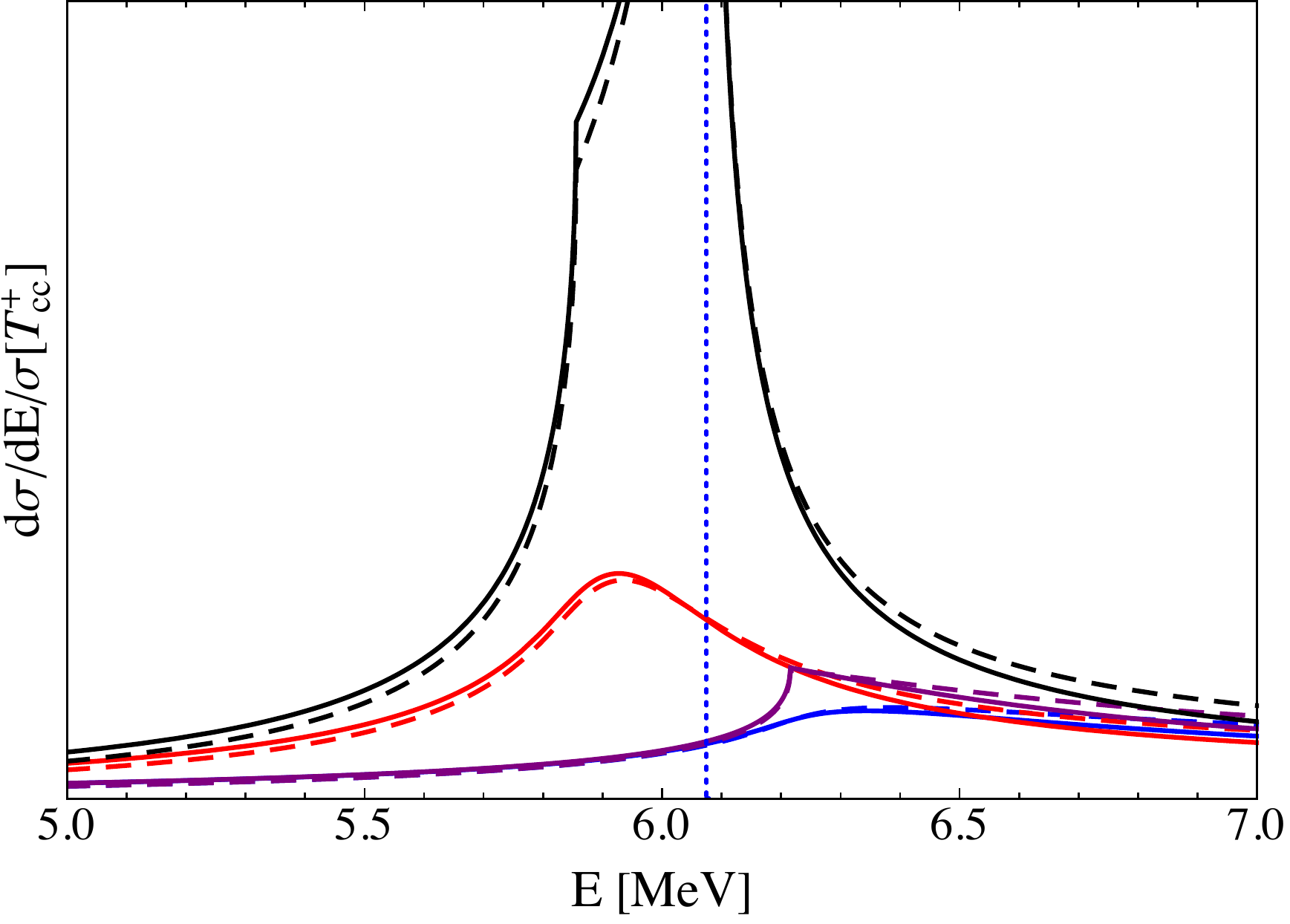} 
\caption{
Differential cross sections $d\sigma/dE$ divided by $\sigma[T_{cc}^+,\mathrm{no}\, \pi]$
as  functions of the invariant kinetic energy $E$ for  $T_{cc}^+\pi^+$. 
The cross sections are calculated using the complete triangle amplitude $T_+(q^2,\gamma^2)$ (solid curves)
and the logarithmic approximation  in Eq.~\eqref{Gtriangle:2} (dashed curves).
The four cases of $(|\varepsilon_T| ,\, \Gamma_{*+} )$ in order of  increasing height of the peak are
(a) (360~keV, 83~keV) (blue curves), (b) (360~keV, 0) (purple curves), 
(c) (0, 83~keV)  (red curves), and (d) (0, 0) (black curves).
The scale on the vertical axis is arbitrary. 
}
\label{fig:Xpi-E}
\end{figure}

The triangle amplitude $T_+(q^2,\gamma^2)$ and the logarithmic approximation  in Eq.~\eqref{Gtriangle:2}
are compared in Fig.~\ref{fig:Xpi-E} 
by showing the differential cross section $d\sigma/dE$ for $T_{cc}^+ \pi^+$
as a function of the invariant energy $E$.
The differential cross section has been divided by $\sigma[T_{cc}^+,\mathrm{no}\, \pi]$
to ensure that the limit as $\varepsilon_T \to 0$ is nonzero.
The logarithmic approximation gives a good fit to the exact curve near the peak 
not only for $|\varepsilon_T| =360$~keV, $\Gamma_{*+}=83$~keV
but also in the limit $\varepsilon_T \to 0$, in the limit $\Gamma_{*+} \to 0$,
and in the simultaneous limits $\varepsilon_T \to 0$, $\Gamma_{*+} \to 0$.
In the limit $\Gamma_{*+} \to 0$,  $d\sigma/dE$ develops a cusp at $E=\delta_{0+} - \varepsilon_T$.
For $|\varepsilon_T| =360$~keV,
the cusp coincides with the peak, as is evident in Fig.~\ref{fig:Xpi-E}.
For $\varepsilon_T = 0$, the cusp at $E=\delta_{0+}$ is
well separated from the $\log^2$ divergence at $E=M_T/(2M) \delta_{0+}$.

The triangle amplitude $T_+(q^2,\gamma^2)$ 
has a logarithmic branch point at the complex triangle-singularity energy $E_{\triangle +}$ in Eq.~\eqref{Etriangle+}
and a square-root branch point at the complex energy $E_+$ in Eq.~\eqref{E+}.
These nearby singularities both approach the real axis in the simultaneous limits $\varepsilon_T \to 0$, $\Gamma_{*+} \to 0$.
The leading behavior of $T_+(q^2,\gamma^2)$ near the square-root and logarithmic  singularities have the forms 
 $A + B \sqrt{E - E_+}$ and $C + D \log(E - E_{\triangle +})$, respectively,
where $A$, $B$, $C$ and $D$ are complex constants that depend on $\varepsilon_T$ and  $\Gamma_{*+}$.
In the simultaneous limits $\varepsilon_T \to 0$,  $\Gamma_{*+} \to 0$, the differential cross section $d\sigma/dE$ 
has a cusp at $E=\delta_{0+}$ and  a $\log^2$ divergence at $E=M_T/(2M) \delta_{0+}$ 
that are both well described by the leading singularities.
If $|\varepsilon_T|$ is increased to 360~keV or $\Gamma_{*+}$ is increased to 83~keV, 
the  peak in $d\sigma/dE$ is not described well by the leading singularities.
The leading square-root singularity gives a cross section that is mononotically increasing.
The leading logarithmic singularity gives a cross section with a peak whose  height is larger by a factor of 3 or more
and whose position is at an energy lower by at least 0.3~MeV  than that from the complete triangle amplitude.
An accurate description of the peak requires the logarithm in  Eq.~\eqref{Gtriangle:2},  which involves an
interplay between the two singularities.

The triangle amplitude $T_+(q^2,\gamma^2)$  at large $q^2$ 
can be expanded in powers of $1/q$. The expansion to next-to-leading order is
\begin{eqnarray}
T_+(q^2,\gamma^2) \longrightarrow \left( \frac{M}{M_*} \log \frac{\sqrt{M_T/m}+1}{\sqrt{M_T/m}-1}
+ \frac{\sqrt{M_T m}}{M_*}  \right) \frac{1}{q} - \frac{2i M_T m \gamma}{M_*^2 q^2}.
\label{Ginfinity}
\end{eqnarray}
The numerical value of the dimensionless prefactor of $1/q$ is 0.724.


\section{Triangle Amplitudes in the Coupled-Channel Model} 
\label{sec:TriAmpCC}

In this appendix, we determine the  triangle amplitudes in the coupled-channel model
using the prescriptions in Eqs.~\eqref{psi-sub} and \eqref{psi0+-sub}.

\subsection{Amplitude for $\bm{T_{cc}^+ \pi^+}$}
\label{sec:AmpLargeq:pi+}

The amplitude for the production of $T_{cc}^+ \pi^+$ from the creation of $D^{*+} D^{*+}$ at short distances
is expressed as a loop integral in  Eq.~\eqref{amplitudeXpi0intk}.
Its reduction to the form in Eq.~\eqref{amplitude-Tpi+} defines the triangle amplitude $T_+(q^2, \gamma^2)$.
The first denominator in the loop integral in Eq.~\eqref{amplitudeXpi0intk} can be identified as the
denominator of the universal wavefunction $\psi_T(k_\mathrm{rel})$ given by  Eq.~\eqref{psi-k},
with $\gamma = \sqrt{2\mu(|\varepsilon_T|- i \Gamma_{*+}/2)}$
and the shifted relative momentum  $\bm{k}_\mathrm{rel} = \bm{k} + (\mu/M)  \bm{q}$. 
The universal wavefunction at the origin is ultraviolet divergent.
The regularized wavefunction $\psi_T^{(\Lambda)}(k_\mathrm{rel})$ given by Eq.~\eqref{psi-k:reg} is a simple model with the same momentum dependence as $\psi_T(k_\mathrm{rel})$ at small $k_\mathrm{rel}$ 
but a finite wavefunction at the origin.
The replacement of $\psi_T(k_\mathrm{rel})$  in the loop integral in Eq.~\eqref{amplitudeXpi0intk}
by $\psi_T^{(\Lambda)}(k_\mathrm{rel})$ can be implemented by making the substitution in Eq.~\eqref{psi-sub}.
The resulting  triangle amplitude for $T_{cc}^+  \pi^+$ in the coupled-channel  model  is
\begin{equation}
T^{(\Lambda)}_+(q^2,\gamma^2) =
 \frac{ \sqrt{(\Lambda+\gamma)\Lambda}}{\sqrt{1+Z_{0+}}\, (\Lambda-\gamma)}
\left[ T_+(q^2,\gamma^2)  - T_+(q^2,\Lambda^2)  \right],
\label{G+Lambda-q}
\end{equation}
where $Z_{0+} = (\Lambda+\gamma)\gamma/[(\Lambda+\gamma_{0+})\gamma_{0+}]$
is the relative probability of the $D^{*0} D^+$ channel.
In the expression for $T_+(q^2,\Lambda^2)$, $\sqrt{a+b+c}$ reduces to $i \Lambda$.

\subsection{Amplitude for $\bm{T_{cc}^+ \pi^-}$}
\label{sec:AmpLargeq:pi-}

Charm mesons $D^{*0} D^{*0}$ created at short distances can rescatter into $T_{cc}^+\pi^-$
through the $D^{*0} D^+$ component of the $T_{cc}^+$ wavefunction. 
The amplitude for  the reaction in XEFT can be represented by the Feynman diagram in Fig.~\ref{fig:DDtoXpi}
with an appropriate $D^{*0} D^+$-to-$T_{cc}^+$ vertex.
If that vertex is taken to be the same as the $D^{*+} D^0$-to-$T_{cc}^+$  vertex in Eq.~\eqref{vertex:TtoD*D}, 
the amplitude for producing $T_{cc}^+\pi^-$ with relative momentum $\bm{q}$ in their CM frame  is
\begin{eqnarray}
\mathcal{A}_{T_{cc}^+\,\pi^- +y} (\bm{q}) 
&=& i \left(\mathcal{A}^{ij}_{D^{*0}   D^{*0}+y} \sqrt{M_T m/M_*^2} \right)
 4\pi G_\pi M_* \sqrt{\gamma_T}  \, \varepsilon^{i*}
\nonumber\\
 && \hspace{1cm}
 \times
  \!\!\int\!\! \frac{d^3k}{(2\pi)^3} 
\frac{1}{\big( \bm{k} + (\mu/M)  \bm{q}\big)^2 +\gamma_{0+}^2}\, 
 \frac{q^j  + (m/M_*) k^j }{\bm k^2  - (\mu/\mu_{\pi}) \bm q^2 + M_*E_- },
\label{amplitudeXpi-intk}
\end{eqnarray}
where  $\bm{\varepsilon}$ is the polarization vector for $T_{cc}^+$.
In the first denominator in the integrand,  $\gamma_{0+}$ is the binding momentum of the $D^{*0} D^+$ channel:
$\gamma_{0+}^2 =  2 \mu(\delta - \varepsilon_T)$, where
$\delta = (M_{*0}\!+\!M_+)-(M_{*+}\!+\!M_0)=1.41$~MeV  is the energy difference 
between the $D^{*0}D^+$ and $D^{*+}D^0$ thresholds.
Since the real part  of that denominator is always greater than $2\mu\delta$, we have omitted its imaginary part. 
In the second denominator,  $E_-$ is the complex energy
\begin{equation}
E_- = \delta+\delta_{+-} -\varepsilon_T - i \, \Gamma_{*0} ,
\label{E-}
\end{equation}
where $\delta_{+-} = M_{*0} \!-\! M_+ \!-\! m_- = -2.38$~MeV.

After evaluating the integral over the loop momentum,
the amplitude for producing $T_{cc}^+ \pi^-$ can be reduced to the form
\begin{eqnarray}
\mathcal{A}_{T_{cc}^+\,\pi^- +y} (\bm{q})
= - G_\pi \sqrt{M_T m\gamma_T/4} \, \mathcal{A}^{ij}_{D^{*0}   D^{*0}+y}  \varepsilon^{i*}q^j \,  T_-(q^2,\gamma_{0+}^2).
\label{amplitude-Tpi-}
\end{eqnarray}
The triangle amplitude $T_-(q^2,\gamma_{0+}^2)$ is given by the right side of Eq.~\eqref{Gqfunction} with the coefficients 
\begin{subequations}
\begin{eqnarray}
a &=&  (\mu/\mu_{\pi})  q^2 -  M_*E_- ,
\label{parametera-}
 \\ 
b &=&- 2 (\mu/\mu_{\pi}) (\mu / M) q^2 + M_*E_-
 - \gamma_{0+}^2, 
\label{parameterb-}
\\ 
c &=& (\mu / M)^{2} q^{2}.
\label{parameterc-}
\end{eqnarray}
\label{parameterabc-}%
\end{subequations}
The square root of their sum is $\sqrt{a+b+c} = i\, \gamma_{0+}$.
The triangle amplitude $T_-(q^2,\gamma_{0+}^2)$ can be obtained from $T_+(q^2,\gamma^2)$
by replacing $\gamma^2$ by $ \gamma_{0+}^2$ and replacing $E_+$ by $E_-$.

The first denominator in the integrand in Eq.~\eqref{amplitudeXpi-intk} can be identified as 
the denominator of the simple wavefunction $\psi_{0+}(k_\mathrm{rel})$ for the $D^{*0}D^+$ component of $T_{cc}^+$
given by Eq.~\eqref{psicc-k}, with $\gamma_\mathrm{cc} = \gamma_{0+}$
and the shifted relative momentum $\bm{k}_\mathrm{rel} = \bm{k} + (\mu/M)  \bm{q}$.
The simple wavefunction at the origin is ultraviolet divergent.
The wavefunction $\psi_{0+}^{(\Lambda)}(k_\mathrm{rel})$ given by Eq.~\eqref{psicc-k:reg}
has the same momentum dependence as $\psi_{0+}(k_\mathrm{rel})$ at small $k_\mathrm{rel}$,
and it has a finite  wavefunction  at the origin that is equal to that for $\psi_T^{(\Lambda)}(k_\mathrm{rel})$.
The replacement of $\psi_{0+}(k_\mathrm{rel})$ by $\psi_{0+}^{(\Lambda)}(k_\mathrm{rel})$
can be implemented by making the substitution in Eq.~\eqref{psi0+-sub}
in the amplitude in Eq.~\eqref{amplitudeXpi-intk}.
The resulting triangle  amplitude in the coupled-channel model is
\begin{eqnarray}
T^{(\Lambda)}_-(q^2,\gamma_{0+}^2) = 
- \frac{ \sqrt{(\Lambda+\gamma)\Lambda}}{\sqrt{1+Z_{0+}}\, (\Lambda-\gamma_{0+})}
\left[ T_-(q^2,\gamma_{0+}^2)  - T_-(q^2,\Lambda^2)  \right].
\label{G-Lambda-q}
\end{eqnarray}

\subsection{Amplitude for $\bm{T_{cc}^+ \pi^0}$}
\label{sec:AmpLargeq:pi0}

The amplitude in XEFT  for the production of $T_{cc}^+\pi^0$ from the creation of $ D^{*+}D^{*0}$ at short distances
can be expressed as a loop integral analogous to that for $T_{cc}^+\pi^+$  in  Eq.~\eqref{amplitudeXpi0intk}.
The amplitude can be reduced  to the expression in Eq.~\eqref{amplitude-Tpi0},
which defines the triangle amplitude $T_0(q^2,\gamma^2)$. 
The first denominator  in  Eq.~\eqref{amplitudeXpi0intk} can be identified as the denominator 
of the universal wavefunction $\psi_T(k_\mathrm{rel})$ for the $D^{*+} D^0$ component of the $T_{cc}^+$ .
The replacement of $\psi_T(k_\mathrm{rel})$  in the loop integral
by the regularized wavefunction $\psi_T^{(\Lambda)}(k_\mathrm{rel})$ 
can be implemented by making the substitution in Eq.~\eqref{psi-sub}.
The resulting contribution to the  triangle amplitude for $T_{cc}^+  \pi^0$ in the coupled-channel  model  
from the $D^{*+} D^0$ component of the $T_{cc}^+$ wavefunction is
\begin{eqnarray}
T^{(\Lambda)}_0(q^2,\gamma^2) =
 \frac{\sqrt{(\Lambda+\gamma)\Lambda}}{\sqrt{1+Z_{0+}}\,( \Lambda-\gamma)}
\left[ T_0(q^2,\gamma^2)  - T_0(q^2,\Lambda^2)  \right] .
\label{G0Lambda-q}
\end{eqnarray}

The production of  $T_{cc}^+\pi^0$ can also proceed by the creation of $D^{*0} D^{*+}$ at short distances
and their rescattering into $T_{cc}^+\pi^0$
through the $D^{*0} D^+$ component of the $T_{cc}^+$ wavefunction. 
The amplitude for  the reaction can be represented by the Feynman diagram in Fig.~\ref{fig:DDtoXpi}
with an appropriate $D^{*0} D^+$-to-$T_{cc}^+$ vertex.
If that vertex is taken to be the same as the $D^{*+} D^0$-to-$T_{cc}^+$  vertex in Eq.~\eqref{vertex:TtoD*D}, 
the contribution $\mathcal{A}^\prime_{T_{cc}^+\pi^0 +y} (\bm{q}) $ to the amplitude 
for producing $T_{cc}^+\pi^0$ with relative momentum $\bm{q}$ in their CM frame
 has a form analogous to that in Eq.~\eqref{amplitudeXpi-intk}.  It can be reduced to
\begin{eqnarray}
\mathcal{A}^\prime_{T_{cc}^+\pi^0+y} (\bm{q})
= G_\pi \sqrt{M_T m\gamma_T/8} \, \mathcal{A}^{ij}_{D^{*0}   D^{*+}+y} \varepsilon^{i*} q^j \,  T_0(q^2,\gamma_{0+}^2).
\label{amplitude’-Tpi-}
\end{eqnarray}
The triangle amplitude $T_0(q^2,\gamma_{0+}^2)$ can be obtained from $T_0(q^2,\gamma^2)$
by replacing $\gamma^2$ by $\gamma_{0+}^2$.

The first denominator in the integrand analogous to that in Eq.~\eqref{amplitudeXpi-intk} can be identified as 
the denominator of the simple wavefunction $\psi_{0+}(k_\mathrm{rel})$ for the $D^{*0} D^+$ component of $T_{cc}^+$.
The replacement of $\psi_{0+}(k_\mathrm{rel})$ by $\psi_{0+}^{(\Lambda)}(k_\mathrm{rel})$
can be implemented by making the substitution in Eq.~\eqref{psi0+-sub}.
The  triangle amplitude for $T_{cc}^+  \pi^0$ in the coupled-channel  model  
from the $D^{*0} D^+$ component of the $T_{cc}^+$ wavefunction is
\begin{eqnarray}
T^{\prime (\Lambda)}_0(q^2,\gamma_{0+}^2) = -
 \frac{\sqrt{(\Lambda+\gamma)\Lambda}}{\sqrt{1+Z_{0+}}\, (\Lambda-\gamma_{0+})}
\left[ T_0(q^2,\gamma_{0+}^2)  - T_0(q^2,\Lambda^2)  \right].
\label{G0’Lambda-q}
\end{eqnarray}
It can be obtained from $T^{(\Lambda)}_0(q^2,\gamma^2)$ in Eq.~\eqref{G0Lambda-q}
by replacing $T_0(q^2,\gamma^2)$ by $T_0(q^2,\gamma_{0+}^2)$,
replacing $1/(\Lambda-\gamma)$ by $1/(\Lambda-\gamma_{0+})$, and multiplying by an overall minus sign.

\subsection{Large $q^2$}
\label{sec:AmpLargeq^2}

The behavior of the triangle amplitude $T^\mathrm{(\Lambda)}_+(q^2,\gamma^2)$
defined in Eq.~\eqref{G+Lambda-q} at large $q^2$ can be determined by inserting the asymptotic result
for $T_+(q^2,\gamma^2)$ in Eq.~\eqref{Ginfinity}.
The subtraction cancels the terms that decrease as $1/q$,
so the triangle amplitude decreases as $1/q^2$:
\begin{equation}
T^\mathrm{(\Lambda)}_+(q^2,\gamma^2) \longrightarrow 
i\, \frac{2 \sqrt{(\Lambda+\gamma)\Lambda}\, M_T m}{\sqrt{1+Z_{0+}}\, M_*^2 \,q^2} .
\label{Greg-large}
\end{equation}
This can be expressed in a form with a factor of the regularized wavefunction at the origin 
$\psi_T^{(\Lambda)}(r\!=\!0)$ given by Eq.~\eqref{psi-0:reg}: 
\begin{equation}
T^{(\Lambda)}_+(q^2,\gamma^2) \longrightarrow 
i \frac{4 \mu_{\pi T}}{M_*  \sqrt{\gamma_T/2\pi}}  \frac{\psi_T^{(\Lambda)}(r\!=\!0)}{\sqrt{1+Z_{0+}}} \frac{1}{q^2} .
\label{Greg-largepsi}
\end{equation}
We verify below that this gives the large-$q^2$ limit 
for a general wavefunction with a finite wavefunction at the origin.

\begin{figure}[t]
\includegraphics*[width=0.4\linewidth]{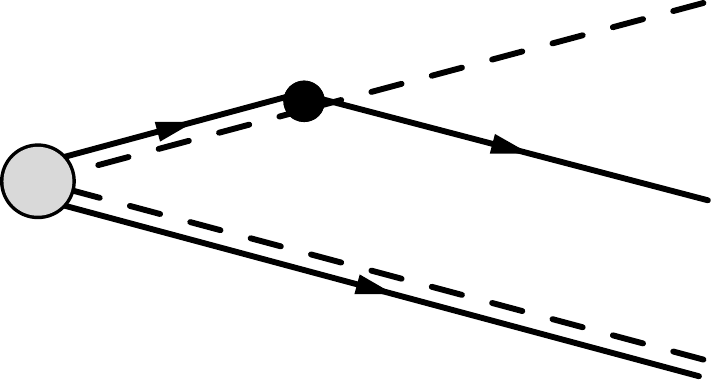} 
\caption{
Feynman diagram in XEFT for $D^{*+} D^{*+}$ created at a point to produce $D^{*+} D^0  \pi^+$.
}
\label{fig:DDtoDDpi}
\end{figure}

The triangle amplitude $T_+(q^2,\gamma^2)$ for $T_{cc}^+ \pi^+$ in Eq.~\eqref{Gqfunction}
was derived from the loop diagram  in Fig.~\ref{fig:DDtoXpi}.
An alternative expression for $T_+(q^2,\gamma^2)$ can be derived from the tree diagram 
for the production of $D^{*+} D^0 \pi^+$ in Fig.~\ref{fig:DDtoDDpi}
along with the wavefunction $\psi(k)$ for the $T_{cc}^+$ bound state.
We take the  momentum of $\pi^+$ in the $D^{*+} D^0 \pi^+$ CM frame to be $\bm{q}$.
We take the  relative momentum of $D^{*+} D^0$ in their CM frame to be $\bm{k}$.
The momentum of $D^{*+}$ and $D^0$ in the $D^{*+} D^0 \pi^+$ CM frame are then 
$-(M_*/M_T) \bm{q} + \bm{k}$ and $-(M/M_T) \bm{q} - \bm{k}$.
The amplitude for producing $T_{cc}^+ \pi^+$ with polarization vector $\bm{\varepsilon}$  plus additional particles $y$ 
can be obtained from the amplitude for producing $D^{*+} D^0 \pi^+$ by multiplying it by the wavefunction $\psi(k)$ 
and integrating over  $\bm{k}$:
\begin{eqnarray}
\mathcal{A}_{T_{cc}^+\,\pi^+ +y} (\bm{q})
&=& i\left(\mathcal{A}^{ij}_{D^{*+}   D^{*+}+y} \sqrt{M_T m/M_*^2} \right) 
\sqrt{8\pi}\, G_\pi \mu_{\pi T} \varepsilon^{i*}
\nonumber\\
 && \hspace{0cm}
 \times   \!\!\int\!\! \frac{d^3k}{(2\pi)^3} \psi(k)
\frac{q^j  + (m/2 \mu) k^j }
{ \big(\bm{q}  + (m/2\mu) \bm{k} \big)^2 - (m/2\mu)M_T(\delta_{0+} - i \, \Gamma_{*+}/2) }.
\label{amplitudeXpiintpsi}
\end{eqnarray}
A similar expression involving the universal wavefunction in Eq.~\eqref{psi-k} can be obtained from 
Eq.~\eqref{amplitudeXpi0intk} by making the momentum shift $\bm{k}   \to \bm{k}  - (\mu/M) \bm{q}$.
The terms in the denominator of the integrand proportional to $\bm{q}^2$, $\bm{q}\!\cdot \!\bm{k}$, and  $\delta_{0+}$
agree with those in Eq.~\eqref{amplitudeXpiintpsi}.
The terms proportional to $\bm{k}^2$, $\varepsilon_T$, and $\Gamma_{*+}$ have different coefficients.
The amplitude therefore agrees with Eq.~\eqref{amplitudeXpiintpsi} through next-to-leading order 
in the expansion in power of $1/q$.

We now consider the amplitude in Eq.~\eqref{amplitudeXpiintpsi} at large $q^2$.
We assume $\psi(k)$ decreases rapidly enough 
 for $k$ beyond some momentum scale $\Lambda$ that its integral over $\bm{k}$ converges.
We take $q^2$ to be much larger than $\Lambda^2$ and also much larger than $m \delta_{0+}$.
In that case, we can take the limit $\bm{k} \to 0$ in the denominator of Eq.~\eqref{amplitudeXpiintpsi}
and in the pion emission factor.
The amplitude reduces to 
\begin{equation}
\mathcal{A}_{T_{cc}^+\,\pi^+ +y} (\bm{q})
\longrightarrow i \left(\mathcal{A}^{ij}_{D^{*+}   D^{*+}+y}\sqrt{M_T m/M_*^2} \right) 
\varepsilon^{i*}  q^j \frac{\sqrt{8\pi}\, G_\pi \mu_{\pi T}}{q^2} \, \psi(r\!=\!0).
\label{amplitudeXpiintpsilarge}
\end{equation}
By comparing this to Eq.~\eqref{amplitude-Tpi+},
we can verify that the large-$q^2$ limit of $T_+(q^2,\gamma^2)$ is given up to a sign by Eq.~\eqref{Greg-largepsi}
with $\psi_T^{(\Lambda)}(r\!=\!0)/\sqrt{1+Z_{0+}}$ replaced by $\psi(r\!=\!0)$.





\begin{thebibliography}{99}
\bibitem{Guo:2017jvc} 
  F.K.~Guo, C.~Hanhart, U.G.~Mei{\ss}ner, Q.~Wang, Q.~Zhao and B.S.~Zou,
Hadronic molecules,
  Rev.\ Mod.\ Phys.\  {\bf 90}, 015004 (2018)
  [arXiv:1705.00141].
  
\bibitem{Ali:2017jda} 
  A.~Ali, J.S.~Lange and S.~Stone,
Exotics: Heavy Pentaquarks and Tetraquarks,
  Prog.\ Part.\ Nucl.\ Phys.\  {\bf 97}, 123 (2017)
  [arXiv:1706.00610].
  
\bibitem{Olsen:2017bmm} 
  S.L.~Olsen, T.~Skwarnicki and D.~Zieminska,
Nonstandard heavy mesons and baryons: Experimental evidence,
  Rev.\ Mod.\ Phys.\  {\bf 90}, 015003 (2018)
  [arXiv:1708.04012].

\bibitem{Karliner:2017qhf} 
  M.~Karliner, J.L.~Rosner and T.~Skwarnicki,
Multiquark States,
  Ann.\ Rev.\ Nucl.\ Part.\ Sci.\  {\bf 68}, 17 (2018)
  [arXiv:1711.10626].
    
\bibitem{Yuan:2018inv} 
  C.Z.~Yuan,
The $XYZ$ states revisited,
  Int.\ J.\ Mod.\ Phys.\ A {\bf 33}, 1830018 (2018)
  [arXiv:1808.01570].
  
\bibitem{Liu:2019zoy}
Y.~R.~Liu, H.~X.~Chen, W.~Chen, X.~Liu and S.~L.~Zhu,
Pentaquark and Tetraquark states,
Prog. Part. Nucl. Phys. \textbf{107}, 237-320 (2019)
[arXiv:1903.11976].

\bibitem{Brambilla:2019esw}
N.~Brambilla, S.~Eidelman, C.~Hanhart, A.~Nefediev, C.P.~Shen, C.E.~Thomas, A.~Vairo and C.Z.~Yuan,
The $XYZ$ states: experimental and theoretical status and perspectives,
Phys.\ Rept.\ \textbf{873}, 1-154 (2020)
[arXiv:1907.07583].

\bibitem{Choi:2003ue} 
  S.K.~Choi {\it et al.} [Belle Collaboration],
Observation of a narrow charmonium-like state in exclusive $B^\pm \to K^\pm \pi^+ \pi^-  J/\psi$ decays,
  Phys.\ Rev.\ Lett.\  {\bf 91}, 262001 (2003)
  [hep-ex/0309032].

\bibitem{LHCb:2021vvq}
R.~Aaij \textit{et al.} [LHCb],
Observation of an exotic narrow doubly charmed tetraquark, Nat. Phys. {\bf 18}, 751–754 (2022)
[arXiv:2109.01038].


\bibitem{LHCb:2021auc}
R.~Aaij \textit{et al.} [LHCb],
Study of the doubly charmed tetraquark $T_{cc}^+$,
Nature Commun. \textbf{13}, no.1, 3351 (2022)
[arXiv:2109.01056].

\bibitem{E598:1974sol}
J.J.~Aubert \textit{et al.} [E598],
Experimental Observation of a Heavy Particle $J$,
Phys.\ Rev.\ Lett.\ \textbf{33}, 1404-1406 (1974).

\bibitem{SLAC-SP-017:1974ind}
J.E.~Augustin \textit{et al.} [SLAC-SP-017],
Discovery of a Narrow Resonance in $e^+ e^-$ Annihilation,
Phys.\ Rev.\ Lett.\ \textbf{33}, 1406-1408 (1974).

\bibitem{Gell-Mann:1964ewy}
M.~Gell-Mann,
A Schematic Model of Baryons and Mesons,
Phys.\ Lett.\ \textbf{8}, 214 (1964).

\bibitem{Zweig:1964}
G.~Zweig,
An SU(3) model for strong interaction symmetry and its breaking. Version 1,
CERN-TH-401;
An SU(3) model for strong interaction symmetry and its breaking. Version 2,
CERN-TH-412.

\bibitem{Aaij:2020qga}
R.~Aaij \textit{et al.} [LHCb],
Study of the lineshape of the $\chi_{c1}(3872)$ state,
Phys.\ Rev.\ D \textbf{102}, 092005 (2020)
[arXiv:2005.13419].

\bibitem{Aaij:2020xjx}
R.~Aaij \textit{et al.} [LHCb],
Study of the $\psi_2(3823)$ and $\chi_{c1}(3872)$ states in $B^+ \rightarrow \left( J\psi\pi^+\pi^-\right)K^+$ decays,
JHEP \textbf{08}, 123 (2020)
[arXiv:2005.13422].

\bibitem{Aaij:2013zoa} 
  R.~Aaij {\it et al.} [LHCb Collaboration],
Determination of the $X(3872)$ meson quantum numbers,
  Phys.\ Rev.\ Lett.\  {\bf 110}, 222001 (2013)
  [arXiv:1302.6269].
  
\bibitem{Braaten:2004rn} 
  E.~Braaten and H.-W.~Hammer,
Universality in few-body systems with large scattering length,
  Phys.\ Rept.\  {\bf 428}, 259 (2006)
  [cond-mat/0410417].

\bibitem{Braaten:2003he}
E.~Braaten and M.~Kusunoki,
Low-energy universality and the new charmonium resonance at 3870~MeV,
Phys.\ Rev.\ D \textbf{69}, 074005 (2004)
[arXiv:hep-ph/0311147].

\bibitem{Meng:2021jnw}
L.~Meng, G.J.~Wang, B.~Wang and S.L.~Zhu,
Probing the long-range structure of the $T_{cc}^+$ with the strong and electromagnetic decays,
Phys.\ Rev.\ D \textbf{104}, L051502 (2021)
[arXiv:2107.14784].

\bibitem{Ling:2021bir}
X.Z.~Ling, M.Z.~Liu, L.S.~Geng, E.~Wang and J.J.~Xie,
Can we understand the decay width of the $T_{cc}^+$ state?,
Phys.\ Lett.\ B \textbf{826}, 136897 (2022)
[arXiv:2108.00947].

\bibitem{Yan:2021wdl}
M.J.~Yan and M.P.~Valderrama,
Subleading contributions to the decay width of the $T_{cc}^+$ tetraquark,
Phys.\ Rev.\ D \textbf{105}, 014007 (2022)
[arXiv:2108.04785].

\bibitem{Fleming:2021wmk}
S.~Fleming, R.~Hodges and T.~Mehen,
$T_{cc}^+$ decays: differential spectra and two-body final states,
Phys.\ Rev.\ D \textbf{104}, 116010 (2021)
[arXiv:2109.02188].

\bibitem{Ren:2021dsi}
H.~Ren, F.~Wu and R.~Zhu,
Hadronic molecule interpretation of $T^+_{cc}$ and its beauty-partners,
Adv. High Energy Phys. \textbf{2022}, 9103031 (2022)
[arXiv:2109.02531].

\bibitem{Feijoo:2021ppq}
A.~Feijoo, W.~H.~Liang and E.~Oset,
$D^0 D^0 \pi^+$ mass distribution in the production of the $T_{cc}$ exotic state,
Phys.\ Rev.\ D \textbf{104}, 114015 (2021)
[arXiv:2108.02730].

\bibitem{Dai:2021wxi}
L.Y.~Dai, X.~Sun, X.W.~Kang, A.P.~Szczepaniak and J.S.~Yu,
Pole analysis on the doubly charmed meson in $D^0D^0\pi^+$ mass spectrum,
Phys. Rev. D \textbf{105}, L051507 (2022)
[arXiv:2108.06002].



\bibitem{Albaladejo:2021vln}
M.~Albaladejo,
$T_{cc}^{+}$ coupled channel analysis and predictions,
Phys. Lett. B \textbf{829}, 137052 (2022)
[arXiv:2110.02944].


\bibitem{Du:2021zzh}
M.L.~Du, V.~Baru, X.K.~Dong, A.~Filin, F.K.~Guo, C.~Hanhart, A.~Nefediev, J.~Nieves and Q.~Wang,
Coupled-channel approach to $T_{cc}^+$ including three-body effects,
Phys. Rev. D \textbf{105}, 014024 (2022)
[arXiv:2110.13765].

\bibitem{Qin:2020zlg}
Q.~Qin, Y.~F.~Shen and F.~S.~Yu,
Discovery potentials of double-charm tetraquarks,
Chin. Phys. C \textbf{45}, 103106 (2021)
[arXiv:2008.08026].

\bibitem{Jin:2021cxj}
Y.~Jin, S.Y.~Li, Y.R.~Liu, Q.~Qin, Z.G.~Si and F.S.~Yu,
Colour and baryon number fluctuation of preconfinement system in production process and $T_{cc}$ structure,
Phys.\ Rev.\ D \textbf{104}, 114009 (2021)
[arXiv:2109.05678].


\bibitem{Cho:2013rpa}
S.~Cho and S.~H.~Lee,
Hadronic effects on the X(3872) meson abundance in heavy ion collisions,
Phys. Rev. C \textbf{88}, 054901 (2013)
[arXiv:1302.6381].

\bibitem{MartinezTorres:2014son}
A.~Martinez Torres, K.~P.~Khemchandani, F.~S.~Navarra, M.~Nielsen and L.~M.~Abreu,
On $X(3872)$ production in high energy heavy ion collisions,
Phys. Rev. D \textbf{90}, 114023 (2014)
[erratum: Phys. Rev. D \textbf{93}, 059902 (2016)]
[arXiv:1405.7583].

\bibitem{Abreu:2016qci}
L.~M.~Abreu, K.~P.~Khemchandani, A.~Martinez Torres, F.~S.~Navarra and M.~Nielsen,
$X(3872)$ production and absorption in a hot hadron gas,
Phys. Lett. B \textbf{761}, 303-309 (2016)
[arXiv:1604.07716].

\bibitem{Zhang:2020dwn}
H.~Zhang, J.~Liao, E.~Wang, Q.~Wang and H.~Xing,
Deciphering the Nature of X(3872) in Heavy Ion Collisions,
Phys. Rev. Lett. \textbf{126}, 012301 (2021)
[arXiv:2004.00024].

\bibitem{Wu:2020zbx}
B.~Wu, X.~Du, M.~Sibila and R.~Rapp,
$X(3872)$ transport in heavy-ion collisions,
Eur. Phys. J. A \textbf{57}, 122 (2021)
[erratum: Eur. Phys. J. A \textbf{57}, 314 (2021)]
[arXiv:2006.09945].

\bibitem{Chen:2021akx}
B.~Chen, L.~Jiang, X.~H.~Liu, Y.~Liu and J.~Zhao,
X(3872) Production in Relativistic Heavy-Ion Collisions,
[arXiv:2107.00969].

\bibitem{Hong:2018mpk}
J.~Hong, S.~Cho, T.~Song and S.~H.~Lee,
Hadronic effects on the $cc\bar{q}\bar{q}$ tetraquark state in relativistic heavy ion collisions,
Phys. Rev. C \textbf{98}, 014913 (2018)
[arXiv:1804.05336].

\bibitem{Fontoura:2019opw}
C.~E.~Fontoura, G.~Krein, A.~Valcarce and J.~Vijande,
Production of exotic tetraquarks $QQ\bar{q}\bar{q}$ in heavy-ion collisions at the LHC,
Phys. Rev. D \textbf{99}, 094037 (2019)
[arXiv:1905.03877].


\bibitem{Hu:2021gdg}
Y.~Hu, J.~Liao, E.~Wang, Q.~Wang, H.~Xing and H.~Zhang,
The production of doubly charmed exotic hadrons in heavy ion collisions,
Phys.\ Rev.\ D \textbf{104},  L111502 (2021)
[arXiv:2109.07733].



\bibitem{Abreu:2021jwm}
L.~M.~Abreu, F.~S.~Navarra, M.~Nielsen and H.~P.~L.~Vieira,
Interactions of the doubly charmed state $T_{cc}^+$ with a hadronic medium,
Eur. Phys. J. C \textbf{82}, 296 (2022)
[arXiv:2110.11145].

\bibitem{Karplus:1958zz}
R.~Karplus, C.M.~Sommerfield and E.H.~Wichmann,
Spectral Representations in Perturbation Theory. 1. Vertex Function,
Phys.\ Rev.\ \textbf{111}, 1187 (1958).

\bibitem{Landau:1959}
L.D.~Landau, 
On analytic properties of vertex parts in quantum field theory,
Nuclear Physics {\bf 13}, 181 (1959).

\bibitem{Szczepaniak:2015eza} 
  A.P.~Szczepaniak,
Triangle Singularities and $XYZ$ Quarkonium Peaks,
  Phys.\ Lett.\ B {\bf 747}, 410 (2015)
  [arXiv:1501.01691].
 
\bibitem{Liu:2015taa} 
  X.H.~Liu, M.~Oka and Q.~Zhao,
Searching for observable effects induced by anomalous triangle singularities,
  Phys.\ Lett.\ B {\bf 753}, 297 (2016)
  [arXiv:1507.01674].
  
\bibitem{Braaten:2019yua}
E.~Braaten, L.-P.~He and K.~Ingles,
Production of $X(3872)$ Accompanied by a Pion in $B$ Meson Decay,
Phys.\ Rev.\ D \textbf{100}, 074028 (2019)
[arXiv:1902.03259].

\bibitem{Braaten:2019sxh}
E.~Braaten, L.-P.~He and K.~Ingles,
Production of $X(3872)$ Accompanied by a Soft Pion at Hadron Colliders,
Phys.\ Rev.\ D \textbf{100}, 094006 (2019)
[arXiv:1903.04355].

\bibitem{Guo:2019qcn}
F.K.~Guo,
Novel Method for Precisely Measuring the $X(3872)$ Mass,
Phys.\ Rev.\ Lett.\ \textbf{122}, 202002 (2019)
[arXiv:1902.11221].

\bibitem{Sakai:2020ucu}
S.~Sakai, E.~Oset and F.K.~Guo,
Triangle singularity in the $B^-\to K^-\pi^0X(3872)$ reaction and sensitivity to the $X(3872)$ mass,
Phys.\ Rev.\ D \textbf{101}, 054030 (2020)
[arXiv:2002.03160].

\bibitem{Sakai:2020crh}
S.~Sakai, H.J.~Jing and F.K.~Guo,
Possible precise measurements of the $X(3872)$ mass with the $e^+e^-\to\pi^0\gamma X(3872)$ and $p\bar p\to\gamma X(3872)$ reactions,
Phys.\ Rev.\ D \textbf{102}, 114041 (2020)
[arXiv:2008.10829].

\bibitem{Dubynskiy:2006cj}
S.~Dubynskiy and M.B.~Voloshin,
$e^+ e^- \to \gamma X(3872)$ near the $D^* \bar D^*$ threshold,
Phys.\ Rev.\ D \textbf{74}, 094017 (2006)
[arXiv:hep-ph/0609302].

\bibitem{Braaten:2019gfj}
E.~Braaten, L.-P.~He and K.~Ingles,
Triangle Singularity in the Production of $X(3872$) and a Photon in $e^+e^-$ Annihilation,
Phys.\ Rev.\ D \textbf{100}, 031501 (2019)
[arXiv:1904.12915].

\bibitem{Braaten:2019gwc}
E.~Braaten, L.-P.~He and K.~Ingles,
Production of $X(3872)$ and a Photon in $e^+e^-$ Annihilation,
Phys.\ Rev.\ D \textbf{101}, 014021 (2020)
[arXiv:1909.03901].

\bibitem{Molina:2020kyu}
R.~Molina and E.~Oset,
Triangle singularity in $B^-\rightarrow K^-X(3872);X\rightarrow \pi ^0\pi ^+\pi ^-$ and the X(3872) mass,
Eur.\ Phys.\ J.\ C \textbf{80}, 451 (2020)
[arXiv:2002.12821].

\bibitem{Nakamura:2019nwd}
S.X.~Nakamura,
Triangle singularity appearing as an $X(3872)$-like peak in $B\to (J/\psi\pi^+\pi^-) K\pi$,
Phys.\ Rev.\ D \textbf{102}, 074004 (2020)
[arXiv:1912.11830].

\bibitem{Braaten:2020iye}
E.~Braaten, L.-P.~He, K.~Ingles and J.~Jiang,
Charm-meson triangle singularity in ${e^+e^-}$ annihilation into ${ D^{*0} \bar{D}^0 + \gamma }$,
Phys.\ Rev.\ D \textbf{101}, 096020 (2020)
[arXiv:2004.12841].

\bibitem{Suzuki:2005ha}
M.~Suzuki,
The X(3872) boson: Molecule or charmonium,
Phys.\ Rev.\ D \textbf{72}, 114013 (2005)
[arXiv:hep-ph/0508258].

  
\bibitem{Fleming:2007rp} 
  S.~Fleming, M.~Kusunoki, T.~Mehen and U.~van Kolck,
Pion interactions in the $X(3872)$,
  Phys.\ Rev.\ D {\bf 76}, 034006 (2007)
  [hep-ph/0703168].


\bibitem{Braaten:2015tga} 
  E.~Braaten,
Galilean-invariant effective field theory for the $X(3872)$,
  Phys.\ Rev.\ D {\bf 91},  114007 (2015)
  [arXiv:1503.04791]. 

\bibitem{Braaten:2020nmc}
E.~Braaten, L.-P.~He and J.~Jiang,
Galilean-invariant effective field theory for the $X(3872)$ at next-to-leading order,
Phys.\ Rev.\ D \textbf{103}, 036014 (2021)
[arXiv:2010.05801].

\bibitem{Braaten:2010mg} 
  E.~Braaten, H.-W.~Hammer and T.~Mehen,
Scattering of an Ultrasoft Pion and the $X(3872)$,
  Phys.\ Rev.\ D {\bf 82}, 034018 (2010)
  [arXiv:1005.1688].

\bibitem{Butenschoen:2013pxa}
M.~Butenschoen, Z.~G.~He and B.~A.~Kniehl,
NLO NRQCD disfavors the interpretation of X(3872) as $\chi_{c1}(2P)$,
Phys.\ Rev.\ D \textbf{88}, 011501 (2013)
[arXiv:1303.6524].

\bibitem{Meng:2013gga}
C.~Meng, H.~Han and K.T.~Chao,
X(3872) and its production at hadron colliders,
Phys.\ Rev.\ D \textbf{96}, 074014 (2017)
[arXiv:1304.6710].

\bibitem{Butenschoen:2019npa}
M.~Butenschoen, Z.G.~He and B.A.~Kniehl,
Deciphering the $X(3872)$ via its polarization in prompt production at the CERN LHC,
Phys.\ Rev.\ Lett.\ \textbf{123}, 032001 (2019)
[arXiv:1906.08553].

\bibitem{Bignamini:2009sk}
C.~Bignamini, B.~Grinstein, F.~Piccinini, A.D.~Polosa and C.~Sabelli,
Is the X(3872) Production Cross Section at Tevatron Compatible with a Hadron Molecule Interpretation?,
Phys.\ Rev.\ Lett.\ \textbf{103}, 162001 (2009)
[arXiv:0906.0882].

\bibitem{Artoisenet:2009wk}
P.~Artoisenet and E.~Braaten,
Production of the X(3872) at the Tevatron and the LHC,
Phys.\ Rev.\ D \textbf{81}, 114018 (2010)
[arXiv:0911.2016].

\bibitem{Albaladejo:2017blx}
M.~Albaladejo, F.K.~Guo, C.~Hanhart, U.G.~Mei\ss{}ner, J.~Nieves, A.~Nogga and Z.~Yang,
Note on X(3872) production at hadron colliders and its molecular structure,
Chin.\ Phys.\ C \textbf{41}, 121001 (2017)
[arXiv:1709.09101].

\bibitem{Braaten:2018eov}
E.~Braaten, L.~P.~He and K.~Ingles,
Estimates of the $X(3872)$ Cross Section at a Hadron Collider,
Phys. Rev. D \textbf{100}, 094024 (2019)
[arXiv:1811.08876].

\bibitem{Esposito:2020ywk}
A.~Esposito, E.G.~Ferreiro, A.~Pilloni, A.D.~Polosa and C.A.~Salgado,
The nature of $X(3872)$ from high-multiplicity $pp$ collisions,
Eur.\ Phys.\ J.\ C \textbf{81}, 669 (2021)
[arXiv:2006.15044].

\bibitem{LHCb:2020sey}
R.~Aaij \textit{et al.} [LHCb],
Observation of Multiplicity Dependent Prompt $\chi_{c1}(3872)$ and $\psi(2S)$ Production in $pp$ Collisions,
Phys.\ Rev.\ Lett.\ \textbf{126}, 092001 (2021)
[arXiv:2009.06619].


\bibitem{Braaten:2020iqw}
E.~Braaten, L-P.~He, K.~Ingles and J.~Jiang,
Production of $X(3872)$ at High Multiplicity,
Phys.\ Rev.\ D \textbf{103}, L071901 (2021)
[arXiv:2012.13499].

\bibitem{Braaten:2005jj} 
  E.~Braaten and M.~Kusunoki,
Factorization in the production and decay of the $X(3872)$,
  Phys.\ Rev.\ D {\bf 72}, 014012 (2005)
  [hep-ph/0506087].
  
\bibitem{Schmid:1967ojm}
C.~Schmid,
Final-State Interactions and the Simulation of Resonances,
Phys.\ Rev.\ \textbf{154}, 1363 (1967).

\end{thebibliography}
\end{document}